\documentclass[aoas, reqno]{imsart}
\RequirePackage[OT1]{fontenc}
\usepackage{algorithm}
\usepackage{algorithmicx}
\usepackage{algpseudocode}
\usepackage{bm}
\usepackage{extarrows}
\usepackage{fix-cm}
\usepackage{indentfirst}
\usepackage{graphicx}
\usepackage{listings}
\usepackage{multirow}
\usepackage{pdfpages}
\usepackage{subfigure}
\usepackage{todonotes}

\usepackage{enumitem}

\RequirePackage{amsthm,amsmath,amsfonts,amssymb}
\RequirePackage[authoryear]{natbib}

\renewcommand{\vec}{\bm}

\newcommand\cT{\mathcal{T}}

\newcommand\bR{\mathbb{R}}

\newcommand\bE{\mathbb{E}}
\newcommand\robs{r_{\text{obs}}}

\makeatletter
\newenvironment{breakablealgorithm}
  {
  \begin{center}
    \refstepcounter{algorithm}
    \hrule height.8pt depth0pt \kern2pt
    \renewcommand{\caption}[2][\relax]{
      {\raggedright\textbf{\ALG@name~\thealgorithm} ##2\par}%
      \ifx\relax##1\relax 
        \addcontentsline{loa}{algorithm}{\protect\numberline{\thealgorithm}##2}%
      \else 
        \addcontentsline{loa}{algorithm}{\protect\numberline{\thealgorithm}##1}%
      \fi
      \kern2pt\hrule\kern2pt
    }
  }{
    \kern2pt\hrule\relax
  \end{center}
  }
\makeatother
\newtheorem{theorem}{Theorem}[section]

\begin{document}

\begin{frontmatter}
    \title{A Statistical Approach to Estimating Adsorption-Isotherm Parameters in Gradient-Elution Preparative Liquid Chromatography}
    \runtitle{Statistical Approach for An Inverse Chromatography Problem}

\begin{aug}
    \author[A]{\fnms{Jiaji} \snm{Su}},
    \author[B]{\fnms{Zhigang} \snm{Yao}\ead[label=e1,mark]{zhigang.yao@nus.edu.sg or zhigang.yao@cmsa.fas.harvard.edu}}
    \author[A]{\fnms{Cheng} \snm{Li}},
    \and
    \author[C]{\fnms{Ye} \snm{Zhang}} 
    \address[A]{Department of Statistics and Data Science, National University of Singapore, 117546 Singapore.}
    \address[B]{Department of Statistics and Data Science, National University of Singapore, 117546 Singapore and Center of Mathematical Sciences and Applications, Harvard University, 02138 Cambridge USA.\\ Corresponding author: \printead{e1}}
    \address[C]{School of Mathematics and Statistics, Beijing Institute of Technology, 100081 Beijing China and Shenzhen MSU-BIT University, 518172 Shenzhen China}

\end{aug}

\begin{abstract}
    Determining the adsorption isotherms is an issue of significant importance in preparative chromatography. A modern technique for estimating adsorption isotherms is to solve an inverse problem so that the simulated batch separation coincides with actual experimental results.
    However, due to the ill-posedness, the high non-linearity, and the uncertainty quantification of the corresponding physical model, the existing deterministic inversion methods are usually inefficient in real-world applications. To overcome these difficulties and study the uncertainties of the adsorption-isotherm parameters, in this work, based on the Bayesian sampling framework, we propose a statistical approach for estimating the adsorption isotherms in various chromatography systems. Two modified Markov chain Monte Carlo algorithms are developed for a numerical realization of our statistical approach. Numerical experiments with both synthetic and real data are conducted and described to show the efficiency of the proposed new method.
    \end{abstract}
    

\end{frontmatter}

\section{Introduction}
\label{intro}

The separation and purification of mixtures are important processes in many fields, including the fine-chemical, biomedical, pharmaceutical, food, and environmental industries. A popular technique for high qualified separation and purification is liquid chromatography, whose operating principle can be roughly described as follows: a mixed sample is injected into a fluid stream that is pumped through a pipe filled with small porous beads. These beads slow down the components traveling through the column because the components adsorb on the surface of the beads. Since the components adsorb at different rates, they will travel at different speeds through the column and exit the column at different times, and thereby be separated from each other. The foundational monograph by~\cite{GuiochonLin-03} provides a detailed theoretical analysis of preparative liquid chromatography.

In preparative chromatography, the \emph{Adsorption Isotherm} can be considered the most important quantity as it can be used to calculate the specific surface area of materials, mean size of deposited particles, pore size or particle size distribution, etc. Loosely speaking, the adsorption isotherm describes the dependence of the amount of adsorbed substance on the partial pressure of the solution concentration at a given temperature. It can be viewed as an intrinsic physical quantity of a given chromatography system. Various approaches exist for estimating the adsorption isotherm in practice, and can be classified into two categories: experimental methods and computational methods. Most traditional methods, such as frontal analysis (\cite{Lisec-01}) and perturbation peak (\cite{DoseJacobson-91}), belong to the category of experimental methods. They are usually conducted in a series of programmed concentration steps, each step resulting in a so-called breakthrough front giving one point on the adsorption-isotherm curve. Hence, to reduce the experimental costs, over the last two decades, numerous computational methods have been developed for efficiently estimating the adsorption isotherm in various chromatography systems (see, for example, \cite{FelingerZhou-03,ForssenArnell-06,zhang2016regularization,zhang2016,ChengJIIP2017}). These types of methods are used to numerically estimate adsorption-isotherm parameters so that the simulated batch separation coincides with the actual experimental results. Most of such computational approaches require only a few injections of different sample concentrations, and so solute consumption and time requirements are very modest. The mathematical fundamentals of most computational methods are based on the numerical estimation of adsorption-isotherm parameters by solving an inverse problem in Partial Differential Equations (PDEs). However, to the best of our knowledge, all existing approaches belong to the class of deterministic models. Hence, the main goal of this paper is to develop a probabilistic model to estimate the adsorption-isotherm parameters, which will constitute a methodological contribution to the field of chromatography.

It should be noted that the complex structure (highly non-linear) of parameter-to-measurement mapping makes the corresponding optimization problem highly non-convex. Consequently, the global optimal estimator of the adsorption-isotherm parameters cannot be efficiently obtained through conventional optimization solvers. As with the fitting of Gaussian-mixture models, the estimation of adsorption-isotherm parameters is hindered by multiple global solutions, and optimization algorithms are valid only under certain constraints. Therefore, this paper proposes a hybrid method of optimization and sampling to estimate the adsorption-isotherm parameters, and the algorithm is shown to be valid on the Gaussian-mixture data, Gamma-mixture data and the data for the experimental gradient-elution preparative liquid chromatography. We adopt a Bayesian approach and model the solution path obtained from the numerical solver with additive white noise. The framework provides an uncertainty quantification for the model parameters using draws of model parameters from the Bayesian posterior distribution based on modified Markov chain Monte Carlo (MCMC) algorithms.

Bayesian modeling of PDEs and complex dynamic systems has already been studied in some previous literature, though their focus is mostly on the estimation and uncertainty quantification for the PDE solution instead of the finite-dimensional system parameters. For example, \citet{Xunetal13} uses B-splines in a Bayesian hierarchical model to estimate the solution of PDEs. Their model assumption is different from ours since the PDEs in our problem have an existing numerical solver while theirs do not. \citet{Chketal16} proposes a systematic Bayesian calibration method to estimate both the PDE solution and the model parameters, characterizing the uncertainty from both the discretization error in numerical solvers and the random error in observed data. We refer readers to \citet{Cocetal19} for a detailed review on Bayesian methods on PDEs. Although the framework considered in \citet{Chketal16} is very general, their main focus is on using the Gaussian process prior to measure the discretization error in the PDE solution. In contrast, in our chromatography problem, the primary interest is in estimating the finite-dimensional system parameters rather than the PDE solution itself, and the main challenge comes from the lack of identification in these parameters due to the Gaussian-mixture-like solution paths. This unique nonidentification issue cannot be addressed by the generic Bayesian MCMC Algorithm 2 in \citet{Chketal16}. Instead, to decouple the strong dependence between model parameters, we propose a dimension-reduction strategy based on the knowledge of mixture-alike solution paths and then incorporate either a gradient descent or a Langevin dynamics subroutine into the MCMC algorithm, leading to significantly improved mixing of the posterior samples.

Section 2 describes the gradient-elution preparative liquid chromatography model from mathematical and statistical perspectives, while in Section 3 a statistical approach is developed to estimate the adsorption-isotherm parameters. Numerical simulations with synthetic data are presented in Section 4 to demonstrate the robustness of the proposed method. Finally, an experimental gradient data set is tested in Section 5, and a short discussion and concluding remarks are provided in Section 6.

\section{Modeling of preparative liquid chromatography}
In this section, we briefly review the mathematical models used in this paper for chromatographic processes and provide a parameter-to-measurement mapping of the considered inverse problem of estimating adsorption isotherms. Following this, a statistical model with spatial noise terms and corresponding notations is introduced. Finally, we visualize the structure of the target function in optimization, to illustrate the multi-solution and correlation in parameters intuitively.

\subsection{Mathematical background}
\label{Section:MathProblem}
Without loss of generality, and for the convenience of readers who are interested only in our statistical approach that can be used for other types of real-world problems, we consider the following mass-balance equation of a two-component system for a fixed-bed chromatography column with the Danckwerts boundary condition, which is the most commonly used one for column chromatography (\cite{Ruthven-1984,GuiochonLin-03,Horvath-1988,Javeed-2011,Lin2017,zhang2016}).
\begin{equation}
  \left\{\begin{array}{ll}
    \frac{\partial C_{i}}{\partial t}+F \frac{\partial q_{i}}{\partial t}+u \frac{\partial C_{i}}{\partial x}=D_{a} \frac{\partial^{2} C_{i}}{\partial x^{2}}, & x \in \mathcal{X} \equiv[0, L], t \in (0,T] \\
    C_{i}(x, 0)=g_{i}(x), & x \in \mathcal{X}, t=0 \\
    u C_{i}(0, t)-D_{a} \frac{\partial C_{i}(0, t)}{\partial x}=u h_{i}(t), & x=0, t \in (0,T] \\
    D_{a} \frac{\partial C_{i}(L, t)}{\partial x}=0, & x=L, t \in (0,T]
    \end{array}\right.,
    \label{eq:solver}
\end{equation}
where $x$ is distance, $t$ is time, and $i = 1,\,2$ refers to the two components. $C$ and $q$ are the concentrations in the mobile and stationary phases, respectively, $u$ is the mobile phase velocity, and $F$ is the stationary-to-mobile phase ratio. $D_{a}$ is the diffusion parameter. $L$ is the length of the chromatographic column, and $T$ is an appropriate time point slightly larger than the dead time of chromatographic time $T_0 = L/u$. In this paper, we set $T = 1.5 T_0$. In addition, $g(x)$ is the initial condition and $h(t)$ is the boundary condition, which describes the injection profile in the experiment. We adopt a simplified model here for ease of illustration; a more detailed version of gradient-elution liquid chromatography, with further discussion, can be found in Section S1 of the supplementary material.

Throughout this paper, we focus on the case in which the adsorption-energy distribution is bimodal. In this case, the bi-Langmuir adsorption isotherm is usually adopted as follows:
\begin{eqnarray}\label{eq:q}
  q_{1}\left(C_{1}, C_{2}\right)=\frac{a_{I, 1} C_{1}}{1+b_{I, 1} C_{1}+b_{I, 2} C_{2}}+\frac{a_{I I, 1} C_{1}}{1+b_{I I, 1} C_{1}+b_{I I, 2} C_{2}}, \\
  q_{2}\left(C_{1}, C_{2}\right)=\frac{a_{I, 2} C_{2}}{1+b_{I, 1} C_{1}+b_{I, 2} C_{2}}+\frac{a_{I I, 2} C_{2}}{1+b_{I I, 1} C_{1}+b_{I I, 2} C_{2}}. \notag
\end{eqnarray}
where subscripts I and II refer to two adsorption sites with different levels of adsorption energy.

In this paper, the collection of adsorption-isotherm parameters is denoted by
\begin{equation}\label{eq:parameters}
\vec\xi = (a_{I,1},a_{II,1},b_{I,1},b_{II,1}, a_{I,2},a_{II,2}, b_{I,2},b_{II,2})^T.
\end{equation}
Now, we consider the measurement-data structure. In most laboratory and industry environments, the total response $R(\vec \xi,t)$ is observed at the column outlet $x=L$ with
\begin{equation}\label{eq:data}
R(\vec \xi,t) =\sum_{i=1}^2 C_i(L,t),
\end{equation}
where $C(x,t)$ is the solution to problem (\ref{eq:solver}) with the bi-Langmuir adsorption-isotherm model (\ref{eq:q}), and $C_i(L,t)$ represents the concentration of the $i$-th component at the outlet $x=L$. The parameter-to-measurement map $\mathcal{A}$: $\mathbb{R}^{8} \to L^2(\mathcal{T})$ can be expressed as
\begin{equation}\label{eq:operator}
\mathcal{A} (\vec \xi) = R(\vec \xi,t),
\end{equation}
where the model operator $\mathcal{A}$ is defined through (\ref{eq:data}). To be more precise, for a given parameter $\vec\xi$, a bi-Langmuir adsorption-isotherm model can be constructed according to (\ref{eq:q}). Then, the concentration in mobile, i.e. $C$, can be obtained by solving PDE (\ref{eq:solver}). Finally, the experimental data can be collected by using the designed sensor with the physical law (\ref{eq:data}). The aim of this paper is to estimate adsorption-isotherm parameters $\vec\xi$ from the time series database $R (\vec\xi,t)$ and the integrated mathematical model (\ref{eq:operator}) via a statistical approach.

\subsection{A statistical model}
\label{Section:StatModel}
In a liquid-chromatography experiment, a sampler brings the sample mixture into the mobile-phase stream, which carries it into a column, and pumps deliver the desired flow and composition of the mobile phase through the column. The detector located at the end of the column records a signal proportional to the amount of sample component emerging from the column, for time period $\cT = [0,T]$. The signal recorded is the observation of interest.

To build up a statistical model, let $\vec r=( r(t_1),\cdots, r(t_n))^T$ be the observation points of the experiment, collected at discrete time points $\cT_n=\{t_1,\cdots,t_n\}\subseteq\cT$. The liquid-chromatography data measured at time $t$ can be modeled by
\begin{equation}
    r(t)=R(\vec\xi, t) + \epsilon(t),~\, t\in \cT_n,
    \label{eq:StatMod}
\end{equation}
where $\epsilon(t)$ represents the measurement noise. The clean liquid-chromatography data are $R(\vec\xi, t)=\bE[r(t)]$, where $\{C_i(L,t;\,\vec\xi):~i=1, \,2\}$ is the solution of a system of differential equations with static parameters $\vec\xi$; $\vec\xi = (a_{I,1},a_{II,1},b_{I,1},b_{II,1}, a_{I,2},a_{II,2}, b_{I,2},b_{II,2})$ represents the parameter of interest.

Let $\vec R(\vec \xi) = (R(\vec\xi, t_1),\cdots,R(\vec\xi, t_n))^T$ be the collection of the exact chromatography signals with parameter $\vec \xi$. Then, the general framework of the chromatography measurement of $\vec r$ is represented by
\begin{equation}
  \vec r = \vec R(\vec \xi) + \vec \epsilon.
\end{equation}
$\vec \epsilon = (\epsilon(t_1),\cdots,\epsilon(t_n))^T$ stands for the measurement noise, and for simplicity we assume the noises $\epsilon(t)$ are uncorrelated between every pair, with zero mean and identical variances $\sigma^2_\epsilon$.

Throughout this paper, the framework is based on a single observation, but it can be generalized to a case with multiple observations or multiple injection groups. Assuming there is a single observation $\vec \robs = \vec R(\vec \xi^*)+\vec\epsilon^*$ with fixed parameter $\vec \xi^*$, the aim of the paper is to estimate $\vec \xi$ through the posterior distribution as described in the following sections, with all the other parameters apart from $\vec\Theta = \{\vec \xi,\sigma^2_\epsilon\}$ assumed to be known.

\subsection{Difficulties in optimization}
\label{Section:PreStudy}
A natural idea is to estimate the adsorption-isotherm parameters with optimization methods. For example, one may think the minimizer of the $L^2$ distance between $\vec\robs$ and $\vec R (\vec{\xi})$ with respect to $\vec{\xi}$ as a good estimator. However, the non-smooth loss function and the non-unique global minimum prevent us from obtaining valid estimates by optimization.

\begin{figure}[htbp]
  \subfigure[]{
    \includegraphics[width=.47\textwidth]{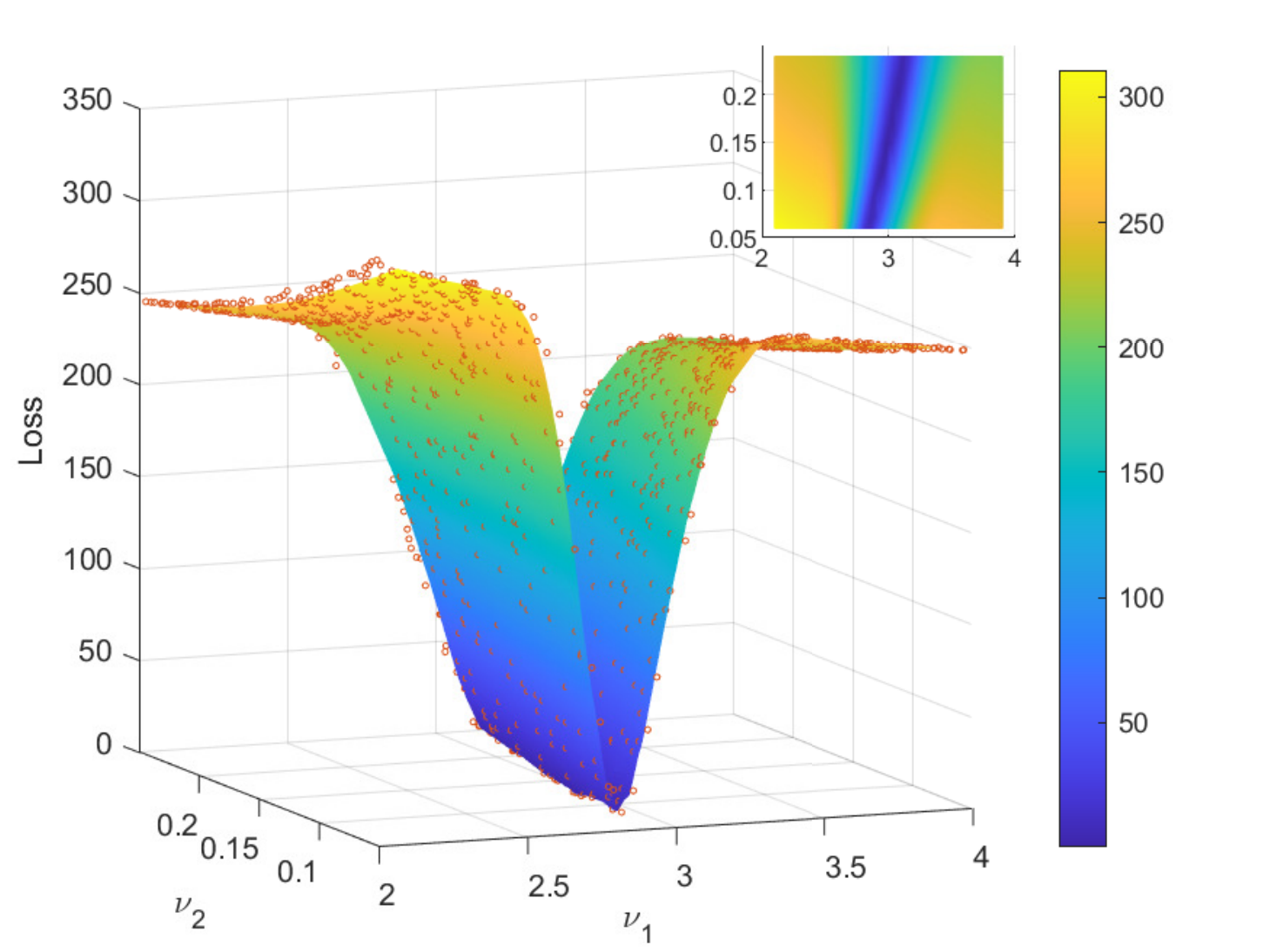}
  }
  \subfigure[]{
    \includegraphics[width=.47\textwidth]{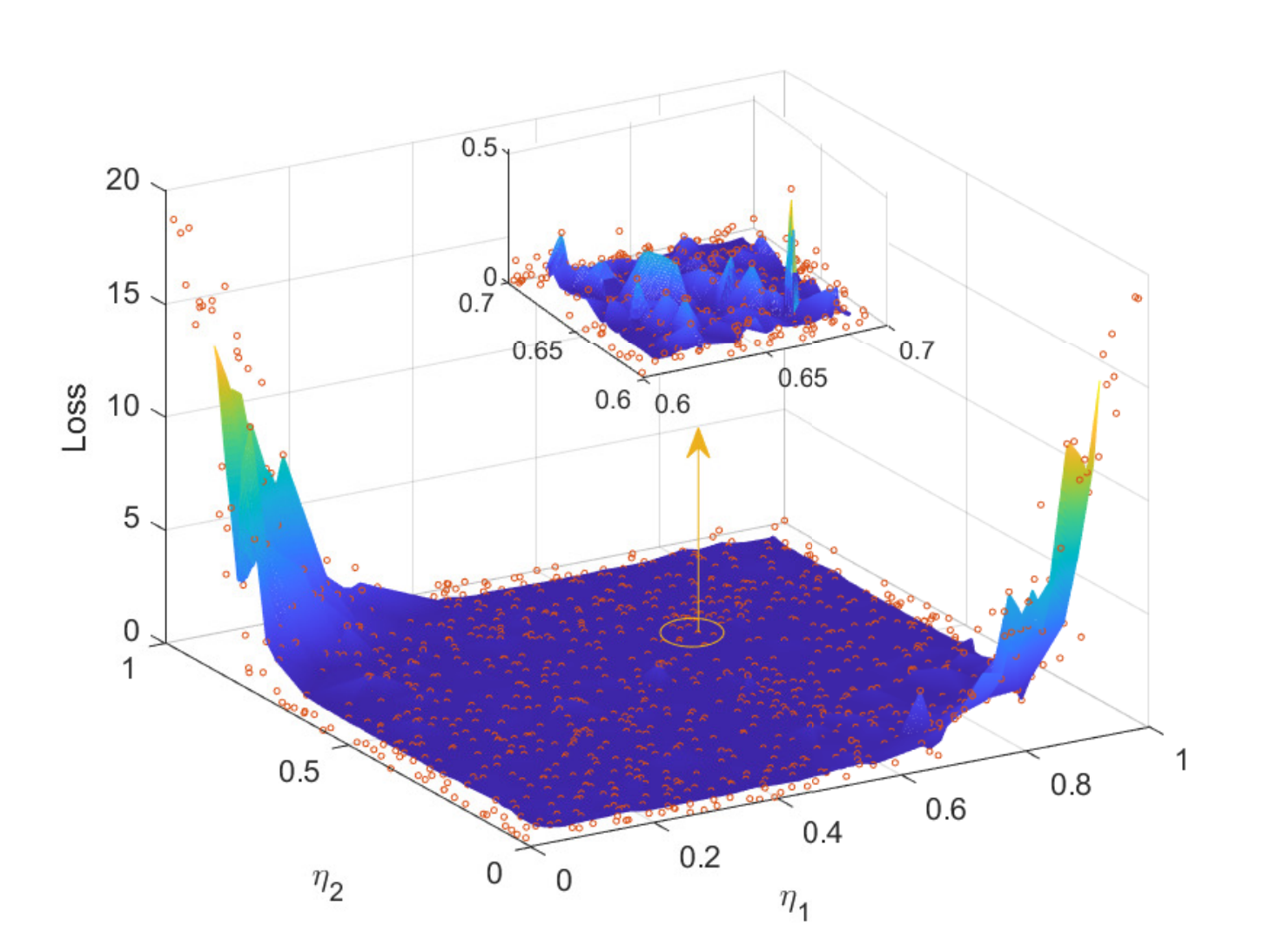}
  }
  \caption{The structure of the loss function $\mathcal{L}(\vec\eta,\vec\nu)$ for the chromatography system. (a) The $L^2$ distance between $\vec\robs$ and $\vec R (g(\vec\eta,\vec\nu))$, where $\vec{\eta}$ is set to be the truth and $\vec\nu$ is sampled in a rectangular area containing the truth. (b) The $L^2$ distance between $\vec\robs$ and $\vec R (g(\vec\eta,\vec{\hat\nu}(\vec \eta)))$, where $\vec{\eta}\in[0,1]^2$ is sampled uniformly and $\vec{\hat\nu}(\vec \eta)=\arg\min_{\vec\nu}\mathcal{L}(\vec\eta,\vec\nu)$ is calculated with gradient descent for each $\vec{\eta}$.}
  \label{fig:RealData_LossStucture}
\end{figure}

\begin{figure}[htbp]
  \subfigure[]{
    \includegraphics[width=.47\textwidth]{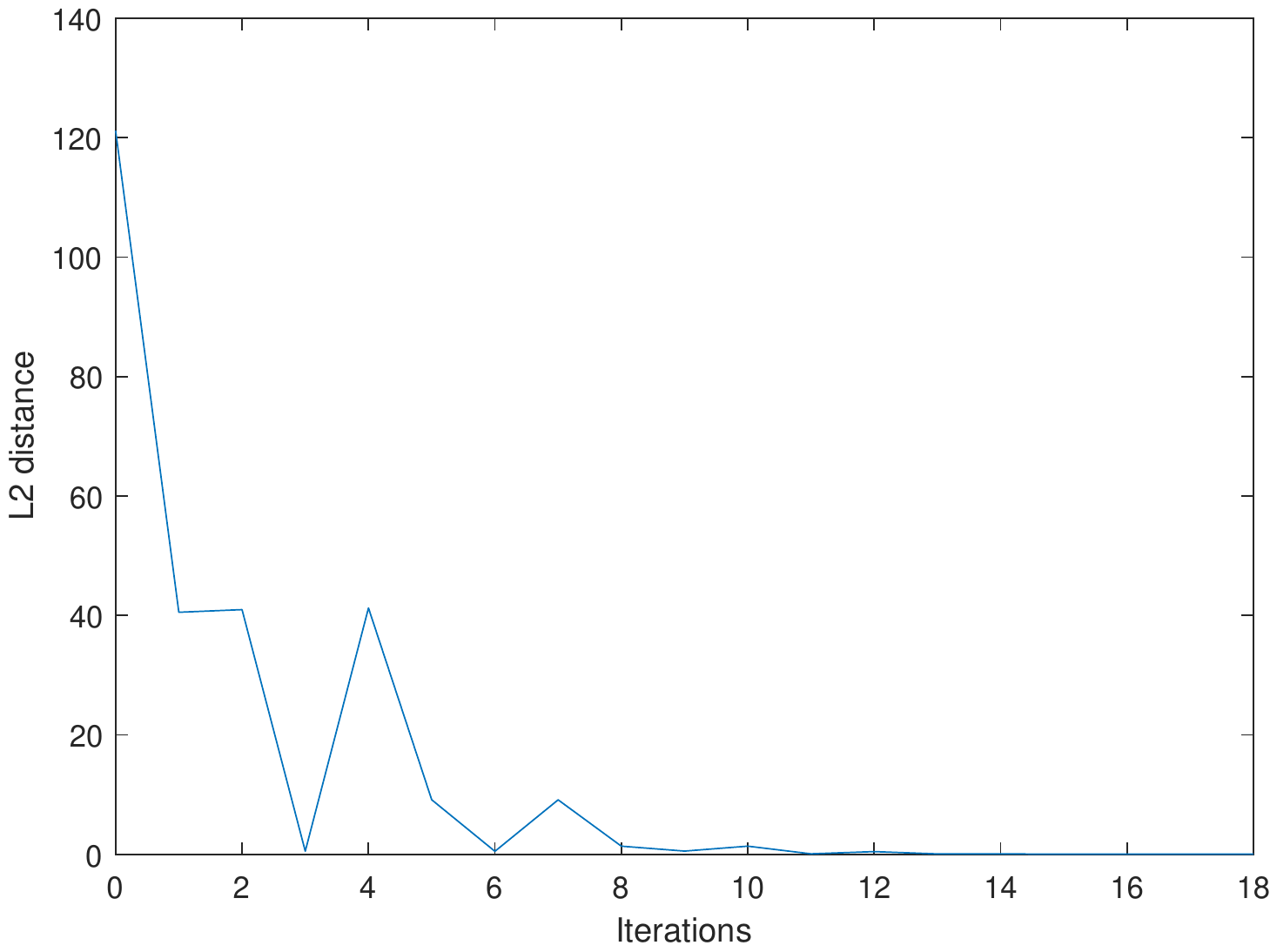}
  }
  \subfigure[]{
    \includegraphics[width=.47\textwidth]{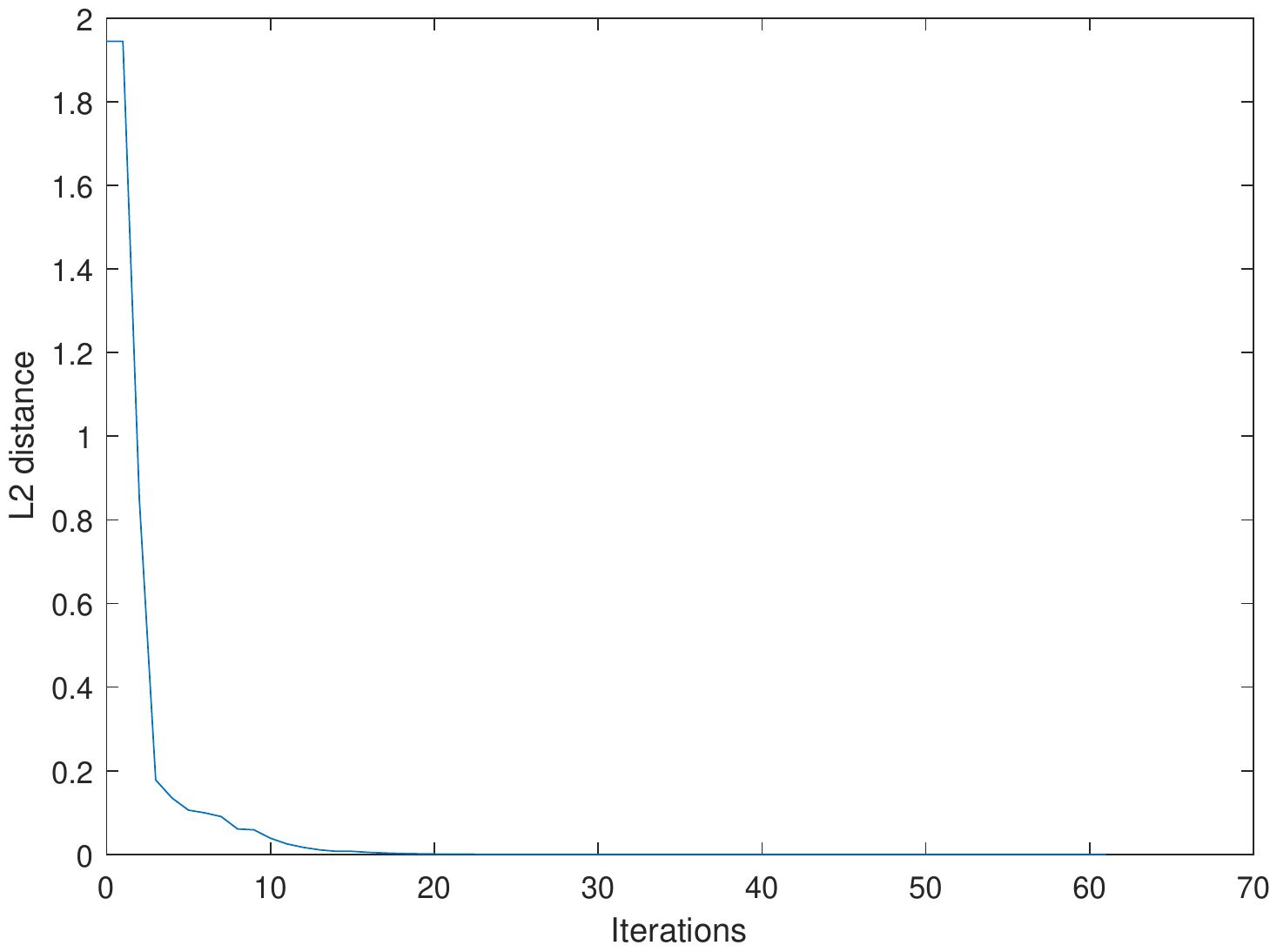}
  }
  \subfigure[]{
    \includegraphics[width=.47\textwidth]{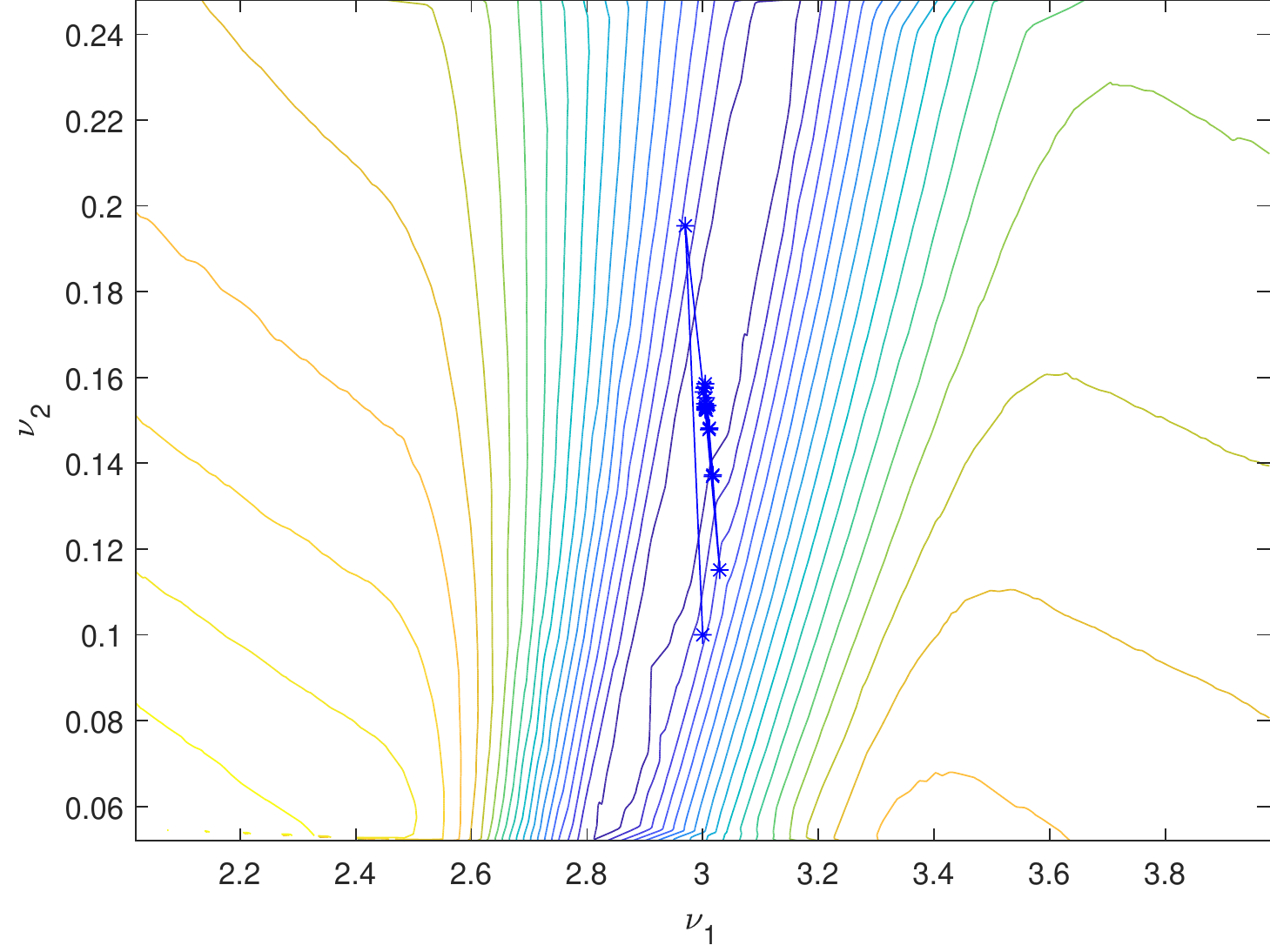}
  }
  \subfigure[]{
    \includegraphics[width=.47\textwidth]{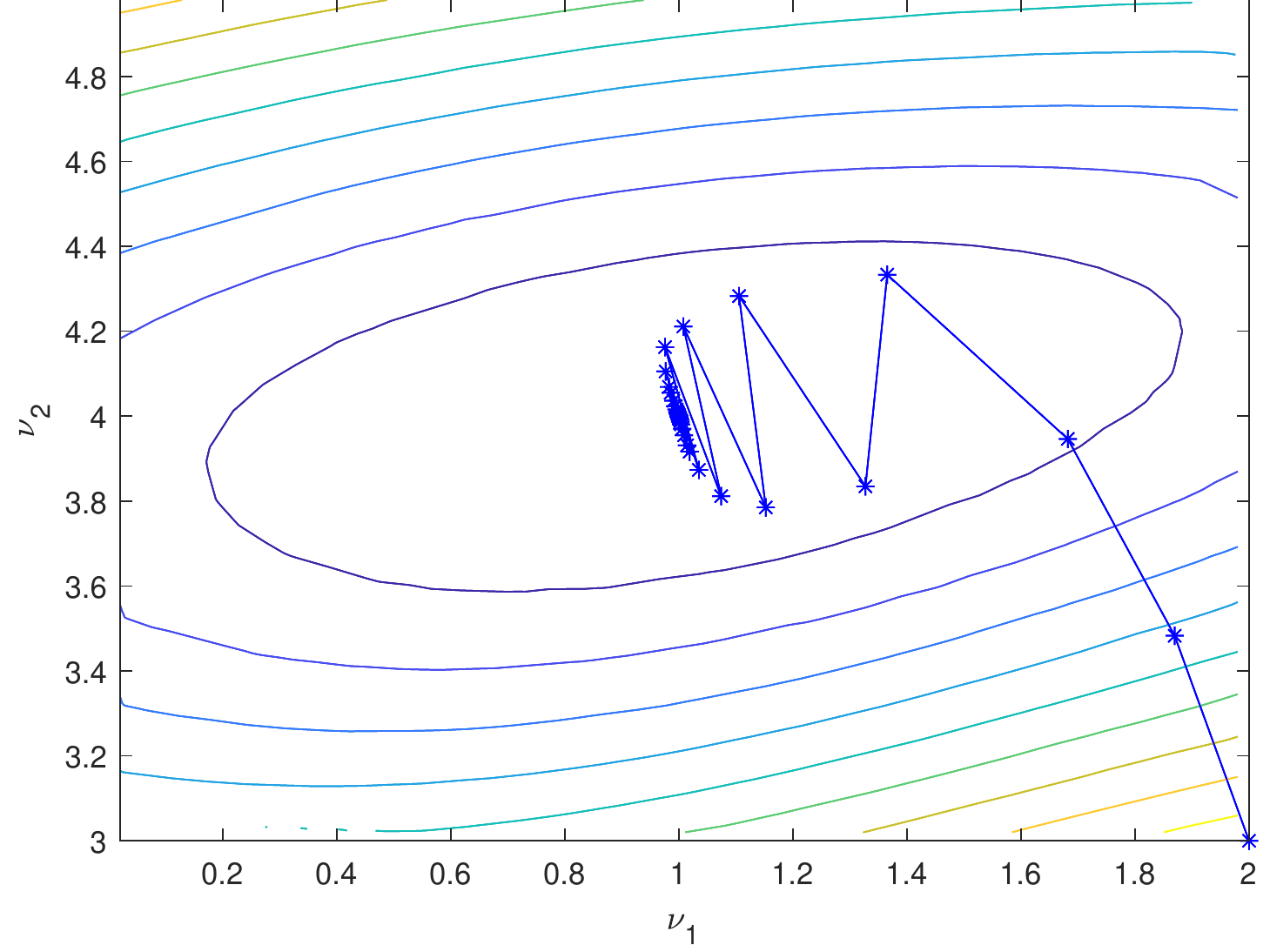}
  }
  \caption{Visualization of the running gradient descent algorithm with the chromatography system (the two left panels) and the two-component Gaussian-mixture model (the two right panels). (a, b) The $L^2$ distance between the observation and simulated signal versus the number of iterations. (c, d) The trajectory of updating $\vec\nu$ in the gradient descent algorithm.}
  \label{fig:GD_plots}
\end{figure}

More specifically, let us consider a four-dimensional parameter $\vec \xi$ by setting the parameters related to the second component as $0$. Since all the elements of $\vec \xi$ are positive, we can decompose $\vec \xi$ into a two-dimensional unbounded parameter $\vec \nu = (a_{I,1}+a_{II,1},b_{I,1}+b_{II,1})^T$ and a two-dimensional bounded parameter $\vec \eta = (a_{I,1}/\nu_1,b_{I,1}/\nu_2)^T$ without losing any information. Let $\vec\xi = g(\vec\eta,\vec\nu)$, and
$$\mathcal{L}(\vec\eta,\vec\nu) = \|\vec\robs - \vec R (g(\vec\eta,\vec\nu))\|_2$$
be the loss function.
While studying the structure of $\mathcal{L}$, we notice that it is convex with respect to $\vec \nu$ for any $\vec{\eta}$, as shown in Fig. \ref{fig:RealData_LossStucture}(a). The topography suggests that an optimal estimator $\vec {\hat\nu}(\vec{\eta}) = \arg\min_{\vec\nu}\mathcal{L}(\vec\eta,\vec\nu)$ can be easily obtained via optimization algorithms for each $\vec{\eta}$, as presented in Fig. \ref{fig:GD_plots}(a)(c). However, if we try to optimize both $\vec{\eta}$ and $\vec \nu$ at the same time, the non-smooth loss function and potential multiple solutions, as illustrated in Fig. \ref{fig:RealData_LossStucture}(b), will make the estimation invalid.

\begin{figure}[htbp]
  \subfigure[]{
    \includegraphics[width=.47\textwidth]{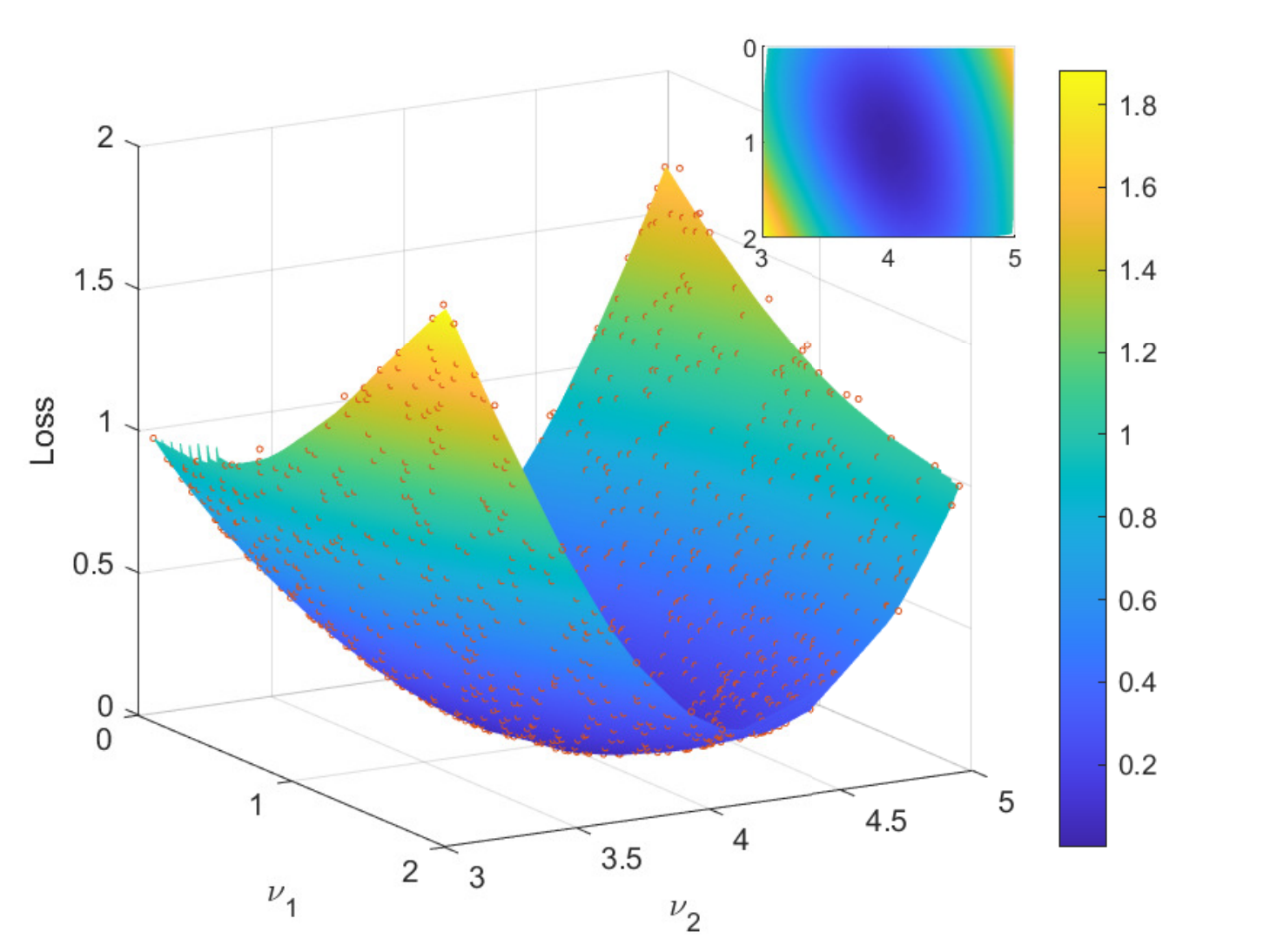}
  }
  \subfigure[]{
    \includegraphics[width=.47\textwidth]{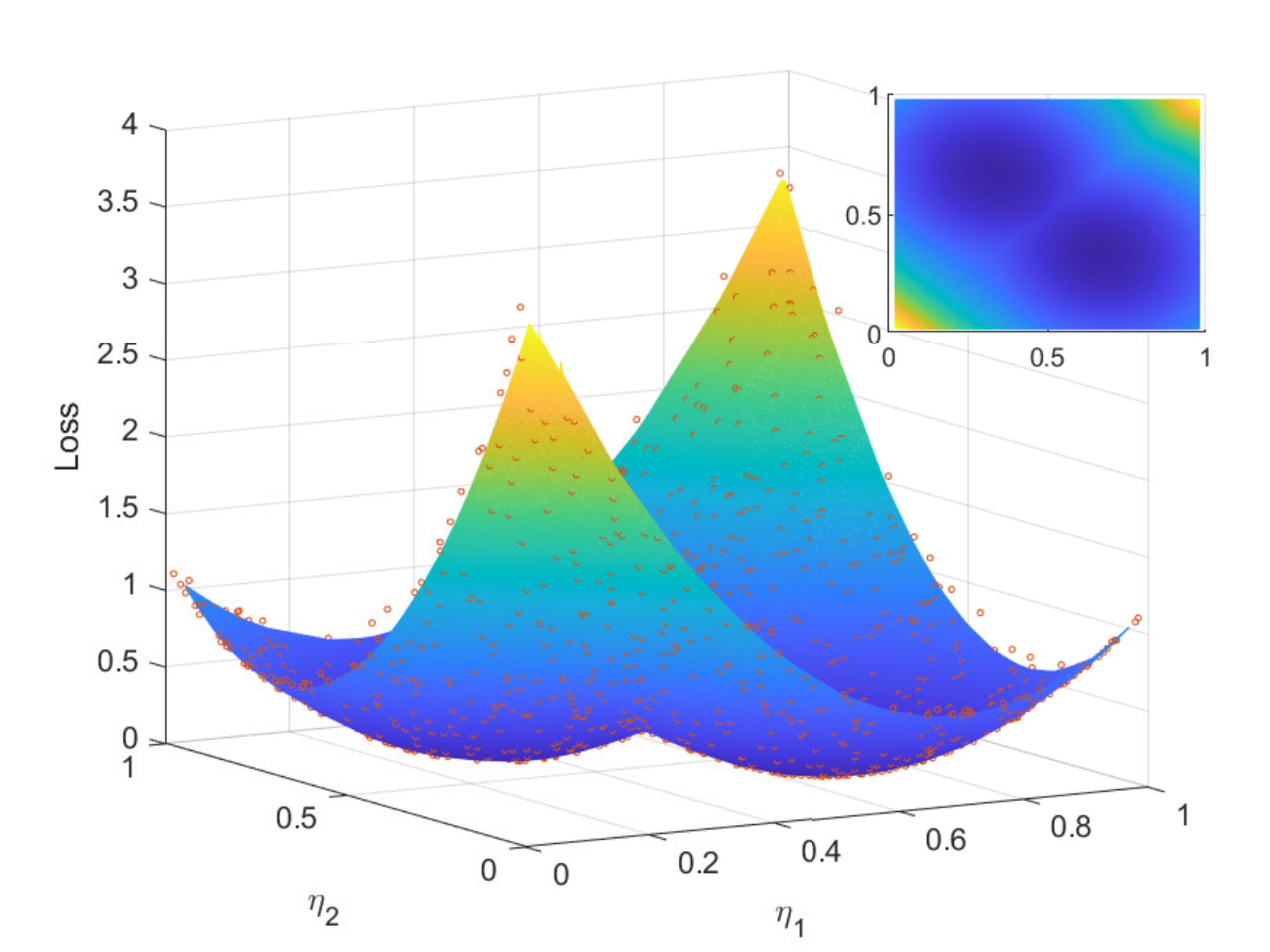}
  }
  \caption{The structure of the loss function for a two-component Gaussian-mixture model. (a) The $L^2$ distance between the observed density function and signal simulated with weights $\vec{\eta}$ and means $\vec \nu$, where $\vec{\eta}$ equals the truth and $\vec\nu$ is sampled in a rectangular area  containing the truth. (b) The $L^2$ distance between the observed density function and signal simulated with weights $\vec{\eta}$ and means $\vec{\hat\nu}(\vec \eta)$, where $\vec{\eta}\in[0,1]^2$ and $\vec{\hat\nu}(\vec \eta)$ is calculated with gradient descent for each $\vec{\eta}$.}
  \label{fig:SimuCase1_LossStucture}
\end{figure}

Similar loss functions can be observed when dealing with mixture models. Let us consider the simplest case. Assume the signal is the density function of a two-component Gaussian-mixture model with weights $\vec\eta$ and mean $\vec\nu$. The corresponding loss function is visualized in Fig. \ref{fig:SimuCase1_LossStucture}. Similar to that of the chromatography system, we can also easily calculate the corresponding optimal means for any fixed weights, but multiple solutions to the problem and saddle points make the optimization unreliable. Unfortunately, this problem will get worse as the number of components grows. Therefore, we usually use Markov chain-based Bayesian methods to make inferences about such parameters of interest. The traversability of these chains and the possible existence of "label switches" also reduce the impact of multiple solutions.

In summary, the adsorption-isotherm parameters are not likely to be directly estimated via gradient-based optimization algorithms, but such optimization methods might lead to a partial solution under certain restricted conditions. Meanwhile, since the mixture models imitate the chromatography system well in terms of the loss function and the former is more computationally efficient, they can be used as toy examples to verify our proposed methods.

\section{Methodology}
In this section, we introduce the main methodology of this paper. The section is organized as follows: first, the traditional Bayes sampling framework is constructed, after which a strategy of dimensionality reduction and restoration is presented. Finally, two algorithms based on this strategy and Metropolis-within-Gibbs sampling are explained.

\subsection{Bayesian approach}
To investigate the parameter of interest, we aim to draw samples from the posterior distribution of $\vec \xi$ given the entire observation $\vec \robs$. In this section, we adopt a Bayesian framework to sample from the posterior distribution $\pi(\vec \theta|\vec\robs)$. Throughout this paper, $p(\cdot)$ is a generic symbol for continuous probability densities, $\pi(\cdot)$ denotes the prior densities, $\pi(\cdot|\vec\robs)$ denotes the posterior densities, and $q(\cdot|\cdot)$ denotes the proposal densities used in Markov chain Monte Carlo (MCMC) algorithms.

For Bayesian inference, we assume that the measurement errors $\{\epsilon(t),t\in\cT\}$ are normally distributed, and impose the following prior distributions on $\vec \xi$ and the error term variance $\sigma^2_\epsilon$:
\begin{equation*}
  \begin{aligned}
    \epsilon(t_1),\ldots,\epsilon(t_n) ~|~ \sigma^2_\epsilon &\overset{\text{i.i.d.}}{\sim} N(0,\sigma^2_\epsilon), \quad \text{for }  i=1,\ldots,n \\
    \sigma^2_\epsilon ~|~ \alpha,\beta & \sim IG(\alpha,\beta)\\
    \pi(\vec \xi ~|~\gamma)  & \propto  \exp \left\{ -\gamma  \|\vec R(\vec \xi) - \vec \robs\|_2^2 \right\}
  \end{aligned}
\end{equation*}
where $N(\mu,\sigma^2)$ represents the normal distribution with mean $\mu$ and variacne $\sigma^2$, $IG(\alpha,\beta)$ represents the inverse gamma distribution with shape parameter $\alpha$ and scale parameter $\beta$, and $\gamma$ is a tuning parameter. We let $\vec \psi= (\alpha,\beta,\gamma)$, which includes all hyperparameters.

Given this prior specification, the posterior distribution of $(\vec\xi,\sigma^2_\epsilon,\vec R(\xi))$ can be written explicitly as
\begin{equation}
  \begin{aligned}
    \pi(\vec \xi,\sigma^2_\epsilon~|~\vec\robs,\vec\psi)
    &\propto \phi(\vec\robs;\vec R(\vec \xi),\sigma^2_\epsilon\mathit{I}_n) \cdot\pi(\sigma^2_\epsilon|\alpha,\beta) \cdot\pi(\vec \xi |\gamma)\\
    &\propto (\sigma^2_\epsilon)^{-(\frac n 2 + \alpha + 1)} \exp\left\{-\sigma^{-2}_\epsilon \left(\frac{E^T E }{2}+\beta\right) -  \gamma E^TE\right\}
  \end{aligned}
  \label{eq:Posterior}
\end{equation}
where $\phi(x;\mu,\Sigma)$ stands for the multivariate normal density for the argument $x$ with mean $\mu$ and covariance matrix $\Sigma$, $I_n$ is the $n\times n$ identity matrix, and $E=\vec R(\vec \xi)-\vec \robs$.

After identifying the posterior distribution, we can use the Metropolis-Hastings algorithm to sample from the posterior. Because $\vec\eta$ and $\vec\nu$ are highly correlated in the posterior, an independent random-walk proposal may not converge quickly, or even get stuck somewhere. For this reason, we consider the Metropolis-within-Gibbs sampler instead. The method is outlined in Algorithm \ref{alg:MH}, in which the acceptance probability is calculated as
\begin{equation*}
  \rho(\vec\theta;\vec\theta^\prime)= \min\left\{
  \frac{\pi(\vec{\xi}^\prime, \sigma^{2\prime}_\epsilon|\vec\robs,\vec\psi)}{\pi(\vec{\xi},
  \sigma^2_\epsilon|\vec\robs,\vec\psi)}
  \frac{q(\vec{\xi}, \sigma_\epsilon^2|\vec{\xi}^\prime, \sigma_\epsilon^{2\prime})}{q(\vec{\xi}^\prime,
  \sigma_\epsilon^{2\prime}|\vec{\xi}, \sigma_\epsilon^2)}, 1\right\}.
\end{equation*}
The hyperparameters $\vec \psi= (\alpha,\beta,\gamma)$ together with the variances of proposal distributions are tuned to make the chain as stable as possible and to have an acceptance rate of approximately $40\%$ to $80\%$ for each element of $(\vec \xi,\sigma^2_\epsilon)$.
Then the sampler will generate a recurrent Markov chain whose stationary distribution is the target posterior distribution $\pi(\vec \xi,\sigma^2_\epsilon~|~\vec\robs,\vec\psi)$.


\begin{breakablealgorithm}
  \caption{Metropolis-within-Gibbs sampler}
  \label{alg:MH}
  \begin{algorithmic}[0]
    \Require
      Observed data $\vec\robs$;
      Hyperparameter $\vec\psi$.
    \Ensure
      $K$ posterior samples $\{\vec \theta^{(k)},k=1,\cdots,K\}$.
    \State Initialize $\vec\theta^{(0)}$;
    \For {$k = 1: K$}
        \State $\vec \theta^\prime \leftarrow \vec\theta^{(k-1)}$;
        \For{$j = 1:\text{length}(\vec\theta)$}
        \State $\vec\theta^\star \leftarrow \vec\theta^\prime$;
        \State Draw $\vec\theta_j^\star \sim q(\vec\theta_j|\vec\theta_j^\prime)$;
        \State Accept $\vec\theta^\prime\leftarrow \vec\theta^\star $ with probability
        $\rho(\vec\theta^\prime;\vec\theta^\star)$; 
        \EndFor
        \State $\vec\theta^{(k)}\leftarrow \vec\theta^\prime$.
    \EndFor
  \end{algorithmic}
\end{breakablealgorithm}

\subsection{Strategy of dimensionality reduction and restoration}
Modeling inverse problems as the posterior distributions will introduce correlation to the parameters of interest. When sampling more than one of the parameters directly from the posterior, the strong correlation might disrupt the motion of the Markov chain. To surmount this problem, we propose a dimension-reduction strategy with two modified MCMC algorithms to stabilize the posterior chains.

{
We first consider the degenerated case, in which the posterior distribution of $\vec\xi$ is almost singular and only supported on a lower dimensional set the size of observed data tends to infinity. Specifically, suppose that the parameter of interest $\vec \xi \in \bR^D$ can be written as $\vec\xi=g(\vec\eta,\vec\nu)$, where $\vec\eta\in \bR^d$ with $d\ll D$, $\vec\nu\in \bR^{D-d}$, and $g(\cdot):\bR^D\to \bR^D$ is a one-to-one mapping determined with pilot study. Roughly speaking, we first set $d$ as the largest dimension that leads to a stable posterior chain of $\vec\eta$. Then the dimensionality is reduced by combining the elements of $\vec\xi$ according to the pilot study. Please see Section S4 of the supplementary material for a detailed discussion on how to choose $\vec\eta,\vec\nu,g$ and some empirical results.

In the extreme case that $\vec\nu$ and $\vec\eta$ are perfectly dependent in the posterior as the size of observed data $n$ tends to infinity, the conditional posterior $\pi(\vec\nu| \vec\eta,\vec\robs,\vec\psi)$ is degenerate to a Dirac distribution $\delta_{h(\vec\eta)}$ for some unknown function $h$, i.e. it is almost sure that $\vec\nu = h(\vec\eta)$. 
Then, the support of the posterior of $\vec\xi$ can be expressed as
$$\{\vec\xi\in \bR^D\mid \vec\xi = g(\vec \eta, h(\vec\eta))\},$$
and the induced posterior distribution of $(\vec\eta,\vec\nu)$ is
$$\pi(\vec\eta, \vec\nu|\vec\robs,\vec\psi)= \pi(\vec\eta|\vec\robs,\vec\psi) \pi(\vec\nu| \vec\eta,\vec\robs,\vec\psi) = \pi(\vec\eta|\vec\robs,\vec\psi) \delta_{h(\vec\eta)}.$$
In such a scenario, we can find the least squares estimator of $\vec\nu$ in terms of $\vec\eta,\vec\robs,\vec\psi$ from the following equation:
\begin{equation}
  \vec{\hat\nu}(\vec \eta) = \arg\min_{\vec{\nu}^\prime}\|\vec R(g(\vec \eta,\vec \nu^\prime))-\vec \robs\|_2 ,
  \label{eq:GD}
\end{equation}
such that $\vec{\hat\nu}(\vec \eta)$ approximates $h(\vec\eta)$ as $n\to\infty$. In what follows, we assume that the optimization problem in \eqref{eq:GD} can be settled by the following gradient descent algorithm (GD) without discussing the optimization accuracy.
\begin{algorithm}
  \caption{Gradient Descent (GD)}
  \label{alg:GD}
  \begin{algorithmic}[0]
  \Require
  Observed data $\vec \robs$;
  Recording time $\cT_n$;
  Argument $\vec \eta$.
  \Ensure
  Restored parameter $\vec\xi$
  \State Initialize complementary information $\vec\nu = \vec\nu_0$, step size $\gamma$.
  \For{iter = 1 : MaxIter}
  \State $L(\vec\nu) <- \|\vec R(g(\vec \eta,\vec\nu)) - \vec\robs \|_2^2$
  \State $\vec G <- $ numerical gradient of $L$ at point $s$
    \If{$\|\vec G\|<10^{-5}$}
      \State Break
    \EndIf
    \For{ i = 1 : 100}
      \State $\vec\nu^\prime = \vec\nu - 0.9^{(i-1)}\gamma \cdot \vec G/\|\vec G\|$
      \If{$L(\vec\nu^\prime) < L(\vec\nu)$}
        \State Break
      \EndIf
    \EndFor
    \State $\vec\nu = \vec\nu^\prime$
  \EndFor
  \\\Return $\vec{\hat\nu} = \vec\nu$ and $\vec{\hat\xi} = g(\vec \eta,\vec{\hat\nu})$.
  \end{algorithmic}
\end{algorithm}

To incorporate the optimizing step in the MCMC algorithm, we keep the overall structure of Algorithm \ref{alg:MH} for sampling $\vec\eta$ and $\sigma_{\epsilon}^2$. In the extreme case of $\vec\nu=h(\vec\eta)$ for some function $h$, instead of sampling $\vec\nu$, we can calculate a $\vec{\hat\nu}(\vec \eta)$ from \eqref{eq:GD} in each MCMC iteration.
Together with the function $g$, $\vec{\hat \xi} = g(\vec \eta,\vec{\hat\nu}(\vec \eta))$ can be obtained for any specific $\vec\eta$.
The detailed algorithm is provided in Algorithm \ref{alg:MGDG}.

Our theoretical justification for this dimension-reduction strategy is as follows. We first introduce a series of technical assumptions, and our theoretical results will depend on different subsets of these assumptions. We define the loss function $L_R(\vec\xi)=\int_{\cT} [R(\vec\xi,t) - R(\vec\xi^*,t)]^2 dt$ associated with the solution function $R$, for any $\vec\xi\in \vec\Xi$.
\begin{itemize}
\item[(A.1)] $(\vec\nu,\vec\eta)$ lies in a compact space $\vec\Theta\subseteq \bR^D$, where $\vec\eta \in \vec\Theta_{\vec\eta} \subseteq \bR^d$ and $\vec\nu\in \vec\Theta_{\vec\nu} \subseteq \bR^{D-d}$. There is a continuous one-to-one mapping such that $\vec\xi=g(\vec\eta,\vec\nu)$ and $\vec\Xi = g(\vec\Theta)$ is also a compact space for $\vec\xi$. The true parameter $\vec\xi^*=g(\vec\eta^*, \vec\nu^*)$ is an interior point of $\vec\Xi$. $\cT$ is a compact set, and $\{\epsilon(t):t\in \cT\}$ are independent and identically distributed as $N(0,\sigma_{\epsilon}^{2*})$ for some $\sigma_{\epsilon}^{2*} <\infty$.
\item[(A.2)]  Let $|\mathcal{T}|$ be the Lebesgue measure of $\mathcal{T}$. As $n\to\infty$,
\begin{align} \label{eq:unif.Rconv}
&\sup_{\vec\xi \in \vec\Xi} \left| \frac{|\mathcal{T}|}{n}\left\|\vec R(\vec\xi)- \vec R(\vec\xi^*) \right\|_2^2 - \int_{\cT} [R(\vec\xi,t) - R(\vec\xi^*,t)]^2 dt\right| \to 0.
\end{align}
\item[(A.3)] There exists a function $h(\vec\eta)$, and positive constants $a_{1R},c_{1R},\kappa_{1R}$, such that for any $(\vec\eta, \vec\nu)\in \vec\Theta$,
\begin{align}\label{eq:iden.nu}
& L_R(g(\vec\eta,\vec\nu)) - L_R(g(\vec\eta, h(\vec\eta)))  \geq \min\left( a_{1R} \|\vec \nu - h(\vec\eta)\|_2^{\kappa_{1R}} , c_{1R}\right).
\end{align}
\item[(A.4)] There exist positive constants $a_{2R},c_{2R},\kappa_{2R}$, such that, for all $\vec\xi\in \vec\Xi$ that satisfies $\|\vec\xi-\vec\xi^*\|_2\leq c_{2R}$, $L_R(\vec\xi)\leq a_{2R} \|\vec\xi - \vec\xi^* \|_2^{\kappa_{2R}}$.
\item[(A.5)] There exist positive constants $a_{3R},c_{3R},\kappa_{3R}$, such that, for all $\vec\xi \in \vec\Xi$, $L_R(\vec\xi)\geq \min\left(a_{3R} \|\vec\xi - \vec\xi^* \|_2^{\kappa_{3R}} , c_{3R}\right)$.
\end{itemize}

Assumption A.1 assumes a compact parameter space and normal errors. Assumption A.2 assumes the uniform convergence of the empirical $L_2$ distance from the function $R(\vec\xi,t)$ to the true function $R(\vec\xi^*,t)$ over the compact parameter space $\vec\Xi$. This assumption usually holds when $\cT_n=\{t_1,\ldots,t_n\}$ are densely distributed in $\cT$ and $R(\vec\xi,t)$ is continuous in both $\vec\xi$ and $t$. Assumption A.3 is the identification condition for $\vec\nu$, where the lower bound guarantees that $h(\vec\eta)$ is the unique minimizer of the loss function $L_R(g(\vec\eta,\vec\nu))$ over the $\vec\nu$ argument for any given $\vec\eta$. This assumption is only used for justifying our later Algorithm \ref{alg:MGDG} in the degenerate case of $\vec\nu=h(\vec\eta)$. Assumption A.4 imposes a local-continuity condition on the loss function $L_R(\vec\xi)$ in a small neighborhood of $\vec\xi^*$. Since we have already assumed that $\vec\xi^*$ is an interior point of $\vec\Xi$ in A.1, the radius $c_{2R}$ can be made small such that the whole ball $\{\vec\xi:\|\vec\xi-\vec\xi^*\|_2\leq c_{2R}\}$ is included in $\vec\Xi$. Assumption A.5 imposes an identification condition on the loss function $L_R(\cdot)$ which guarantees that the true parameter $\vec\xi^*$ is uniquely identified in $\vec\Xi$. This is similar to the identification condition for moment estimation (see, for example, ZE.2 in \citealt{BelChe09}). The power constants $\kappa_{1R}$, $\kappa_{2R}$, and $\kappa_{3R}$ in A.3, A.4, and A.5 are not specified to allow flexibility in the local-continuity property of $L_R(\cdot)$. In Section 2.2 of the supplementary material, we provide some partial verification for these assumptions using the model of simulation case 1 in Section 4.1.



\begin{theorem} \label{thm:nu.sol}
\begin{itemize}
\item[(i)] Suppose that Assumptions A.1, A.2, and A.3 hold. Then, for $\vec{\hat\nu}(\vec \eta)$ defined in \eqref{eq:GD},
\begin{align} \label{eq:nuhat.conv}
& \sup_{\vec\eta\in \vec\Theta_{\vec\eta}} \left\|\vec{\hat\nu}(\vec \eta) - h(\vec\eta)\right\|_2^2 \to 0,
\end{align}
as $n\to\infty$, almost surely.
\item[(ii)] Suppose that Assumptions A.1, A.2, A.3, and A.4 hold. Then, for any $\varepsilon>0$, and for every $\vec\eta \in \vec\Theta_{\vec\eta}$ and every $\sigma_{\epsilon}^2>0$,
\begin{align} \label{eq:post.conv1}
& \Pi(\|\vec\nu-h(\vec\eta)\|_2 > \varepsilon ~|~\sigma_{\epsilon}^2,\vec \robs,\vec\psi) \to 0,
\end{align}
as $n\to\infty$, almost surely, where $\Pi(\cdot| \sigma_{\epsilon}^2,\vec \robs,\vec\psi)$ denotes the conditional posterior measure of $\vec\xi$ given $\sigma_{\epsilon}^2$.
\item[(iii)] Suppose that Assumptions A.1, A.2, A.4, and A.5 hold. Then, for any $\varepsilon>0$, and for every $\vec\eta \in \vec\Theta_{\vec\eta}$ and every $\sigma_{\epsilon}^2>0$,
\begin{align} \label{eq:post.conv2}
& \Pi(\|\vec\xi-\vec\xi^*\|_2 > \varepsilon ~|~\sigma_{\epsilon}^2,\vec \robs,\vec\psi) \to 0,
\end{align}
as $n\to\infty$, almost surely.
\end{itemize}
\end{theorem}

Theorem \ref{thm:nu.sol} provides justification for our Algorithm \ref{alg:MGDG}, in the sense that, in each MCMC iteration, we can numerically solve for $\vec\nu$ in terms of $\vec\eta$ from \eqref{eq:GD}. Part (i) shows that, as the sample size of observations $n$ increases, the empirical solution $\hat{\vec\nu}(\vec\eta)$ becomes increasingly close to $h(\vec\eta)$, which is the unique minimizer of the loss function $L_R(\cdot)$. Part (ii) shows that, under the additional continuity condition on $L_R(\cdot)$, most of the posterior probability mass of $\vec\nu$ is concentrated around $h(\vec\eta)$. Part (iii) gives the stronger result of \textit{posterior consistency} of $\vec\xi$ to the truth $\vec\xi^*$ (Chapter 6, \citealt{GhoVan17}) under the additional identification condition A.5 on $\vec\xi$, which implies that most of the posterior probability mass will concentrate around the true parameter $\vec\xi^*$ asymptotically. The detailed proof of Theorem \ref{thm:nu.sol} can be found in Section S2.1 of the supplementary material. We emphasize that even when the strong assumption of A.3 that assumes $\vec\nu=h(\vec\eta)$ does not hold, Part (iii) of Theorem \ref{thm:nu.sol} on the posterior consistency for the whole parameter vector $\vec\xi$ remains valid as the sample size $n\to\infty$.
}

\subsection{Main algorithms}
After identifying the method of dimensionality reduction and restoration, we propose a new modified Metropolis-embedded gradient-descent-within-Gibbs sampler (MGDG). The detailed algorithm is presented below.

\begin{breakablealgorithm}
  \caption{Metropolis-embedded gradient-descent-within-Gibbs sampler (MGDG)}
  \label{alg:MGDG}
  \begin{algorithmic}[0]
    \Require
      Observed data $\vec \robs$;
      Recording time $\cT_n$;
      Hyperparameter $\vec\psi$.
    \Ensure
      $K$ posterior samples $\{\vec \eta^{(k)},\sigma^{2,(k)}_\epsilon,\hat{\vec \xi}^{(k)},\, k =1:K\}.$
    \State Initialize $\vec \eta^{(0)}, \sigma^{2,(0)}_\epsilon$;
    \State Set $\vec \xi^{(0)} \leftarrow g(\vec \eta^{(0)},\vec{\hat\nu}(\vec \eta^{(0)}))$, and calculate $\vec
    R(\vec\xi^{(0)})$ from the numerical solver;
    \For {$k = 1:K$}
    \State Let $\vec \eta,\,\sigma^{2}_\epsilon\leftarrow \vec \eta^{(k-1)},\sigma^{2,(k-1)}_\epsilon$, $E\leftarrow \vec
    R(\vec \xi)-\vec \robs$;
    \State Propose $\vec\eta^\prime,\,\sigma^{2\prime}_\epsilon\sim q(~\cdot~|\vec \eta,\,\sigma^{2}_\epsilon)$;
    \vspace{5mm}

    \State \textbf{(Sampling of $\sigma^{2}_\epsilon$)}
    \State $\rho_{\sigma^2_{\epsilon}} \leftarrow \min\left\{
    \left(\frac{\sigma_\epsilon^{2\prime}}{\sigma_\epsilon^{2}}\right)^{-\frac{n}{2}}\exp\left\{-\frac{1}{2}\left[(\sigma_\epsilon^{2\prime})^{-1}-(\sigma_\epsilon^{2})^{-1}\right]
    \right\}
    \frac{\pi(\sigma^{2\prime}_\epsilon|\vec\psi)}{\pi(\sigma^{2}_\epsilon|\vec\psi)}
    \frac{q(\sigma^{2}_\epsilon|\sigma^{2\prime}_\epsilon)}{q(\sigma^{2\prime}_\epsilon|\sigma^{2}_\epsilon)}, 1\right\};$
    \If{$U[0,1]<\rho_{\sigma^2_\epsilon}$}
    \State $\sigma_\epsilon^{2,(k)}\leftarrow \sigma_\epsilon^{2\prime}$;
    \Else
    \State $\sigma_\epsilon^{2,(k)}\leftarrow \sigma_\epsilon^{2}$;
    \EndIf
    \State  Let $\sigma_\epsilon^{2}\leftarrow \sigma_\epsilon^{2,(k)}$;
    \vspace{5mm}

    \State \textbf{(Sampling of $\vec\eta$)}
    \For{$j = 1 : d$}
        \State $\vec \eta_{cand} \leftarrow  (\eta_1,\cdots,\eta_{j-1},\eta_j^\prime,,\eta_{j+1}\cdots,\eta_d)$;
        \State $\vec \xi_{cand} \leftarrow  GD(\vec \robs,\vec \eta_{cand})$;
        \State Calculate $\vec R(\vec \xi_{cand})$ from numerical solver, $E^\prime\leftarrow \vec R(\vec \xi_{cand})-\vec
        \robs$;
        \State$\rho_{\vec \eta}\leftarrow  \min\left\{\exp\left\{ -\frac{1}{2\sigma_\epsilon^2} (E^{\prime
        T}E^\prime-E^TE)\right\} \frac{\pi(\vec \eta_{cand}|\vec\psi)}{\pi(\vec \eta|\vec\psi)} \frac{q(\vec \eta|\vec
        \eta_{cand})}{q(\vec \eta_{cand}|\vec \eta)}, 1\right\}$;
        \If{$U[0,1]<\rho_{\vec \eta}$}
        \State $\vec \eta_j^{(k)}\leftarrow \vec \eta_j^\prime$;
        \Else
            \State $\vec \eta_j^{(k)}\leftarrow \vec \eta_j$;
        \EndIf
        \State $\hat{\vec \xi}^{(k)} \leftarrow  GD(\vec \robs,\vec\eta^{(k)})$;
    \EndFor
    \EndFor
    \\\Return $\{\vec \eta^{(k)}, \sigma^{2,(k)}_\epsilon,\hat{\vec \xi}^{(k)},\, k =1:K\}.$
  \end{algorithmic}
\end{breakablealgorithm}


The main strategy of this algorithm is to sample only $\vec \eta\in \bR^d$ and the error term variance $\sigma^2_{\epsilon}$ from the posterior distribution, and $\vec\nu$ is treated as a function of $\vec\eta$ in the posterior. We can restore a posterior sample of
$\vec\xi=g(\vec\eta,\vec\nu)$ using the posterior draws of $\vec\eta$ and $\vec\nu$. The lower dimensionality of $\vec\eta$ helps to improve the mixing of Markov chains and prevent the "label switching" phenomenon that occurs in mixture models (\citealt{Jasetal05}). After reaching the stationary distribution, the samples $\{\hat{\vec \nu}^{(k)}\}$ should be distributed around the
truth $h(\vec\nu)$, as shown in Theorem \ref{thm:nu.sol}.


Now we consider the general case in which $\vec\nu$ is not a function of $\vec\eta$, but may be highly correlated with $\vec\eta$ in the posterior distribution $\pi(\vec\xi,\sigma_{\epsilon}^2|\robs,\vec\psi)$. In such a scenario, we use the Metropolis-adjusted Langevin algorithm (MALA) \citep{RobRos98,RobTwe96} to sample $\vec\nu$. We still use the Metropolis-within-Gibbs algorithm to draw $\vec\eta$ and $\sigma^2_{\epsilon}$, and then use the MALA to draw $\vec\nu$ using the gradient information from the conditional posterior of $\vec\nu$. We call this algorithm the Metropolis-adjusted Langevin-dynamics-within-Gibbs sampler (MALG).
\begin{breakablealgorithm}
  \caption{Metropolis-adjusted Langevin-dynamics-within-Gibbs sampler (MALG)}
  \label{alg:MALG}
  \begin{algorithmic}[0]
    \Require
      Observed data $\vec \robs$;
      Recording time $\cT_n$;
      Hyperparameter $\vec\psi$; Scaling parameter $\tau$; Length of Langevin Monte Carlo $m$.
    \Ensure
      $K$ posterior samples $\{\vec \eta^{(k)}, \vec\nu^{(k)},\sigma^{2,(k)}_\epsilon,\, k =1:K\}.$
    \State Initialize $\vec \eta^{(0)},\vec{\nu}^{(0)},\sigma^{2,(0)}_\epsilon$;
    \State Set $\vec \xi^{(0)} \leftarrow g(\vec \eta^{(0)},\vec{\nu}^{(0)})$, and calculate $\vec R(\vec\xi^{(0)})$ from
    the numerical solver;
    \For {$k = 1:K$}
    \State Let $\vec \xi, \vec \eta,\vec\nu, \sigma^{2}_\epsilon\leftarrow \vec\xi^{(k-1)},\vec
    \eta^{(k-1)},\vec\nu^{(k-1)},\sigma^{2,(k-1)}_\epsilon$;
    \State Let $E\leftarrow \vec R(\vec \xi)-\vec \robs$;
    \vspace{5mm}

    \State \textbf{(Sampling of $\sigma^{2}_\epsilon$)}
    \State Propose $ \sigma^{2\prime}_\epsilon\sim q(\cdot|\,\sigma^{2}_\epsilon)$;
      \State $\rho_{\sigma^2_{\epsilon}} \leftarrow \min\left\{
      \left(\sigma_{\epsilon}^{2\prime}/\sigma_{\epsilon}^2\right)^{-\frac{n}{2}}\exp\left\{-\frac{1}{2}\left[
      (\sigma_{\epsilon}^{2\prime})^{-1}-(\sigma_{\epsilon}^{2})^{-1}\right]E^T E\right\}
      \frac{\pi(\sigma^{2\prime}_\epsilon|\vec\psi)}{\pi(\sigma^{2}_\epsilon|\vec\psi)}
      \frac{q(\sigma^{2}_\epsilon|\sigma^{2\prime}_\epsilon)}{q(\sigma^{2\prime}_\epsilon|\sigma^{2}_\epsilon)},
      1\right\};$
      \If{$U[0,1]<\rho_{\sigma^2_\epsilon}$}
      \State $\sigma_\epsilon^{2,(k)}\leftarrow \sigma_\epsilon^{2\prime}$;
      \Else
      \State $\sigma_\epsilon^{2,(k)}\leftarrow \sigma_\epsilon^{2}$;
      \EndIf
      \State Let $\sigma_\epsilon^{2}\leftarrow \sigma_\epsilon^{2,(k)}$;
      \vspace{5mm}

      \State \textbf{(Sampling of $\vec\nu$)}
      \State $\vec\nu^{(k,0)} \leftarrow \vec\nu^{(k-1)}$;
      \For {$j=1:m$}
      \State $\vec\nu_{cand} \leftarrow \vec\nu^{(k,j-1)} + \tau \nabla_{\vec\nu} \log \pi(\vec\nu^{(k,j-1)} |
      \vec\eta,\sigma_{\epsilon}^{2},\vec R(\vec\xi),\vec \robs) + \sqrt{2\tau} N(0,I_{D-d})$;
      \State $\vec \xi_{cand} \leftarrow  g(\vec\eta,\vec \nu_{cand})$;
      \State Calculate $\vec R(\vec \xi_{cand})$ from the numerical solver, $E^\prime\leftarrow \vec R(\vec
      \xi_{cand})-\vec \robs$;
      \State Calculate $q_{LD}(\vec \nu_{cand}|\vec \nu)$ and $q_{LD}(\vec \nu|\vec \nu_{cand})$, using the formula
      $$q_{LD}(\vec\nu'|\vec\nu) = \exp\left(-\frac{\|\vec\nu'-\vec\nu-\tau \nabla_{\vec\nu} \log \pi(\vec\nu|
      \vec\eta,\sigma_{\epsilon}^{2},\vec R(\vec\xi),\vec \robs) \|}{4\tau}\right).$$
      \State$\rho_{\vec \nu}\leftarrow  \min\left\{\exp\left\{ -\frac{1}{2\sigma_{\epsilon}^{2}} \left(E^{\prime
      T}E^\prime-E^TE\right)\right\} \frac{\pi(\vec \xi_{cand}|\vec\psi)}{\pi(\vec \xi|\vec\psi)} \frac{q_{LD}(\vec
      \nu|\vec \nu_{cand})}{q_{LD}(\vec \nu_{cand}|\vec \nu)}, 1\right\}$;
          \If{$U[0,1]<\rho_{\vec \nu}$}
          \State $\vec \nu^{(k,j)}\leftarrow \vec \nu_{cand}$;
          \Else
              \State $\vec \nu^{(k,j)} \leftarrow \vec \nu^{(k,j-1)}$;
          \EndIf
      \EndFor
      \State Let $\vec \nu \leftarrow  \vec\nu^{(k)} \leftarrow \vec\nu^{(k,m)}$;
      \vspace{5mm}

      \State \textbf{(Sampling of $\vec\eta$)}
      \State Propose $\vec\eta' \sim q(\cdot|\vec \eta)$; Let $\vec\xi_{cand}\leftarrow g(\vec\eta',\vec\nu)$;
      \State Calculate $\vec R(\vec \xi_{cand})$ from the numerical solver, $E^\prime\leftarrow \vec R(\vec
      \xi_{cand})-\vec \robs$;
      \State $\rho_{\vec \eta}\leftarrow  \min\left\{\exp\left\{ -\frac{1}{2\sigma_{\epsilon}^{2}} \left(E^{\prime
      T}E^{\prime}-E^T
      E\right)\right\} \frac{\pi(\vec \eta_{cand}|\vec\psi)}{\pi(\vec \eta|\vec\psi)} \frac{q(\vec \eta|\vec
      \eta_{cand})}{q(\vec \eta_{cand}|\vec \eta)}, 1\right\}$;
          \If{$U[0,1]<\rho_{\vec \eta}$}
          \State $\vec \eta^{(k)}\leftarrow \vec \eta^\prime$;
          \Else
              \State $\vec \eta^{(k)}\leftarrow \vec \eta$;
          \EndIf

    \EndFor
    \\\Return $\{\vec \eta^{(k)}, \vec \nu^{(k)},  \sigma^{2,(k)}_\epsilon,~ k =1:K\}.$
  \end{algorithmic}
\end{breakablealgorithm}
where the values of the proposal densities, $q_{LD}(\vec \nu^{\prime}|\vec \nu)$ and $q_{LD}(\vec \nu|\vec \nu^{\prime})$, are calculated from
$$q_{LD}(\vec\nu'|\vec\nu) = \exp\left(-\frac{\|\vec\nu'-\vec\nu-\tau \nabla_{\vec\nu} \log \pi(\vec\nu|
\vec\eta,\sigma_{\epsilon}^{2},\vec R(\vec\xi),\vec \robs) \|}{4\tau}\right).$$
Because the log-likelihood term $\log \pi(\vec\nu | \vec\eta,\sigma_{\epsilon}^{2},\vec R(\vec\xi),\vec
\robs)$ is proportional to $\|\vec R(g(\vec \eta,\vec \nu))-\vec \robs\|_2^2$, the gradient term is the same gradient $\nabla_{\vec\nu} \|\vec R(g(\vec \eta,\vec \nu))-\vec \robs\|_2^2$ as used in Algorithm \ref{alg:MGDG}. After choosing
suitable values for the stepsize $\tau$ and the sub-chain length $m$, this algorithm will return MCMC samples of all parameters, including $\vec\nu$.

\section{Simulation study}
In this section, we propose three numerical solvers based on mixture models to confirm the robustness of our algorithm from three aspects: initialization, dimension of $\vec\xi$ and skewness of $\vec R(\vec\xi)$. To reach this conclusion, repeated sampling is performed in all three simulation cases to account for the effect of the initialization; the second case is designed by adding dimensions to $\vec\xi$ on the basis of the first case; and the third case is constructed by introducing skewness to the first case. Although these numerical solvers have relatively simple structures, their parameters are highly correlated, which mimics the problem with the data for real experimental gradient-elution preparative liquid chromatography. Meanwhile, the calculation can be implemented through vectorization, which makes it possible to have larger sample sizes and more repeated trials. We only present the key plots in this section and leave other plots to Section S6 of the supplementary material.

\subsection{Simulation Case 1}
\label{Section:SimuCase1}
For the first case, we considered a Gaussian-mixture model with two components. Let the parameter of interest be $\vec{\xi}\in \mathbb{R}_+^4$, with a single observation $\vec \robs = \vec R(\vec \xi^*) + \vec \epsilon$ with  $\vec\xi^* = (\frac{1}{3},\frac{2}{3},\frac{8}{3},\frac{4}{3})$ and noise variance $\sigma^{2*}_\epsilon = 0.001$ from the following numerical solver:
\begin{equation}
  R(\vec\xi,t) = \sum_{i=1}^2 \frac{\xi_{2i-1}}{(\xi_{2i-1}+\xi_{2i})}\frac{1}{\sqrt{2\pi}}e^{-\frac{(t-(\xi_{2i-1}+\xi_{2i}))^2}{2}}, \quad t \in \cT.
  \label{eq:sim1}
\end{equation}
To perform the dimension reduction, we selected $\vec \eta \in \bR^2$ and $\vec\nu \in \bR^2$ as follows:
\begin{equation*}
  \eta_i = \frac{\xi_{2i-1}}{\xi_{2i-1}+\xi_{2i}},\quad \nu_i = {\xi_{2i-1}+\xi_{2i}},\qquad i = 1,2.
\end{equation*}
These two parameters can be regarded as weights and means in the Gaussian-mixture model. Because $\vec \eta \in [0,1]^2$, the MGDG algorithm can be initialized  by uniformly sampling $\{\vec \eta_i \}_{i=1}^{1000} \sim \text{Uniform}(0,1)^2$ and setting $\vec \eta^{(0)}$ as the one minimizing the $L^2$ norm between $\vec R(g(\vec \eta_i,\vec{\hat\nu}(\vec \eta_i)))$ and $\vec \robs$, i.e.
\begin{equation*}
  \vec \eta^{(0)} = \arg \min_{\vec \eta_i} \|\vec R(g(\vec \eta_i,\vec{\hat\nu}(\vec \eta_i)))-\vec \robs\|_2,
\end{equation*}
with $\vec{\hat\nu(\vec \eta )}$ defined in (\ref{eq:GD}). The noise variance $\sigma^2_\epsilon$ is initialized with a random value from its prior distribution.

Let $TN(\mu,\sigma^2,l,u)$ stand for the truncated normal density on interval $[l,u]$ with mean $\mu$ and variance $\sigma^2$. The other hyperparameters, prior distributions, and proposal distributions are summarized in Table \ref{tab:simu1}. Since the two elements of $\vec \eta$ can always be exchanged in the sampling, we set the smaller one as $\eta_1$ and the larger one as $\eta_2$ after each iteration.

\begin{table*}[htbp]
  \caption{Hyperparameters in Case 1: the prior distributions, proposal distributions, and other parameters.}
  \label{tab:simu1}
  \begin{tabular}{ll}
  \hline
  $\eta_i^\prime \mid \eta_i \sim TN(\eta_i,0.02^2,0,1)$
   & Proposal distribution of $\eta_i$ for $i=1,2$\\
  $\pi(\vec \eta)\propto \exp\{-\gamma\|\vec R(g(\vec \eta,\vec{\hat\nu}(\vec \eta)))-\vec \robs\|^2_2\}$
   & Prior of $\vec \eta$\\
  $\cT=[-2,7]$
   & Recording time\\
  $\cT_n\subseteq\cT,\,n =50$
   & Equally spaced time points used in calculation\\
  $B = 500$
   & Number of burn-in samples\\
  $\vec\psi = (\alpha,\beta,\gamma)$
   & $(2,\|\vec R(g(\vec \eta^{(0)},\vec{\hat\nu}(\vec \eta^{(0)})))-\vec \robs\|_2^2/n,8)$\\
  $(\tau, m ) = (0.001,200)$
   & Step size and chain length of sampling $\vec\nu$ in MALG\\
  \hline
  \end{tabular}
\end{table*}

\begin{figure}[htbp]
  \subfigure[MGDG]{
    \includegraphics[width=.3\textwidth]{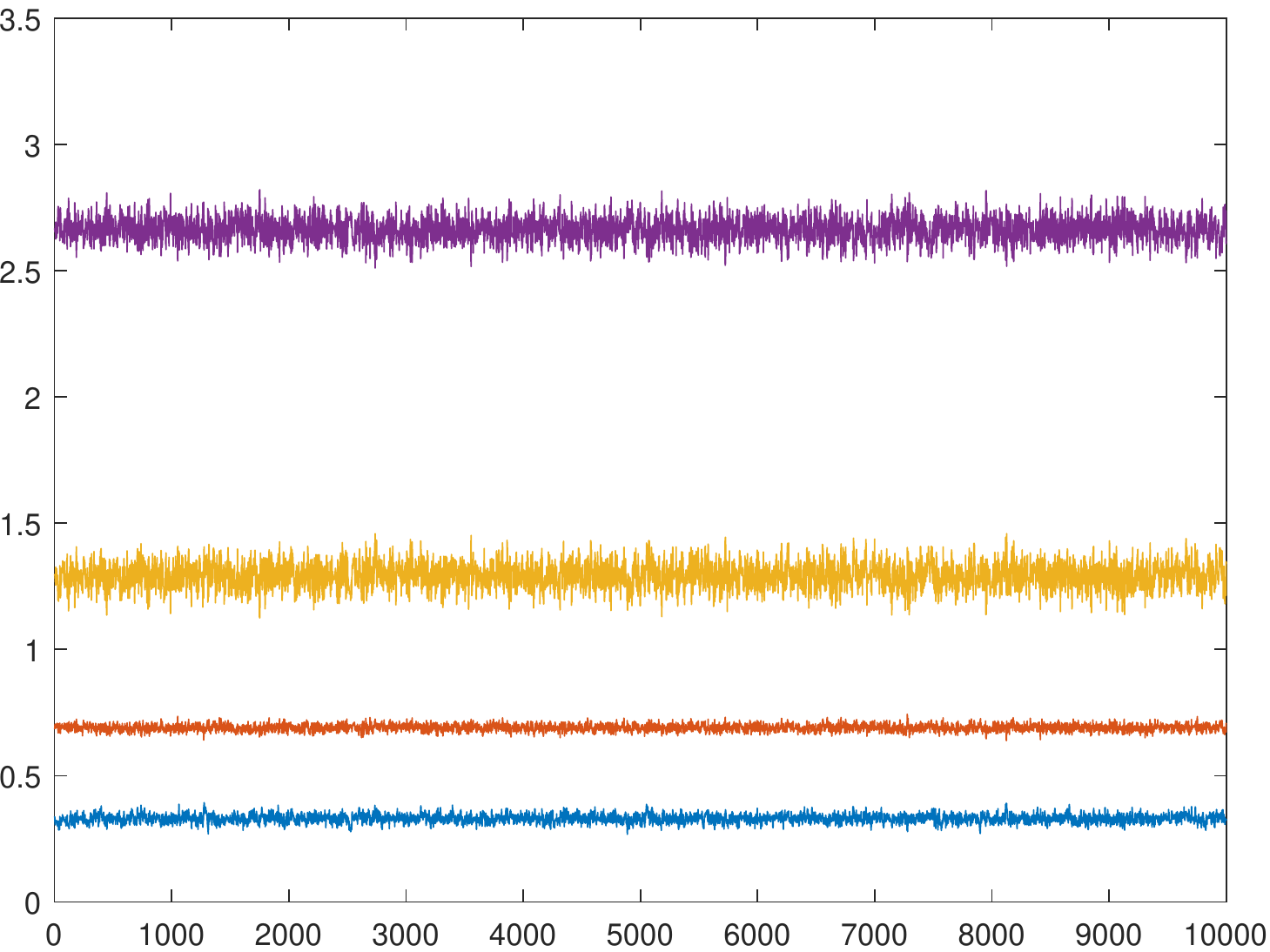}
  }
  \subfigure[MALG]{
    \includegraphics[width=.3\textwidth]{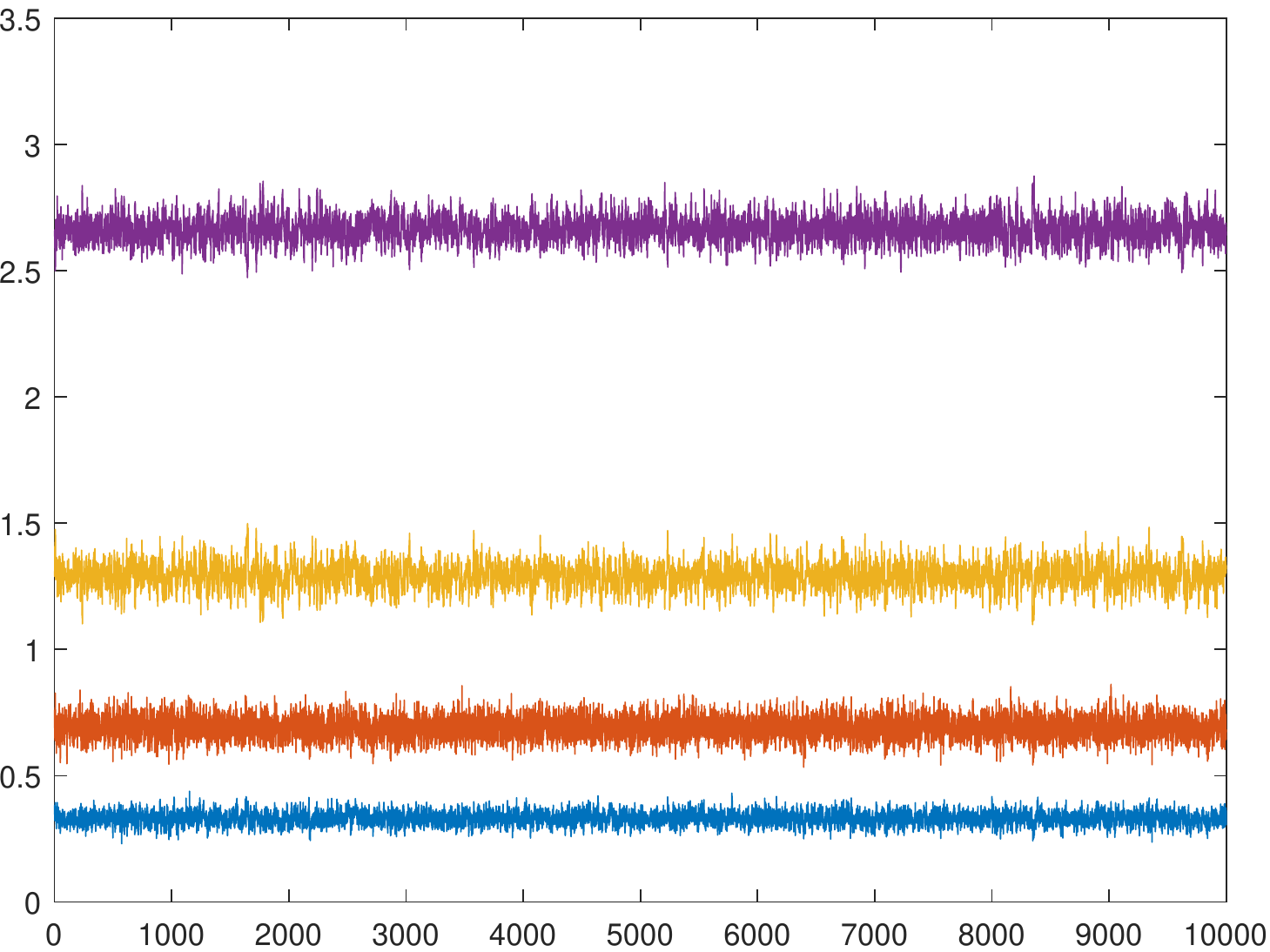}
  }
  \subfigure[Metropolis-within-Gibbs]{
    \includegraphics[width=.3\textwidth]{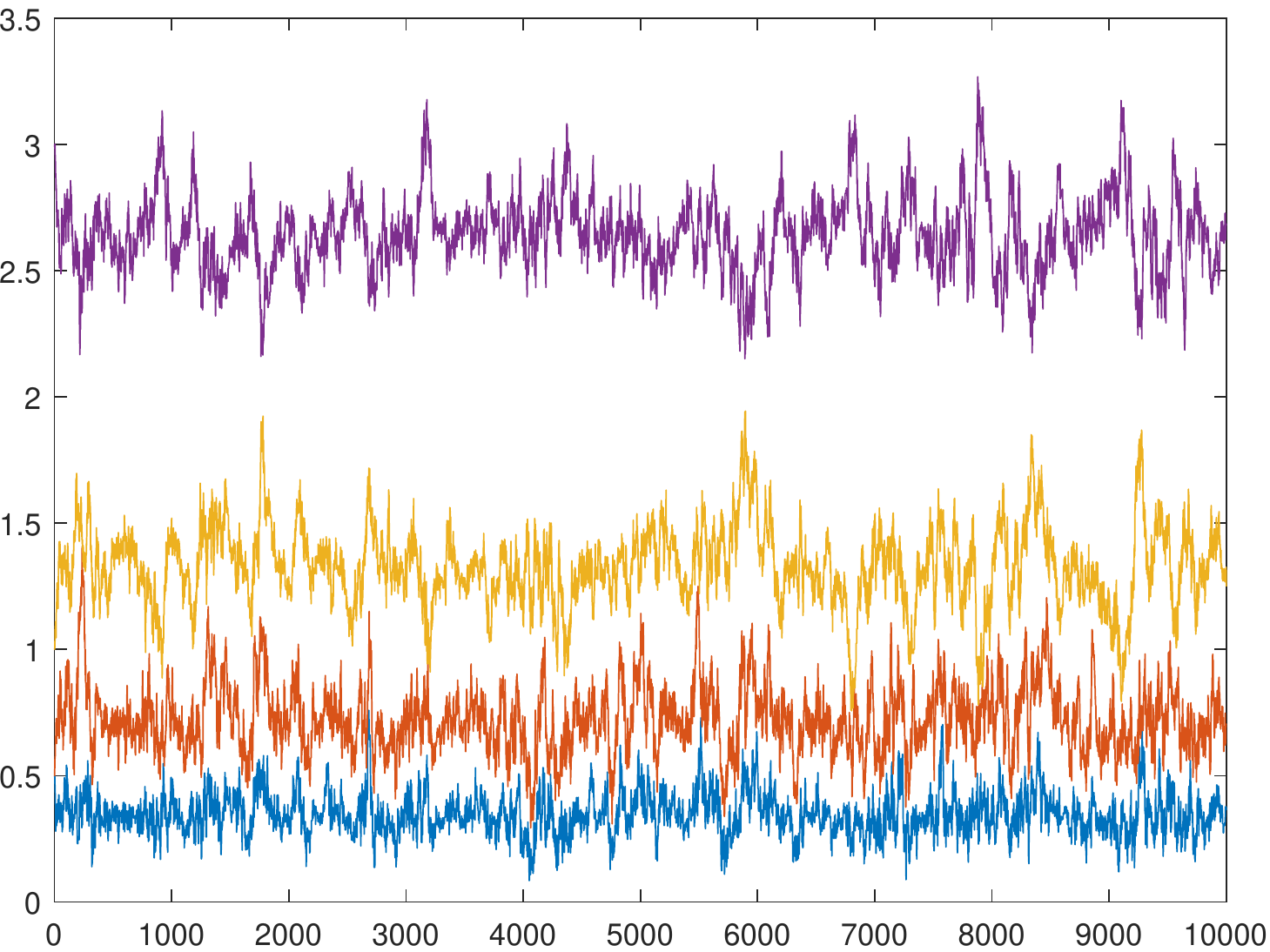}
  }
  \caption{An example of sampling result of $\vec{\hat\xi}$ from (a) MGDG, (b) MALG, and (c) Metropolis-within-Gibbs sampler. In each panel, the four solid lines represent the trace of the four elements in $\vec{\hat\xi}$ with sample size $K = 1\times 10^4$.}
  \label{fig:simu1:trace}
\end{figure}

Fig. \ref{fig:simu1:trace} shows an example of the samples from our algorithms and from the Metropolis-within-Gibbs sampler. It can be observed that traces of both MGDG and MALG are distributed closely around the ground truth. Such distribution suggests that our method can stably sample from the correlated posterior and that the sample is reliable in inferring $\vec\xi$. The performance of our method, in this case, differs from that of the Metropolis-within-Gibbs sampler (\cite{MHsampling}), for which the sample is less stationary and more biased in inferring $\vec\xi$. A possible reason is that the elements of $\vec\xi$ are highly correlated in the posterior, which makes the Metropolis-within-Gibbs sampler not very reliable.

\begin{figure}[htbp]
  \subfigure[$\vec{\eta} - \vec{\eta}^*$]{
    \includegraphics[width=.3\textwidth]{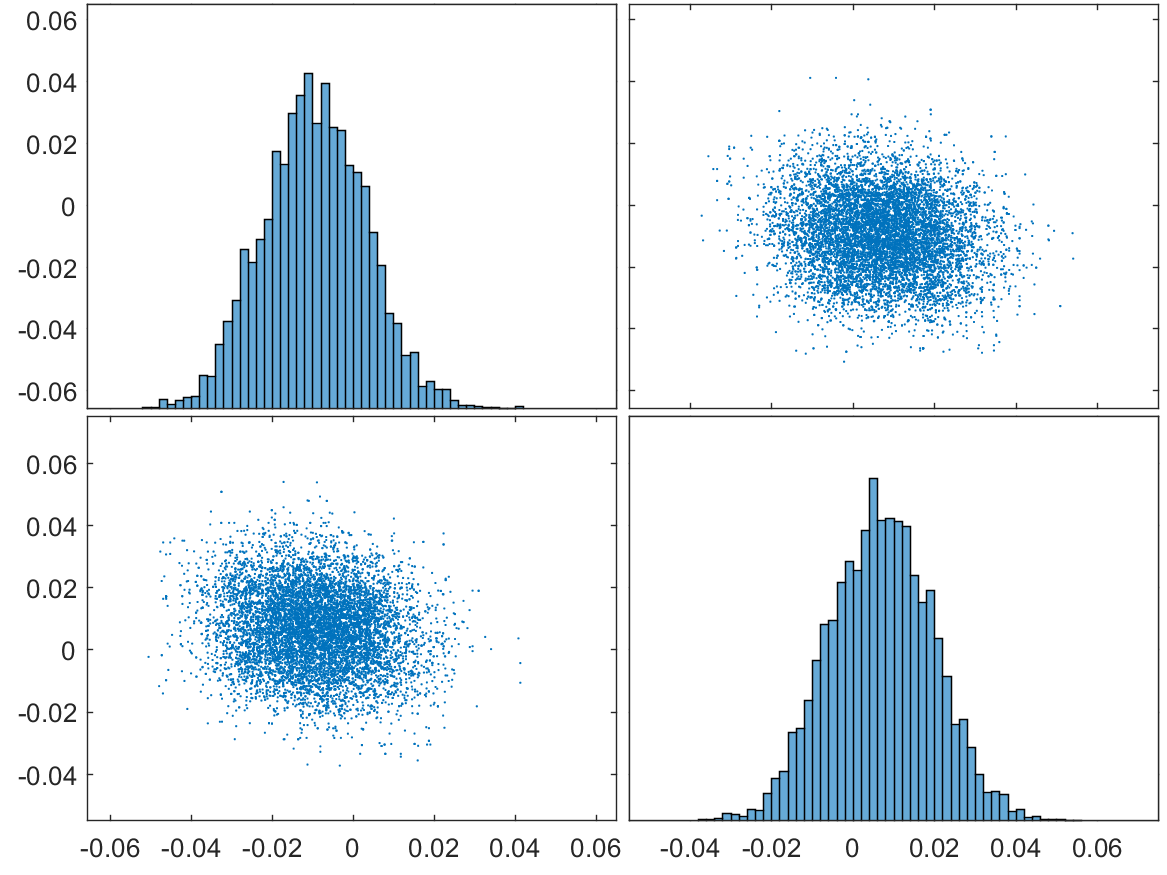}
  }
  \subfigure[$\vec{\hat{\xi}} - \vec{\xi}^*$]{
    \includegraphics[width=.3\textwidth]{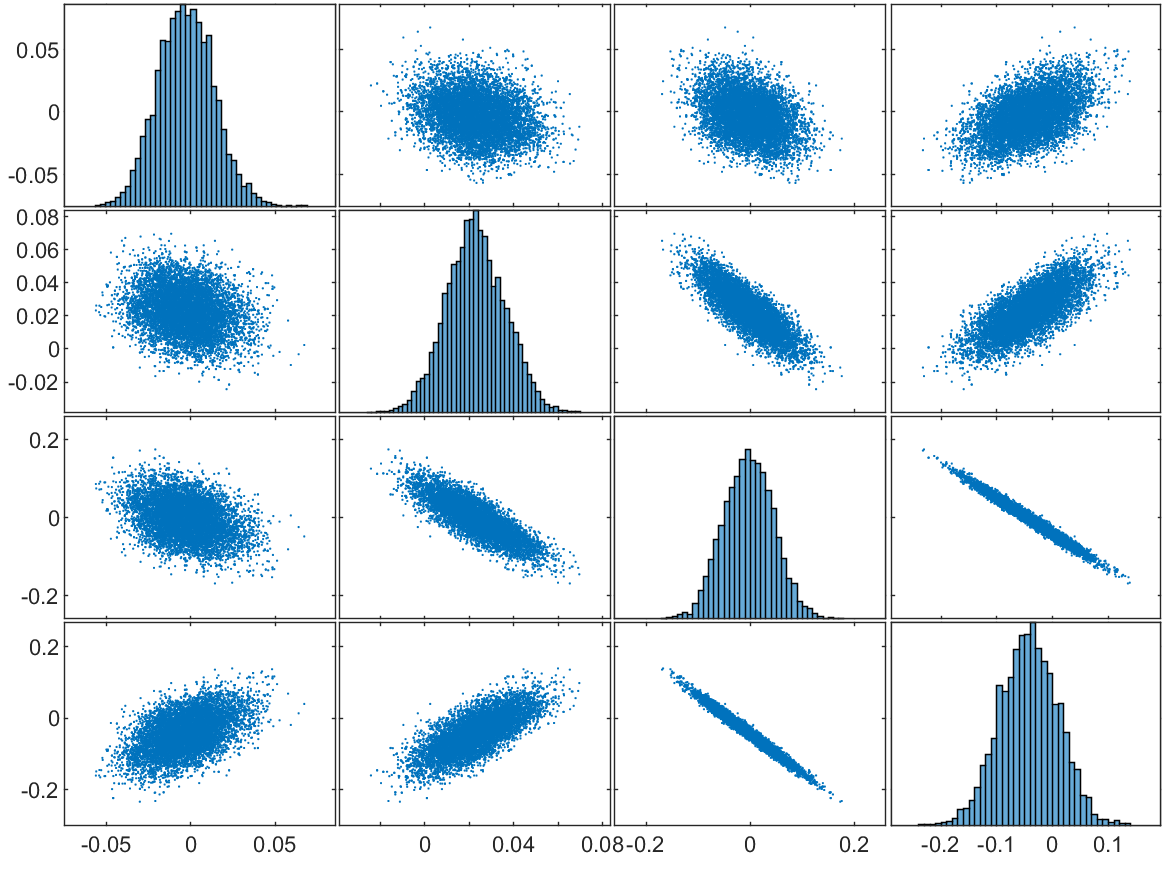}
  }\\
  \subfigure[$\vec{\eta} - \vec{\eta}^*$]{
    \includegraphics[width=.3\textwidth]{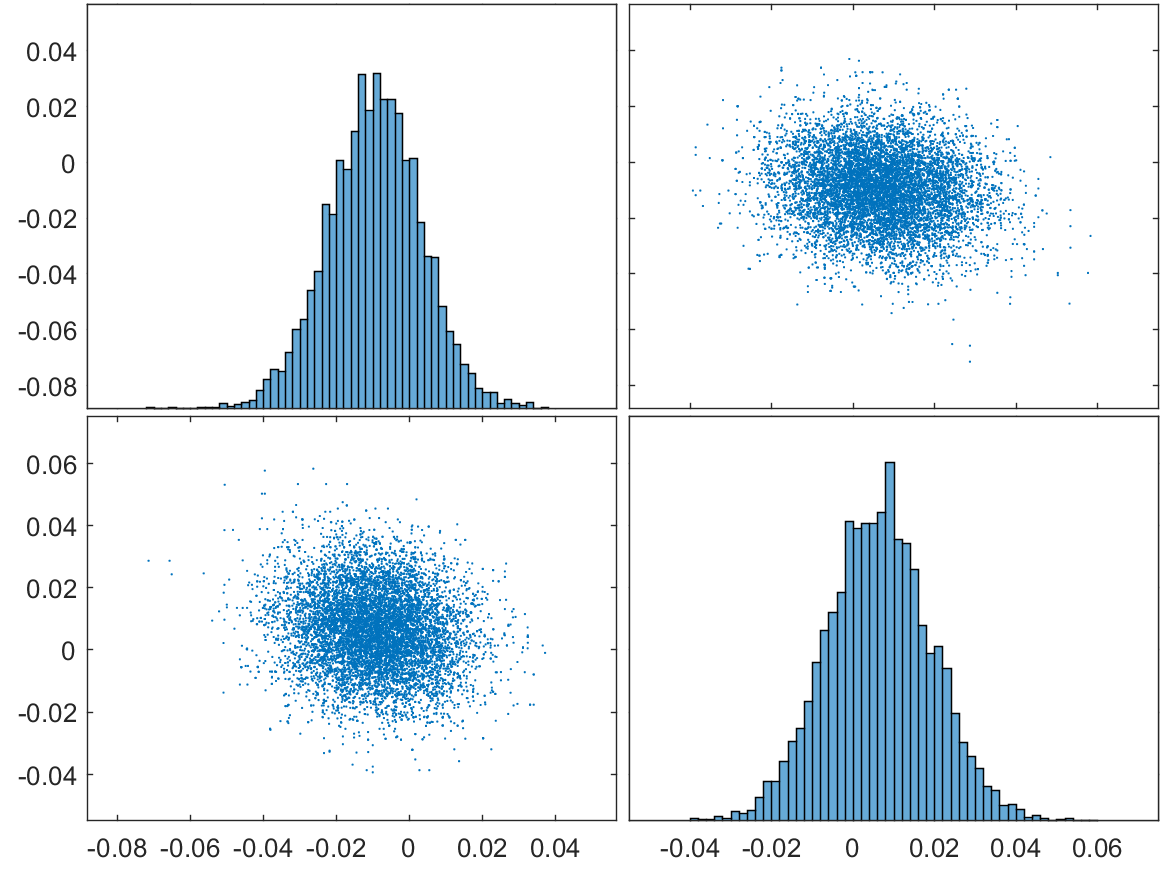}
  }
  \subfigure[$\vec{\nu} - \vec{\nu}^*$]{
    \includegraphics[width=.3\textwidth]{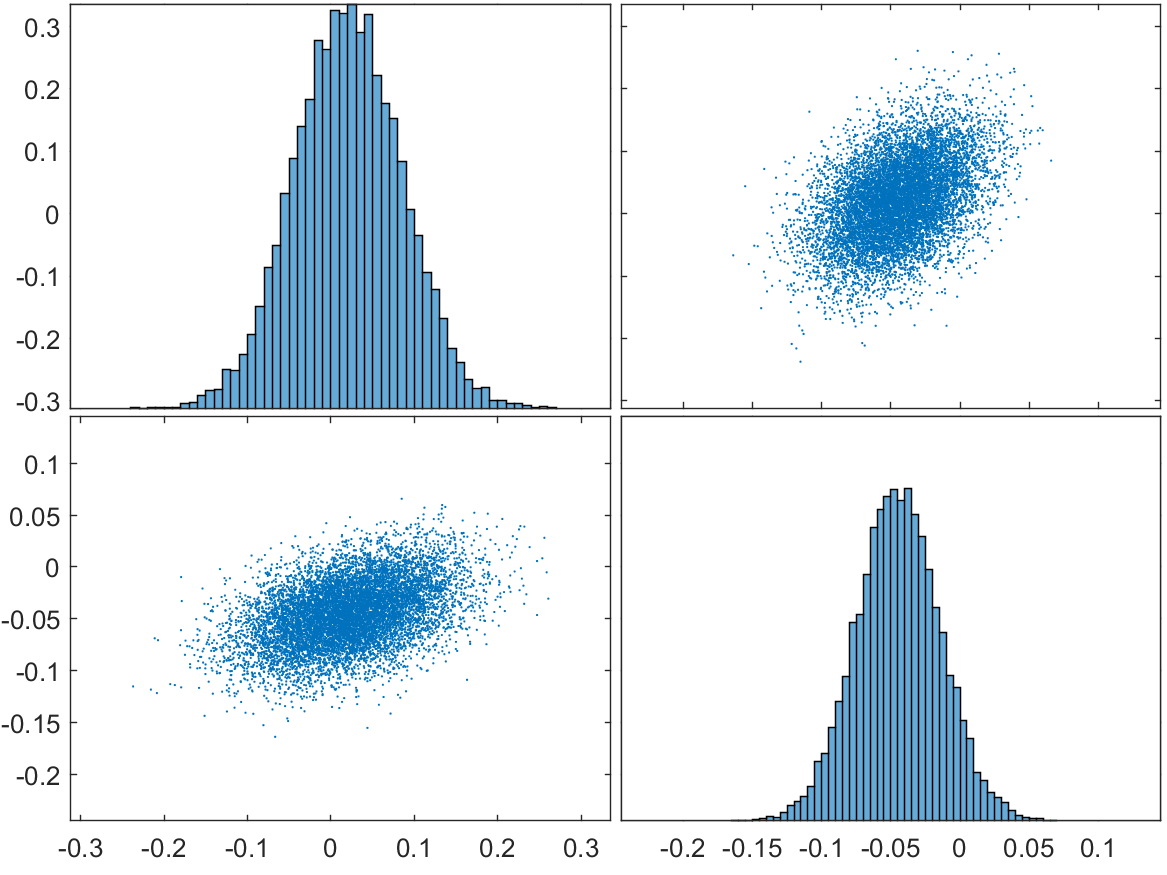}
  }
  \subfigure[$\vec{\hat{\xi}} - \vec{\xi}^*$]{
    \includegraphics[width=.3\textwidth]{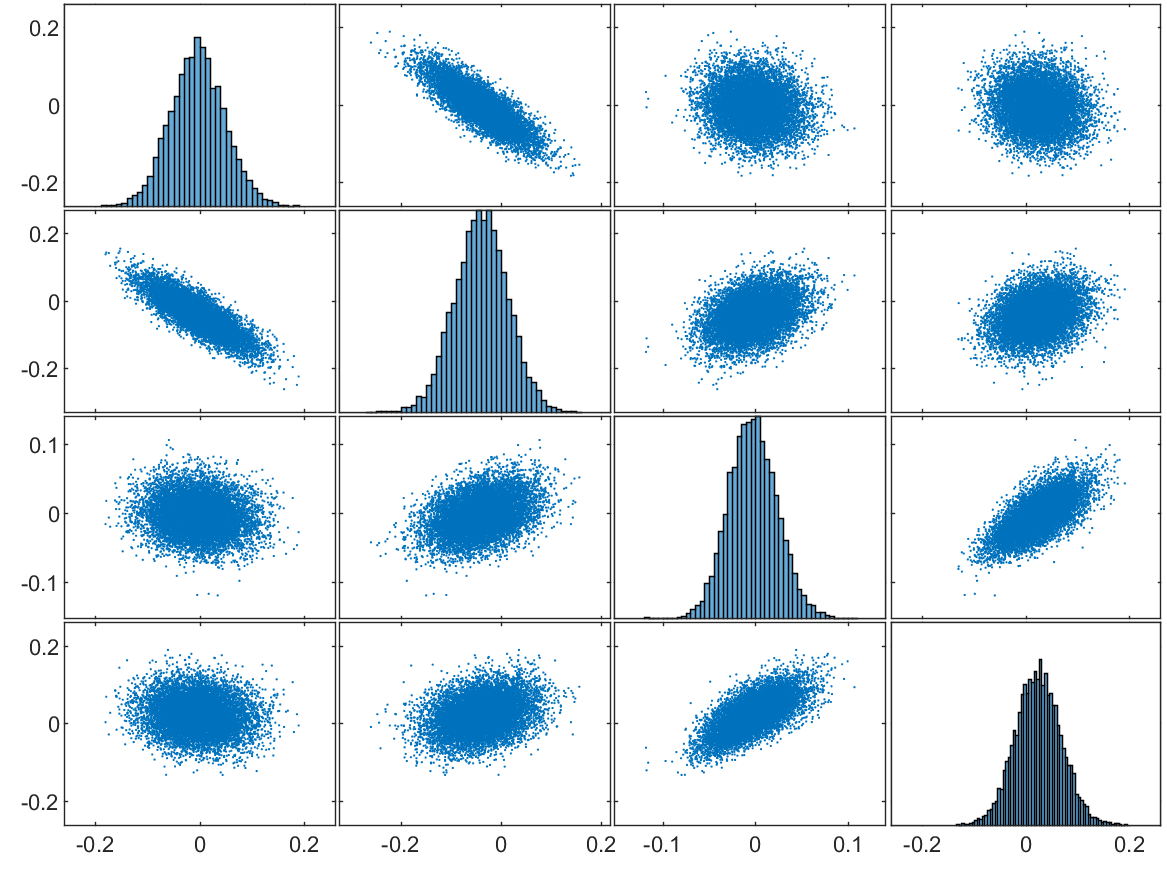}
  }
  \caption{Distribution of residuals in one running trial of Case 1 with sample size $K = 1\times 10^4$. (a,b)Scatter-plot matrix of result from MGDG. (c,d,e) Scatter-plot matrix of result from MALG.}
  \label{fig:simu1:PlotMatrix}
\end{figure}

The algorithms are evaluated by the residuals in one sampling trial, i.e. $\vec{\eta} - \vec{\eta}^*$, $\vec{\nu} - \vec{\nu}^*$, and $\vec{\hat{\xi}} - \vec{\xi}^*$. The corresponding scatter-plot matrices with sample size $K = 1\times 10^4$ are presented in Fig. \ref{fig:simu1:PlotMatrix}. These plots suggest that these residuals are approximately normally distributed around 0, and that the bias is in the same order as the threshold in the gradient descent algorithm. The correlation between $\vec{\eta}$ is not significant, but the variables in the same group in $\vec \xi$ have a strong linear correlation.

\begin{figure}[htbp]
  \subfigure[$\eta_1$]{
    \includegraphics[width=.3\textwidth]{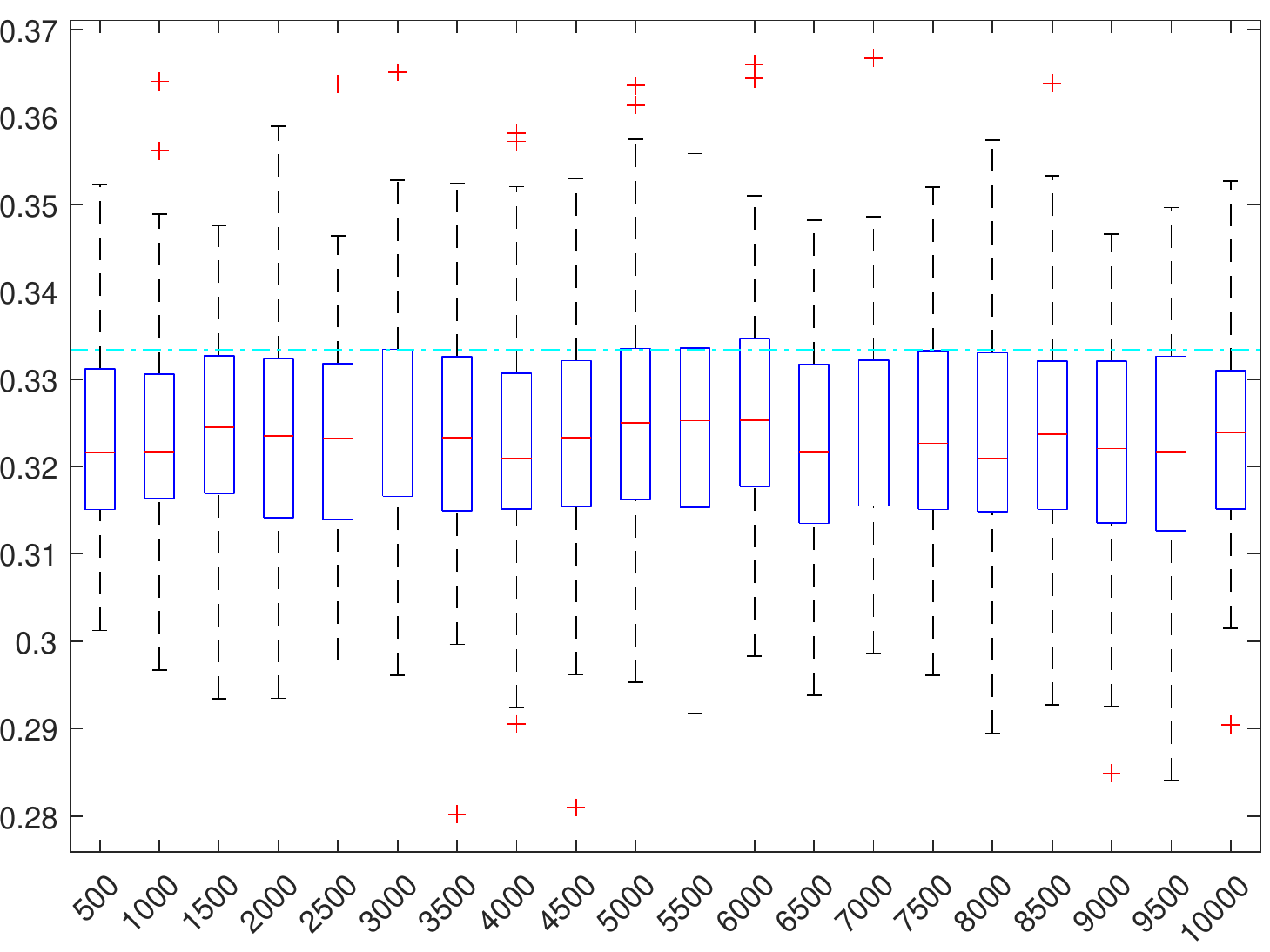}
  }
  \subfigure[$\hat\xi_1$]{
    \includegraphics[width=.3\textwidth]{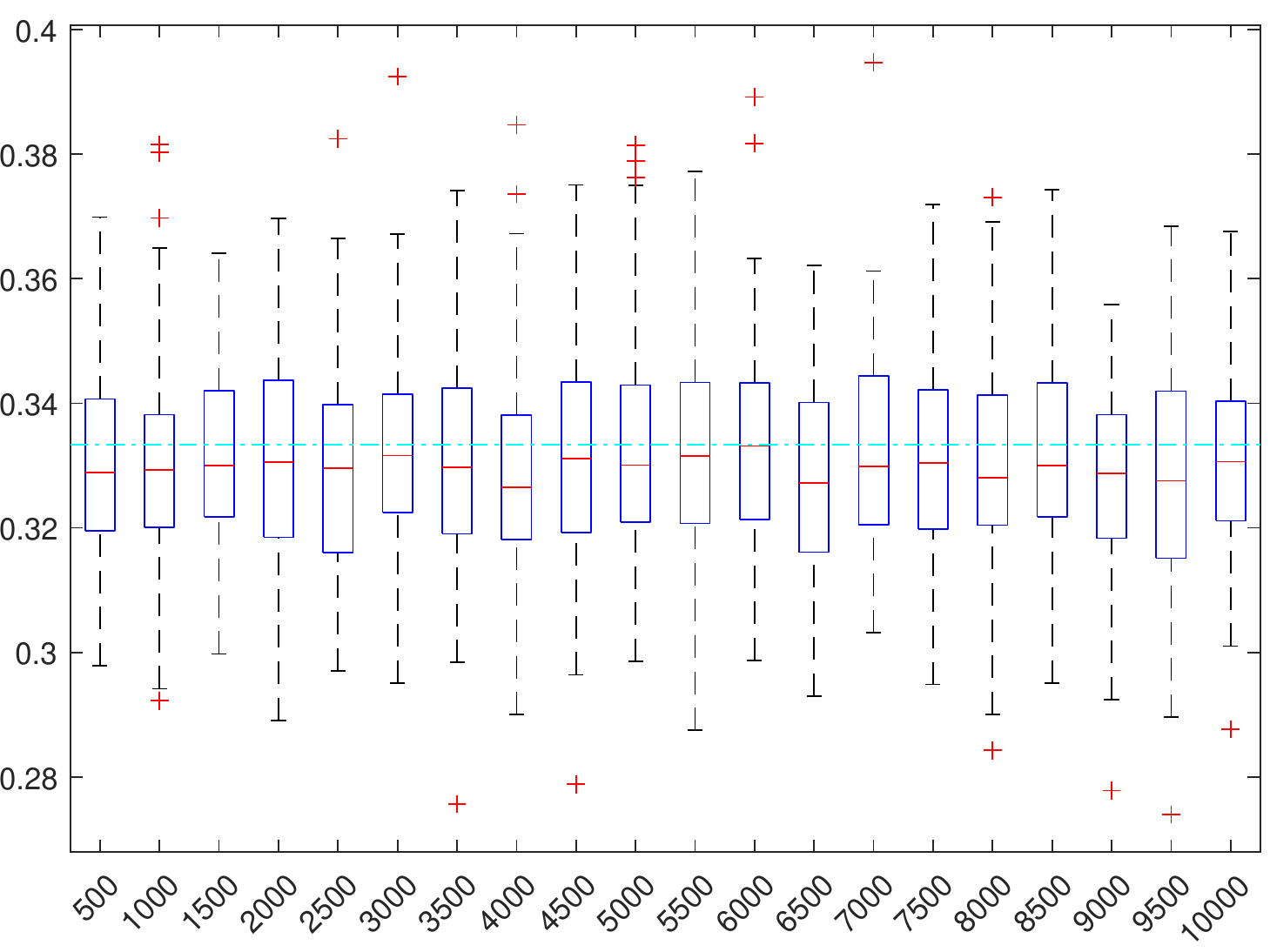}
  }
  \subfigure[$\hat\xi_2$]{
    \includegraphics[width=.3\textwidth]{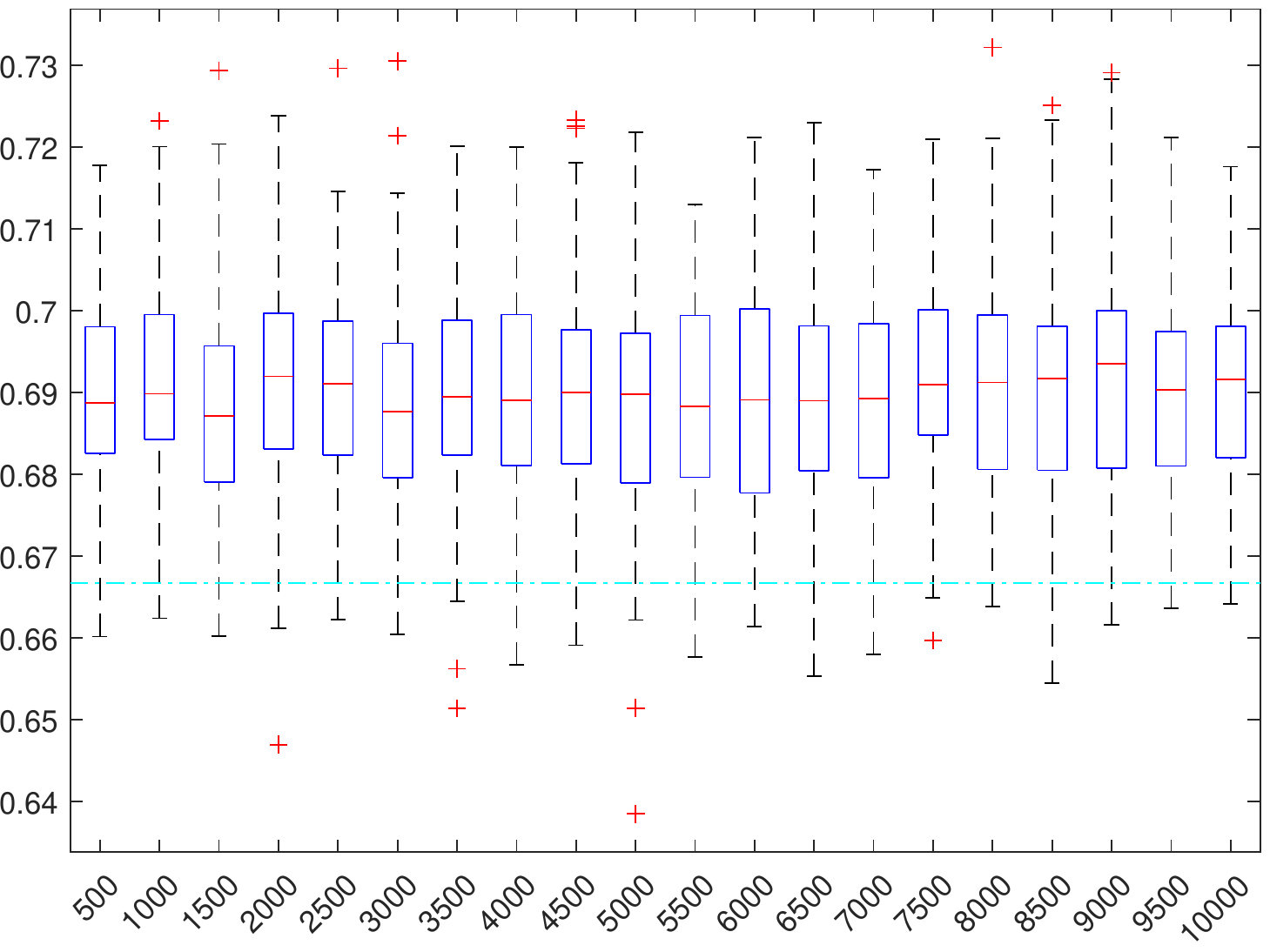}
  }
  \subfigure[$\eta_2$]{
    \includegraphics[width=.3\textwidth]{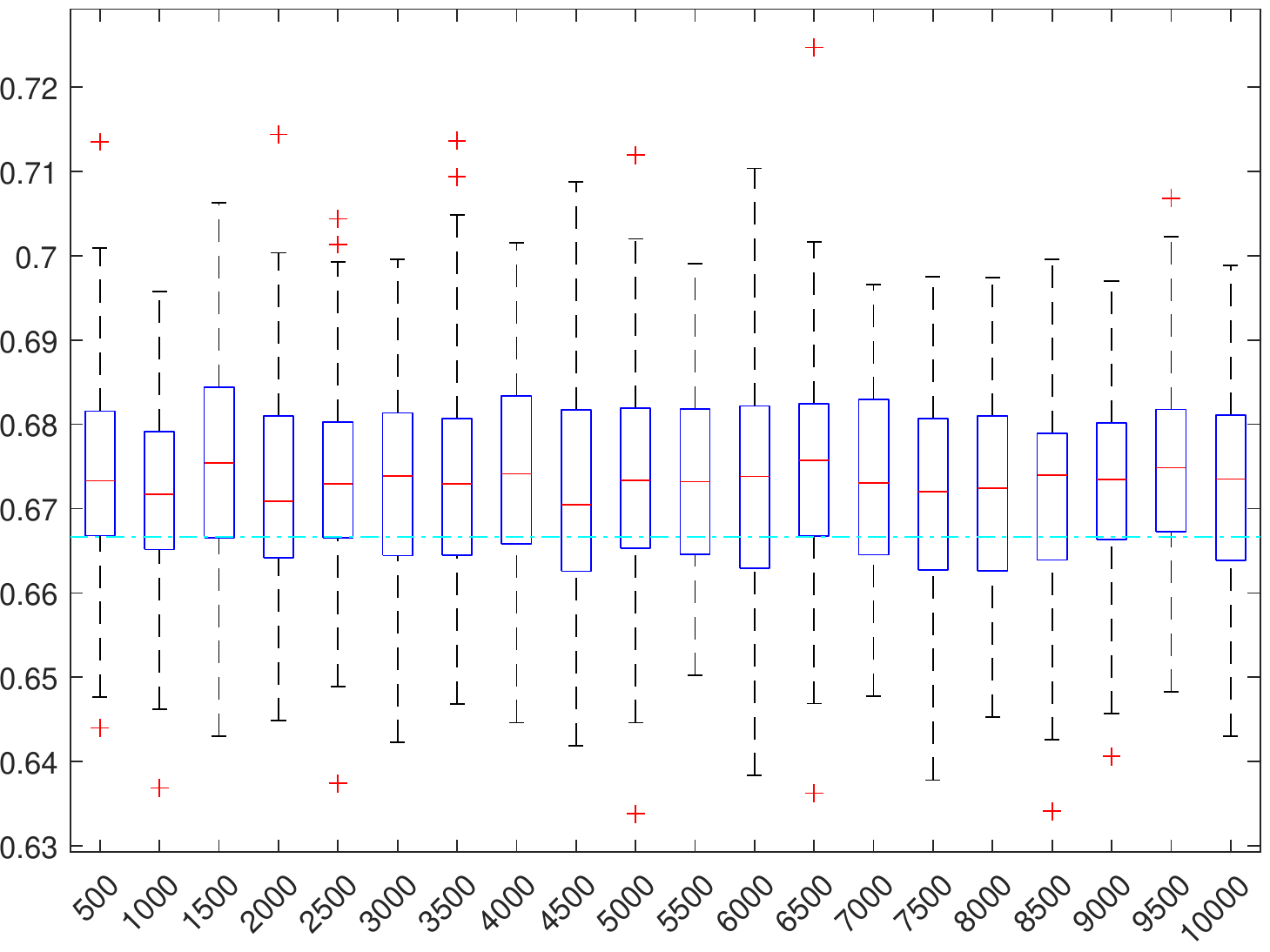}
  }
  \subfigure[$\hat\xi_3$]{
    \includegraphics[width=.3\textwidth]{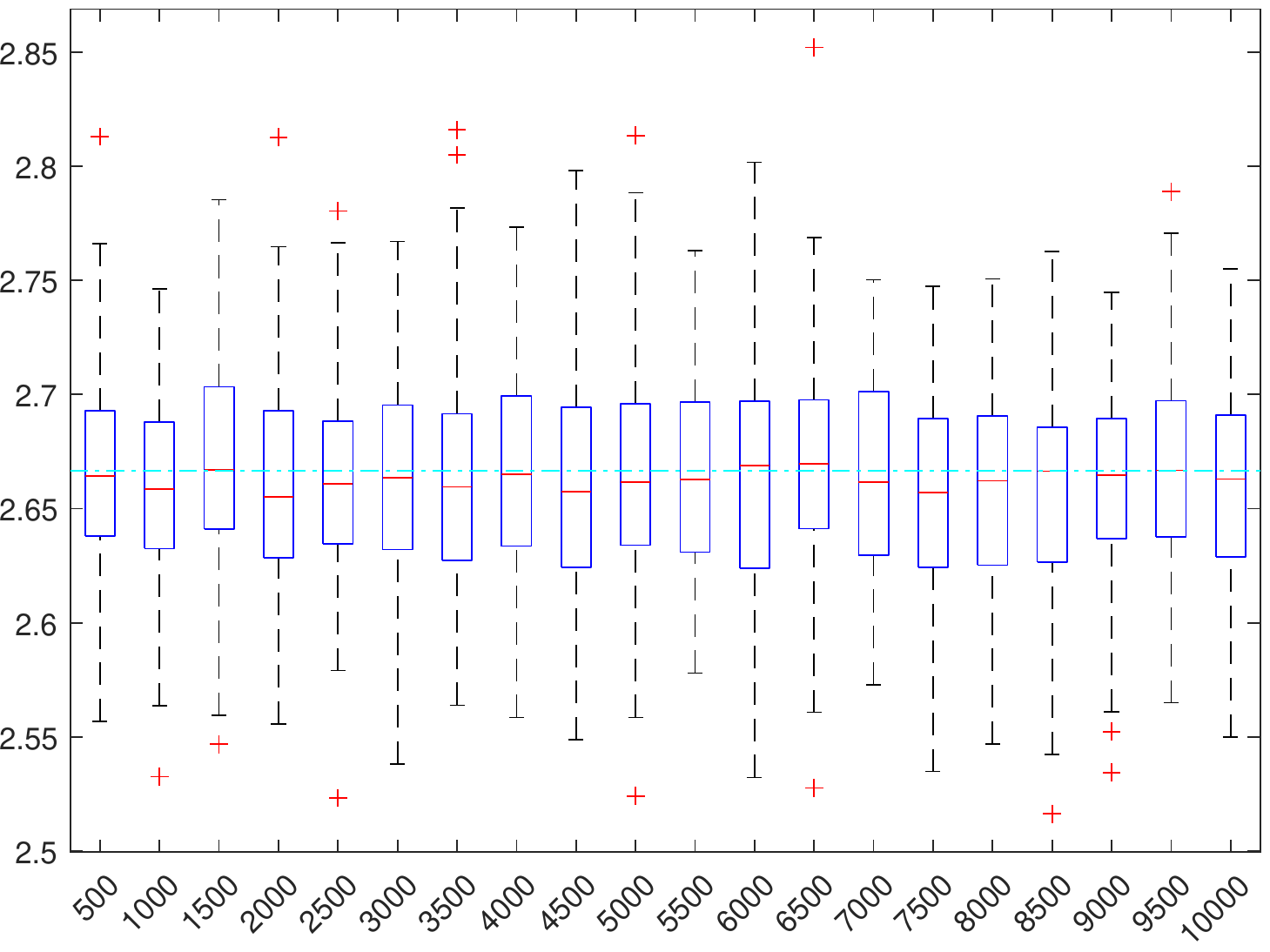}
  }
  \subfigure[$\hat\xi_4$]{
    \includegraphics[width=.3\textwidth]{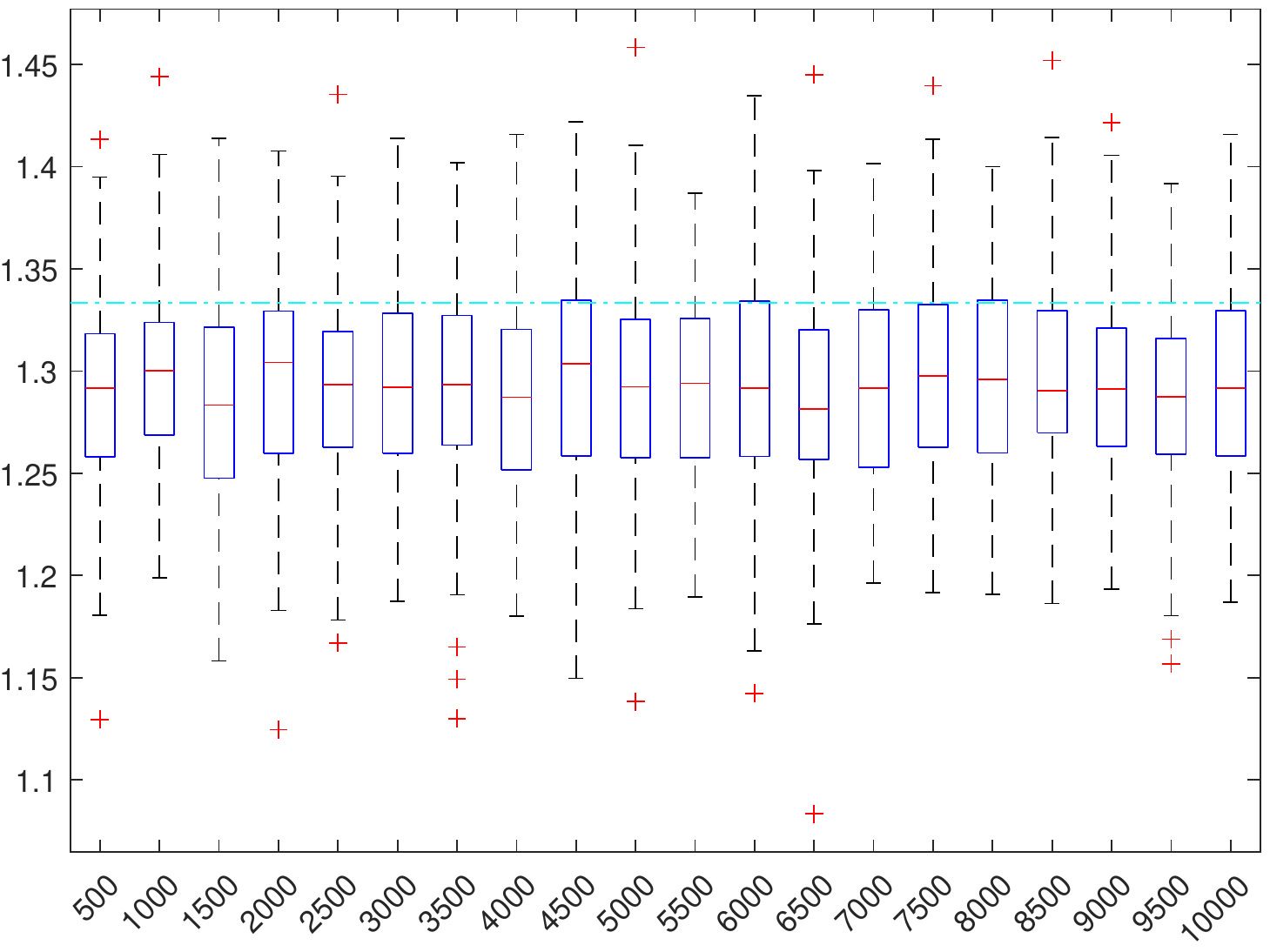}
  }
  \caption{Box plots of repeating sampling $\vec \eta$ and $\vec{\hat\xi}$ with sample size $K=1\times 10^4$ for 100 times via MGDG algorithm in Case 1. Each of the boxes represents the distribution of 100 sampled values of a corresponding variable, with the sampling iterations as on the x-axis, while the ground truth is depicted by the dotted cyan line.}
  \label{fig:simu1:asym}
\end{figure}

To validate the robustness of our MGDG algorithm, it was run 100 times, and the result is summarized in Fig. \ref{fig:simu1:asym}. The two graphs in the first column show the boxplots of sampled $\vec \eta$ in some iterations.
These two panels suggest that the distribution of sampled $\vec \eta$ in multiple repeated trials is very stable, and that the overall error is controlled within an acceptable range. Therefore, the restored $\hat{\vec\xi}$ should also have a stable distribution that is not far away from $\vec\xi^*$, which is consistent with the remaining four panels, where the boxplots of $\hat{\vec\xi}$ have similar quantiles, and the distance between $\hat{\vec\xi}$ and the box is in the same order as the gradient descent threshold. In general, the MGDG algorithm can robustly estimate the posterior distribution in multiple repeated experiments. The other algorithm, MALG, performed similarly, and the results can be found in Section S6 of the supplementary material.

Overall, our method is able to deal with the solver defined in (\ref{eq:sim1}). The algorithms provide samples of $\vec \eta$ that are stably distributed around $\vec \eta^*$. With these samples, $\hat{\vec \xi}$ can be restored; it is also close to the ground truth $\vec\xi^*$. Compared with the Metropolis-within-Gibbs sampler, our MGDG algorithm is more reliable for estimating $\vec\xi$ in this case.

\subsection{Simulation Case 2}
To test our algorithm with more local minimums on the basis of Case 1, another attempt with a 4-component Gaussian-mixture model and $\vec\xi \in \mathbb{R}_+^8$ was performed. In this scenario, the observation was $\vec \robs = \vec R(\vec \xi^*)+ \vec \epsilon$ with  $\vec\xi^* = (\frac{1}{6},\frac{5}{6},\frac{5}{2},\frac{5}{2},\frac{16}{3},\frac{8}{3},9,3)$ and noise variance $\sigma^{2*}_\epsilon = 0.001$ from the following numerical solver:
\begin{equation}
  R(\vec\xi,t) = \sum_{i=1}^4 \frac{\xi_{2i-1}}{(\xi_{2i-1}+\xi_{2i})}\frac{1}{\sqrt{2\pi}}e^{-\frac{(t-(\xi_{2i-1}+\xi_{2i}))^2}{2}}, \quad t \in \cT.
  \label{eq:sim2}
\end{equation}
To perform the dimension reduction, we selected $\vec \eta \in \bR^4$ and $\vec\nu \in \bR^4$ as follows:
\begin{equation*}
  \eta_i = \frac{\xi_{2i-1}}{(\xi_{2i-1}+\xi_{2i})},\quad \nu_i = {(\xi_{2i-1}+\xi_{2i})},\qquad i = 1,\cdots,4,\\
\end{equation*}
where $\vec \eta \in [0,1]^4$. Similarly, these parameters can be regarded as the weights and means in the Gaussian-mixture model, and the MGDG algorithm was initialized by uniformly sampling $\{\vec \eta_i \}_{i=1}^{1000} \sim \text{Uniform}(0,1)^4$ and setting $\vec \eta^{(0)}$ as the sample minimizing the $L^2$ norm between $\vec R(g(\vec \eta_i,\vec{\hat\nu}(\vec \eta_i)))$ and $\vec \robs$. The noise-term variance $\sigma^2_\epsilon$ was initialized with a random value from its prior distribution. The other hyperparameters, prior distributions, and proposal distributions are summarized in Table \ref{tab:simu2}. Given that each of the elements in $\vec \eta = (\eta_1,\eta_2,\eta_3,\eta_4)$ plays the same role in (\ref{eq:sim2}), we sort them in ascending order after each iteration in sampling.

\begin{table*}[htbp]
  \caption{Hyperparameters in Case 2: the prior distributions, proposal distributions, and other parameters.}
  \label{tab:simu2}
  \begin{tabular}{ll}
  \hline
  $\eta_i^\prime \mid \eta_i \sim TN(\eta_i,0.02^2,0,1)$
   & Proposal distribution of $\eta_i$ for $i=1,\cdots,4$\\
  $\pi(\vec \eta)\propto \exp\{-\gamma\|\vec R(g(\vec \eta,\vec{\hat\nu}(\vec \eta)))-\vec \robs\|^2_2\}$
   & Prior of $\vec \eta$\\
  $\cT=[-4,15]$
   & Recording time\\
  $\cT_n\subseteq\cT,\,n =100$
   & Equally spaced time points used in calculation\\
  $B = 500$
   & Number of burn-in samples\\
  $\vec\psi = (\alpha,\beta,\gamma)$
   & $(2,\|\vec R(g(\vec \eta^{(0)},\vec{\hat\nu}(\vec \eta^{(0)})))-\vec \robs\|_2^2/n,10)$\\
  $(\tau, m ) = (0.001,200)$
   & Step size and chain length of sampling $\vec\nu$ in MALG\\
  \hline
  \end{tabular}
\end{table*}

\begin{figure}[htbp]
  \subfigure[$\vec{\eta} - \vec{\eta}^*$]{
    \includegraphics[width=.3\textwidth]{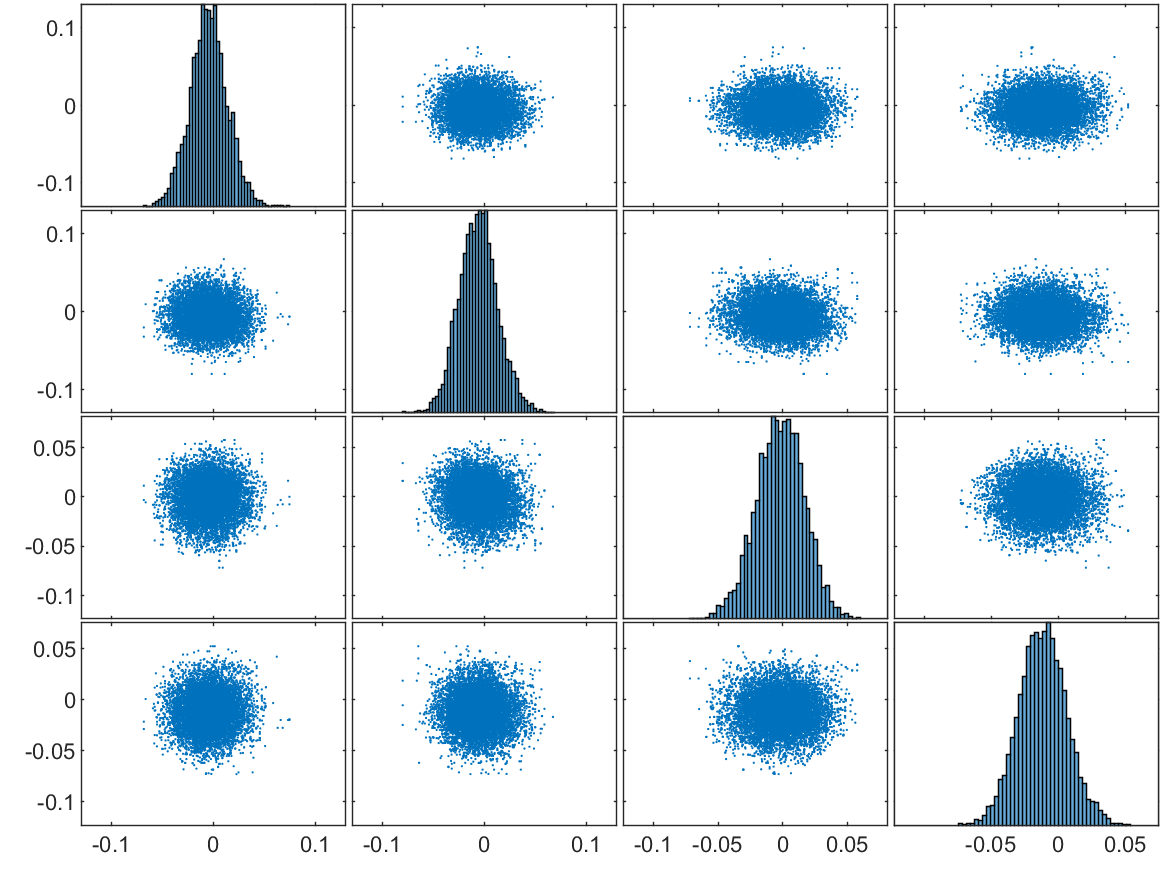}
  }
  \subfigure[$\vec{\hat{\xi}} - \vec{\xi}^*$]{
    \includegraphics[width=.3\textwidth]{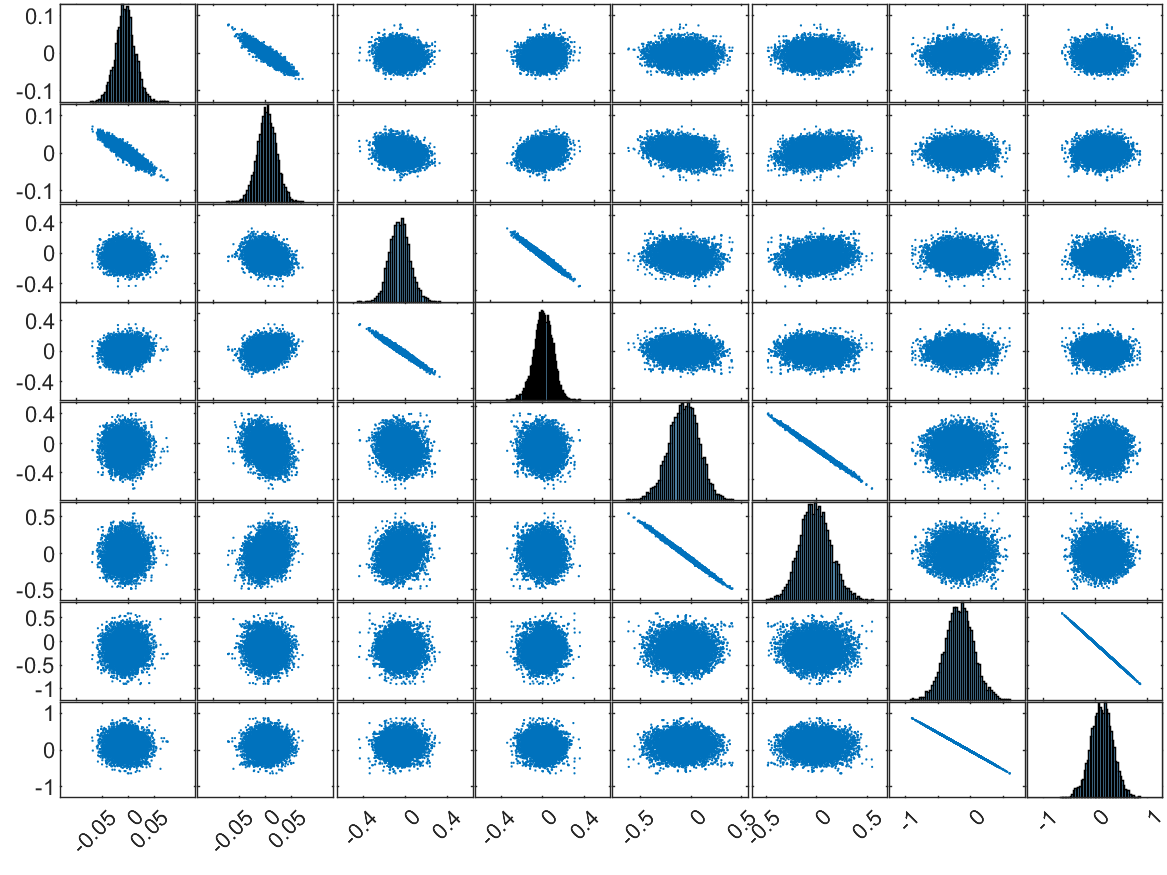}
  }\\
  \subfigure[$\vec{\eta} - \vec{\eta}^*$]{
    \includegraphics[width=.3\textwidth]{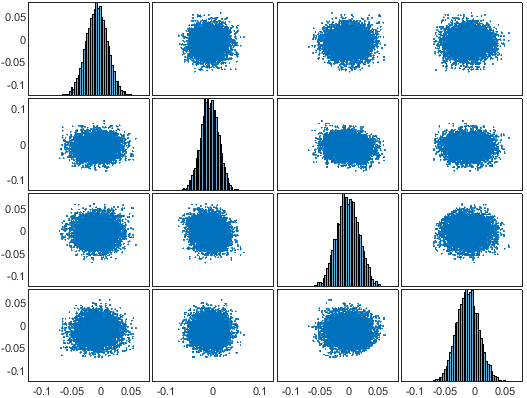}
  }
  \subfigure[$\vec{\nu} - \vec{\nu}^*$]{
    \includegraphics[width=.3\textwidth]{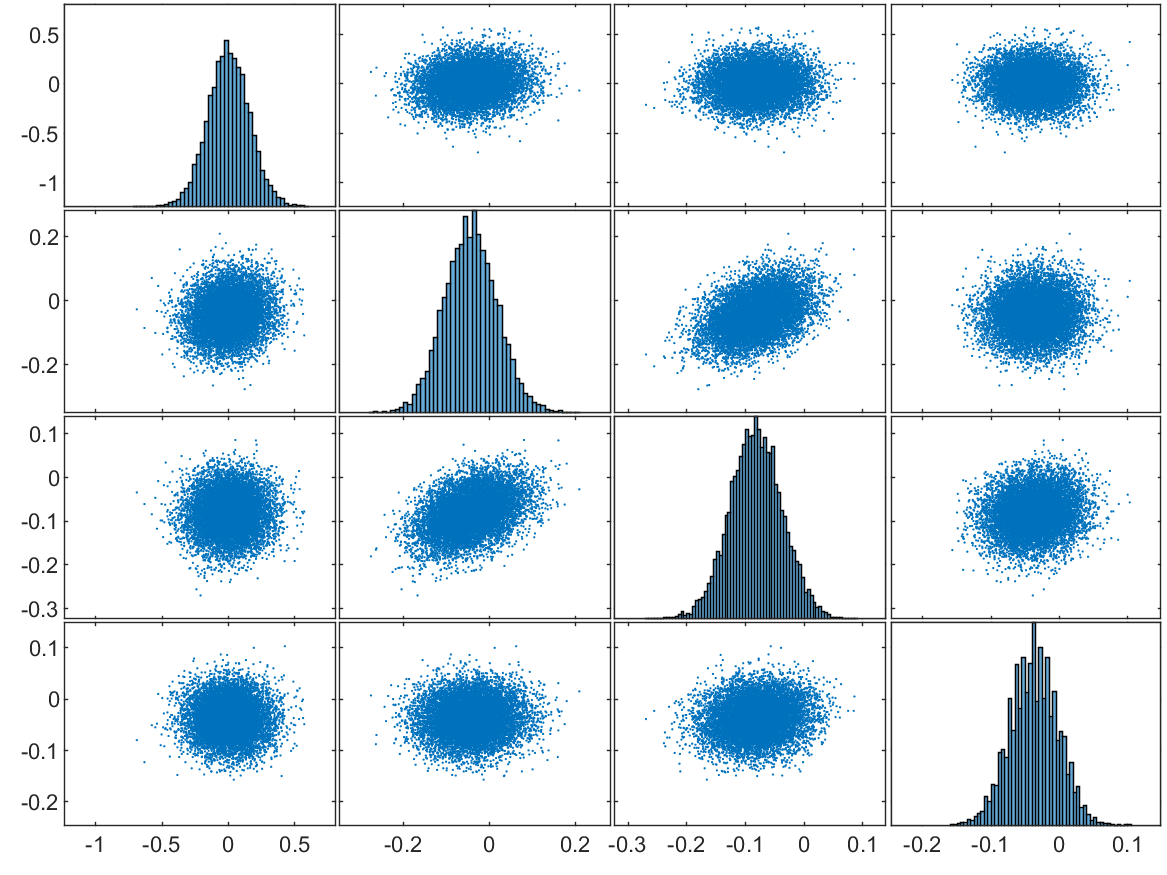}
  }
  \subfigure[$\vec{\hat{\xi}} - \vec{\xi}^*$]{
    \includegraphics[width=.3\textwidth]{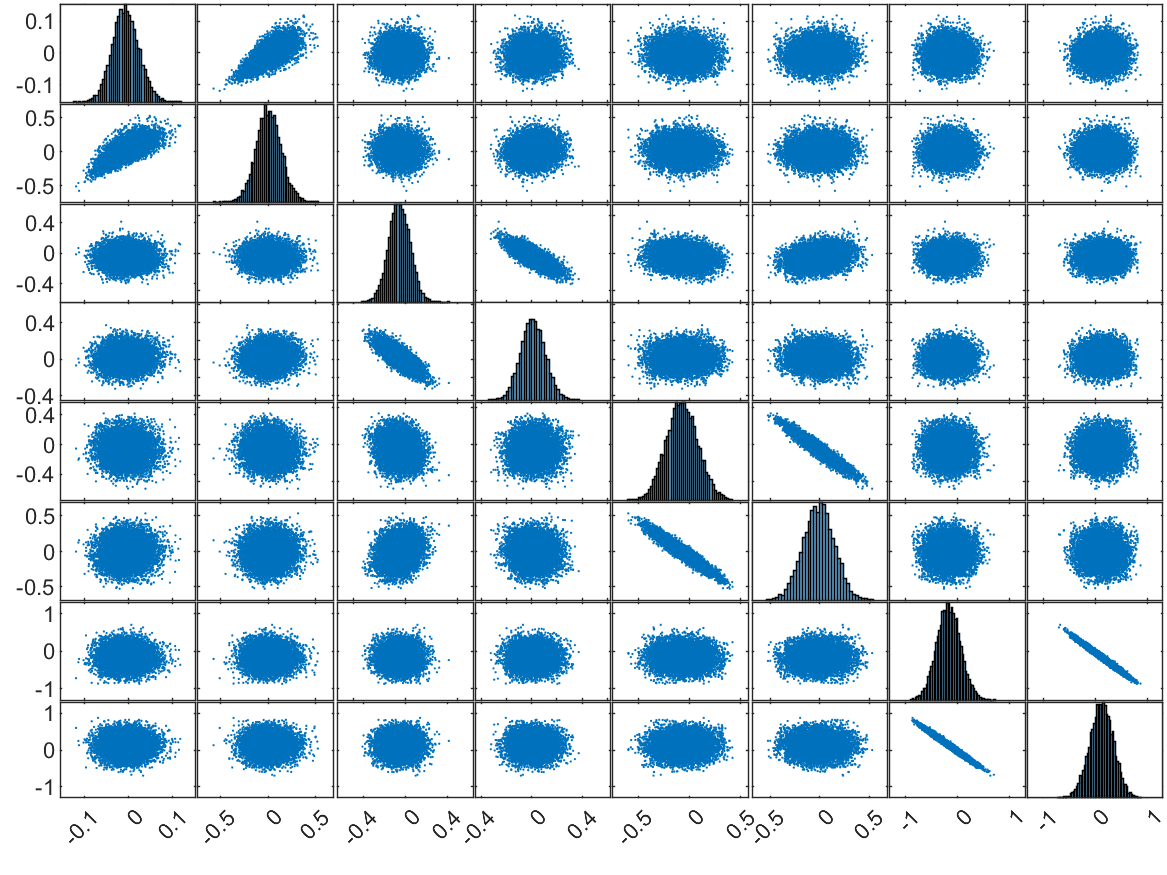}
  }
  \caption{Distribution of residuals in one running trial of Case 2 with sample size $K = 1\times 10^4$. (a,b)Scatter-plot matrix of result from MGDG. (c,d,e) Scatter-plot matrix of result from MALG.}
  \label{fig:simu2:PlotMatrix}
\end{figure}

Fig. \ref{fig:simu2:PlotMatrix} shows the scatter-plot matrix of the residuals in a trial of MGDG and MALG with sample size $K = 1\times 10^4$. These panels suggest that all of the elements are unimodally distributed around $0$, and that the correlations among the elements of $\vec{\eta}$ and $\vec \nu$ are not very significant, although some of the adjacent elements in $\hat{\vec\xi}$ have almost linear correlations. Overall, both $\vec{\eta}$ and $\vec{\hat\xi}$ have nearly normal distributions, and the distance from their center to 0 is within an acceptable range. The algorithms were also repeated 100 times to investigate the robustness with the solver defined in (\ref{eq:sim2}), and the boxplots can be found in Section S6 of the supplementary material.

In general, our algorithms are still able to effectively estimate the parameters in (\ref{eq:sim2}) with acceptable bias. After increasing the dimensions by adding components to the solver, the exchangeable components give $\|\vec R(g(\vec \eta,\vec{\hat\nu}(\vec \eta)))-\vec \robs\|_2$ more local minima. Our algorithm is not much affected by these local solutions, and is able to provide samples that are normally distributed around the ground truth.

\subsection{Simulation Case 3}
\label{Section:SimuCase3}
To complement Case 1, we adopted a Gamma-mixture model to study the performance of our algorithm on skewed observations and steep loss functions. Let the parameter of interest be $\vec{\xi}\in \mathbb{R}_+^4$, and the single observation $\vec \robs = \vec R(\vec \xi^*) + \vec \epsilon$ with  $\vec\xi^* = (4,\frac{3}{4},2,\frac{1}{4})$ and $\sigma^{2*} = 0.001$ from the following numerical solver:
\begin{equation}
  R(\vec\xi,t) = \sum_{i=1}^2 \frac{1}{\Gamma (\xi_{2i-1})\xi_{2i}^{\xi_{2i-1}} } t^{\xi_{2i-1}-1} e^{-\frac{t}{\xi_{2i}}}, \quad t \in \cT.
  \label{eq:sim3}
\end{equation}
To perform the dimension reduction, we selected $\vec \eta \in \bR^2$ and $\vec\nu \in \bR^2$ as follows:
\begin{equation*}
  \eta_i = \xi_{2i},\quad \nu_i = \xi_{2i-1},\qquad i = 1,2.
\end{equation*}
These two parameters can be regarded as the shape and scale parameters in a Gamma-mixture model. With the prior knowledge that $\vec \eta \in [0,1]^2$, we can still uniformly sample $\{\vec \eta_i \}_{i=1}^{1000} \sim \text{Uniform}(0,1)^2$ and initialize $\vec \eta^{(0)}$ as the minimizer of $\|\vec R(g(\vec \eta_i,\vec{\hat\nu}(\vec \eta_i)))-\vec \robs\|_2$/.
The noise variance $\sigma^2_\epsilon$ was initialized by sampling from its prior distribution. The other hyperparameters, prior distributions, and proposal distributions are summarized in Table \ref{tab:simu3}. Although the two elements of $\vec \eta$ can always be exchanged in the sampling, we do not sort them in this case.

\begin{table*}[htbp]
  \caption{Hyperparameters in Case 3: the prior distributions, proposal distributions, and other parameters.}
  \label{tab:simu3}
  \begin{tabular}{ll}
  \hline
  $\eta_i^\prime \mid \eta_i \sim TN(\eta_i,\sigma^2_{q,i},0,1)$
   & Proposal distribution of $\eta_i$ for $i=1,2$\\
  $(\sigma_{q,1},\sigma_{q,2}) = (0.05,0.15)$
   & Propose standard derivation for MALG\\
  $(\sigma_{q,1},\sigma_{q,2}) = (0.08,0.33)$
    & Propose standard derivation for MGDG\\
  $\pi(\vec \eta)\propto \exp\{-\gamma\|\vec R(g(\vec \eta,\vec{\hat\nu}(\vec \eta)))-\vec \robs\|^2_2\}$
   & Prior of $\vec \eta$\\
  $\cT=[0,10]$
   & Recording time\\
  $\cT_n\subseteq\cT,\,n =200$
   & Equally spaced time points used in calculation\\
  $B = 500$
   & Number of burn-in samples\\
  $\vec\psi = (\alpha,\beta,\gamma)$
   & $(2,\|\vec R(g(\vec \eta^{(0)},\vec{\hat\nu}(\vec \eta^{(0)})))-\vec \robs\|_2^2/n,8)$\\
  $(\tau, m ) = (0.0002,200)$
   & Step size and chain length of sampling $\vec\nu$ in MALG\\
  \hline
  \end{tabular}
\end{table*}

\begin{figure}[htbp]
  \subfigure[Input observation]{
    \includegraphics[width=.47\textwidth]{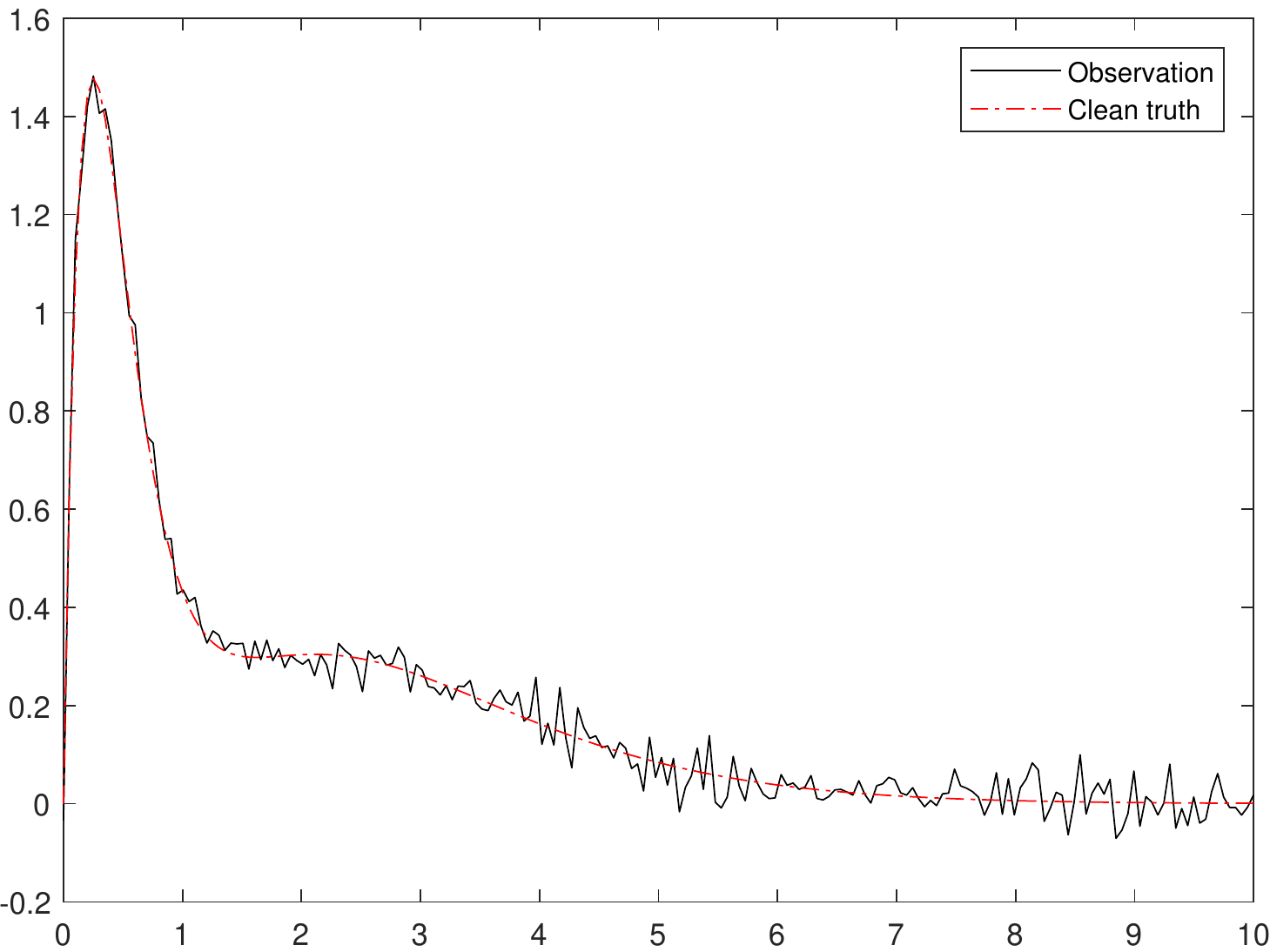}
    \label{fig:simu3:Input}
  }
  \subfigure[An example of MGDG result]{
    \includegraphics[width=.47\textwidth]{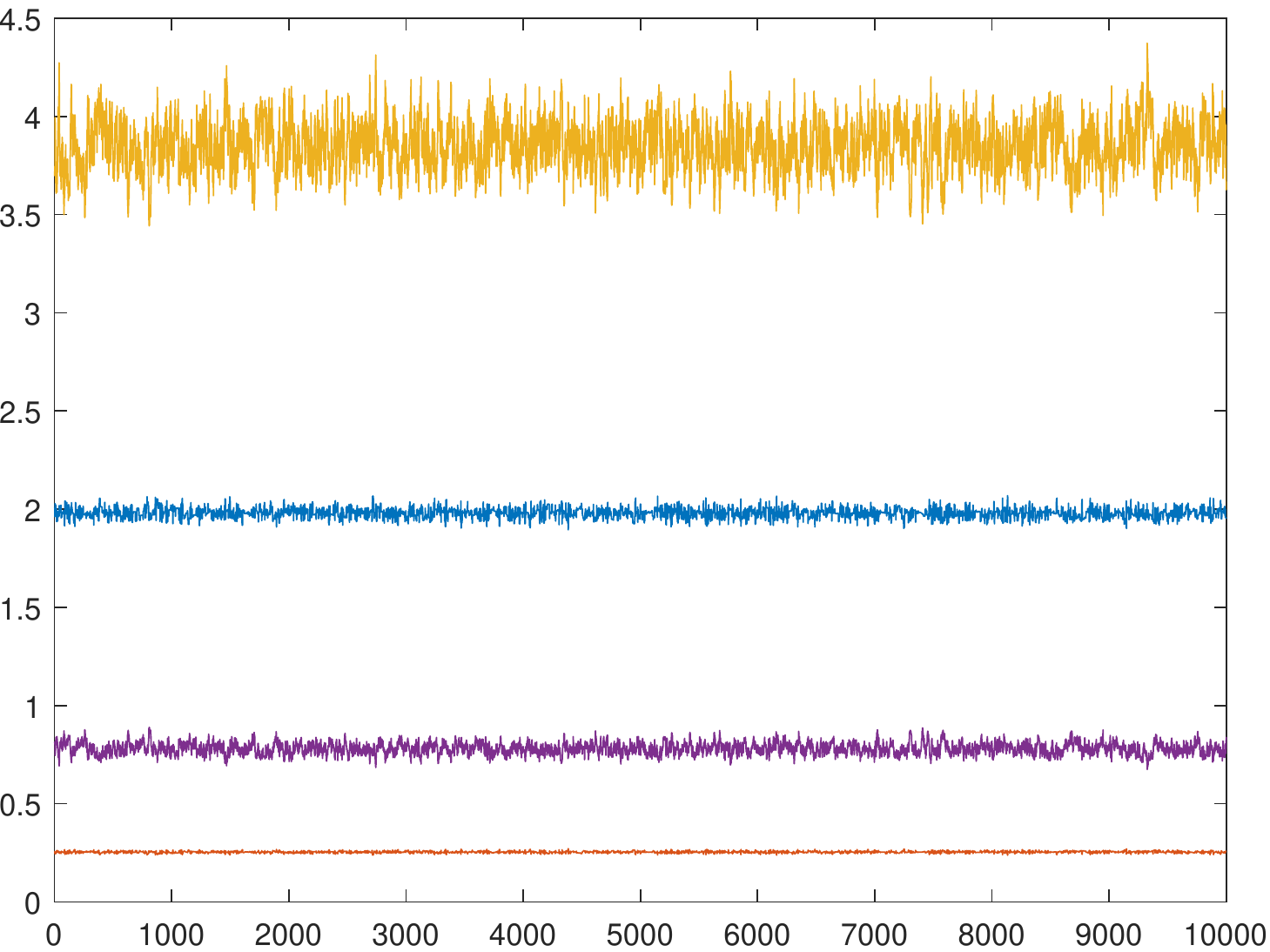}
    \label{fig:simu3:trace}
  }
  \caption{Input data and sampling result from the MGDG algorithm in Case 3. (a) An illustration of the input $\vec\robs$ (solid black line) and clean $\vec R(\vec \xi^*)$ without noise (dotted red line) (b) MGDG result $\vec{\hat\xi}$ with sample size $K=1\times 10^4$ in one trial.}
\end{figure}

Fig. \ref{fig:simu3:Input} roughly shows the shape and noise level of the input data in this simulation case. With such a skewed observation, the $L^2$ distance between $\vec R(\vec\xi)$ and $\vec \robs$ is no longer as symmetrical as in the previous cases. The huge slope on one side makes the gradient descent more difficult, and the rejection rate during the sampling process increases. As shown in 
Fig. \ref{fig:simu3:trace}, although the trajectory of the sample is still stable, it is far less smooth than in the previous cases.

\begin{figure}[htbp]
  \subfigure[$\vec{\eta} - \vec{\eta}^*$]{
    \includegraphics[width=.3\textwidth]{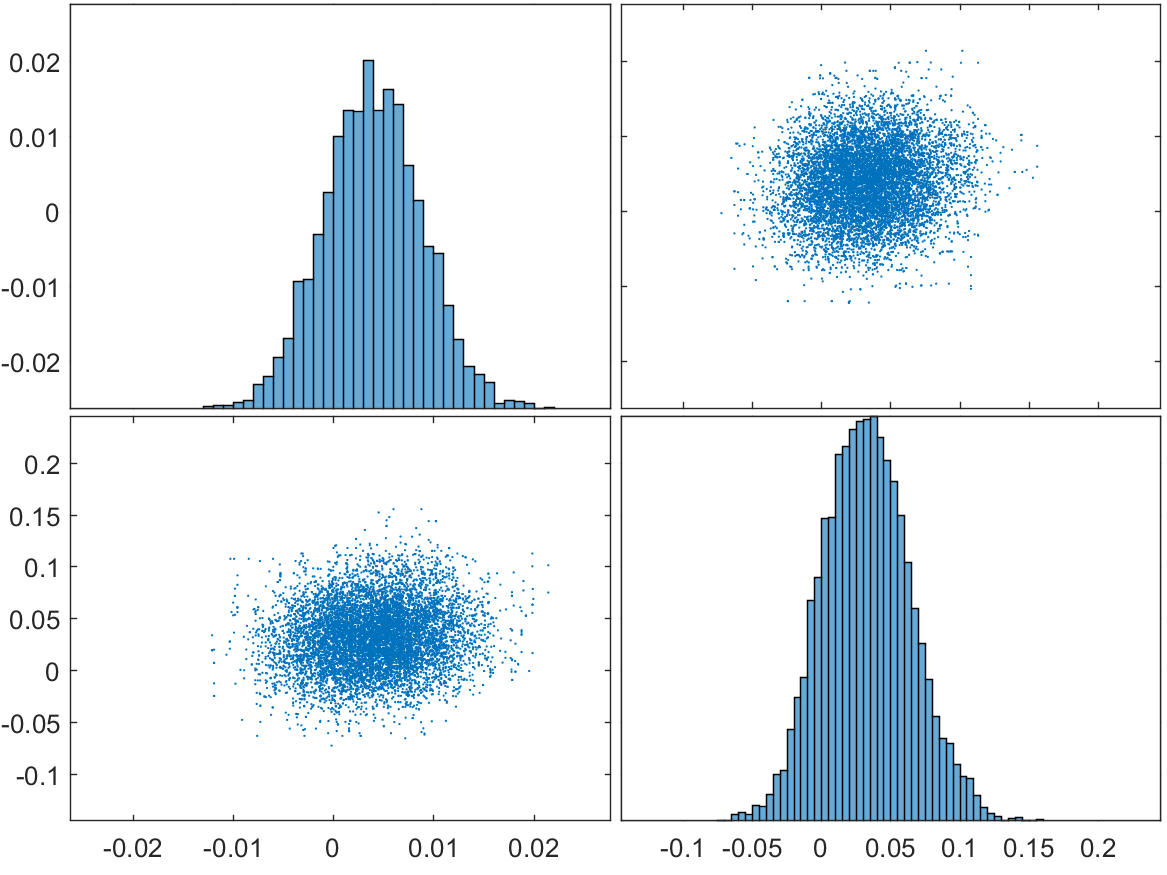}
  }
  \subfigure[$\vec{\hat{\xi}} - \vec{\xi}^*$]{
    \includegraphics[width=.3\textwidth]{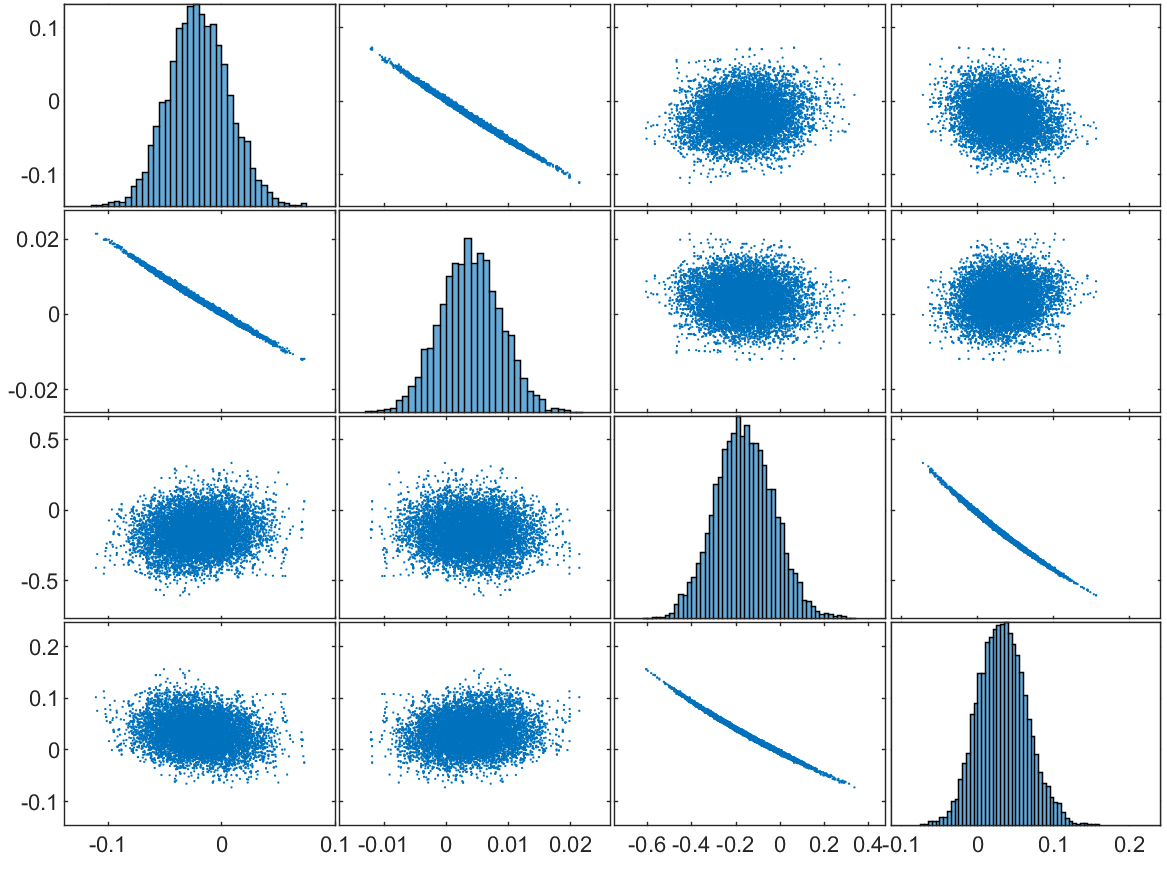}
  }\\
  \subfigure[$\vec{\eta} - \vec{\eta}^*$]{
    \includegraphics[width=.3\textwidth]{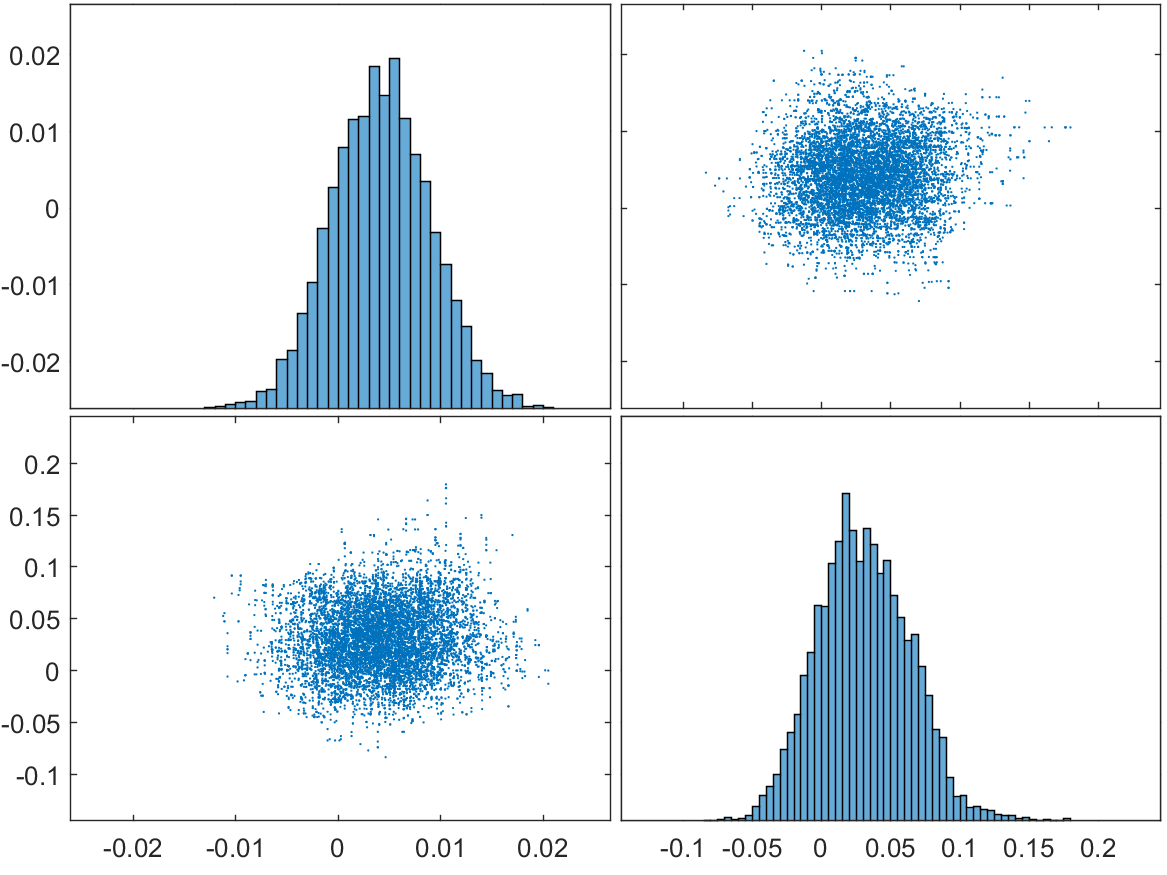}
  }
  \subfigure[$\vec{\nu} - \vec{\nu}^*$]{
    \includegraphics[width=.3\textwidth]{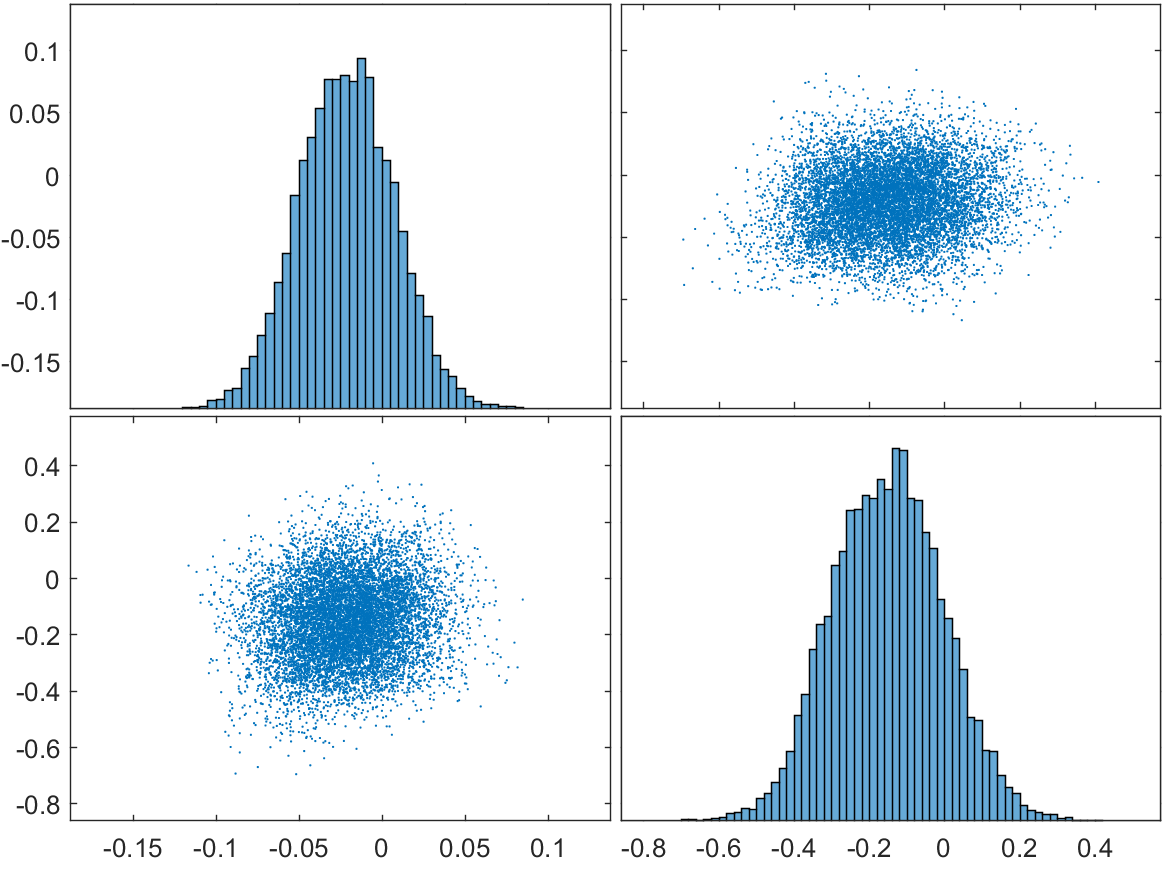}
  }
  \subfigure[$\vec{\hat{\xi}} - \vec{\xi}^*$]{
    \includegraphics[width=.3\textwidth]{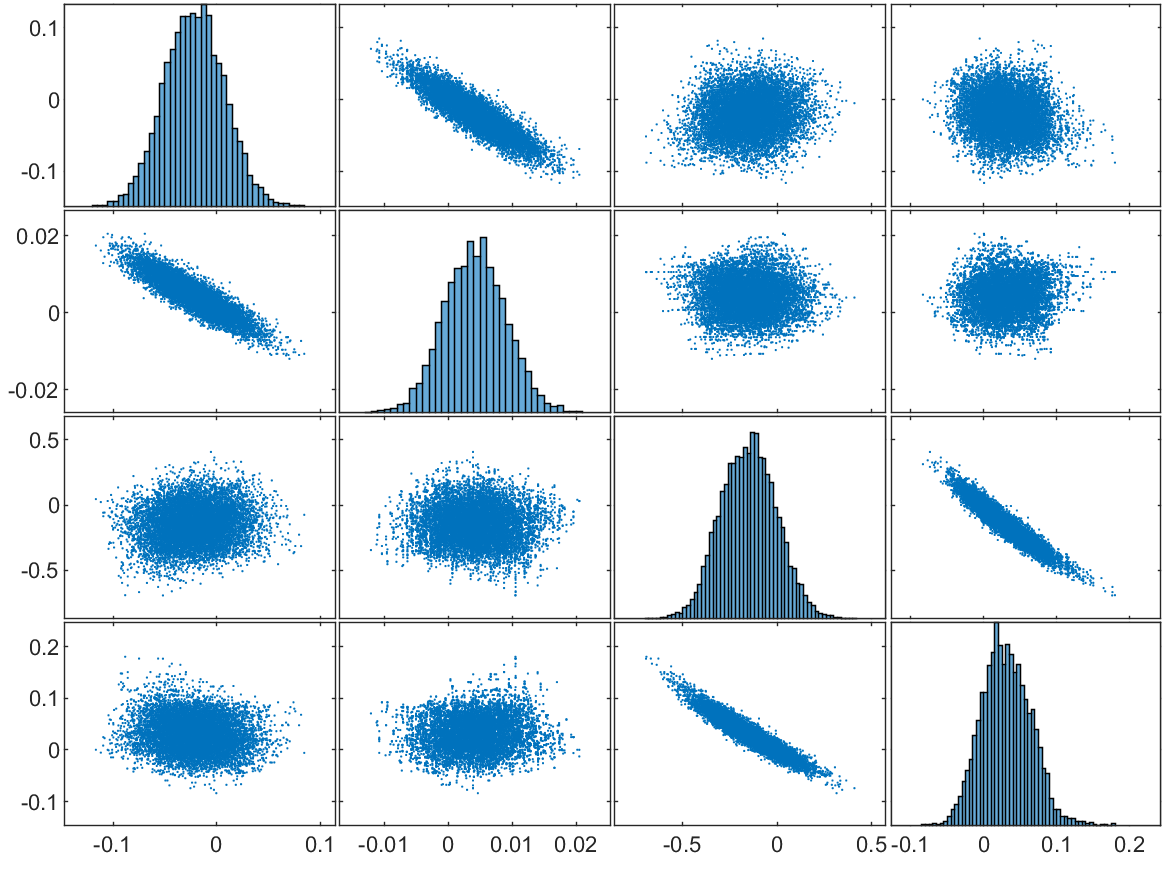}
  }
  \caption{Distribution of residuals in one running trial of Case 3 with sample size $K = 1\times 10^4$. (a,b)Scatter-plot matrix of result from MGDG. (c,d,e) Scatter-plot matrix of result from MALG.}
  \label{fig:simu3:PlotMatrix}
\end{figure}

The estimation in one trial is also evaluated by the residuals of samples, and the corresponding scatter-plot matrices with sample size $K = 1\times 10^4$ are shown in Fig. \ref{fig:simu3:PlotMatrix}. According to these scatter plots, the residuals of each parameter are symmetrically distributed with a single peak, and the centers of the peaks are very close to 0, which is also within the range of the threshold of the gradient descent algorithm. The correlation between $\vec{\eta}$ is not significant, but there seem to be straight lines in the scatter plots, which may be caused by the rejected proposals. As for $\hat{\vec\xi}$, we can clearly see that the residuals of each element are also symmetrically and unimodally distributed around 0, and that the two elements belonging to the same component are highly correlated. In general, in a single experiment, samples can be effectively drawn from the posterior distribution.

According to the results from both algorithms, the bias of parameters related to the second component is much larger, which is because the second peak is flatter in signal and such bias does not change the shape of the observation significantly. We also repeated this sampling experiment 100 times, and the results were similar to that of the original (see Section S6 of the supplementary material). Overall, our method is able to deal with the skew-solver from the Gamma-mixture density function.

\section{Real data application}
In this section, we verify our algorithm with real data. First, we briefly introduce the background of the real experiment, after which the model, parameter settings, and data-generation process are presented in detail. Finally, we evaluate the performance of our method through repeated experiments.

{
\subsection{Experiment background}
Our data consist of gradient separations of a cycloheptanone/cyclohexanone mixture. The experiments were conducted on an Agilent 1200 system (Palo Alto, CA, USA), with a 150 mm $\times$ 4.6 mm Kromasil column (AkzoNobel Eka, Bohus, Sweden) filled with C18-bonded porous silica, with an average particle diameter of 5 $\mu$m. The system contained a 900 $\mu$L-injection-loop auto-sampler, a binary pump system, a diode-array UV detector, and a column thermostat. In all experiments, the column was held at a constant temperature of 22 $^\circ$C, and the flow rate was 1.0mL/min.

The solutes adopted in the experiments were cyclohexanone ($\geq$95\%) and cycloheptanone ($\geq$95\%) from Sigma-Aldrich (Steinheim, Germany), while the solvents used in the pycnometer measurements were dichloromethane ($\geq$99.5\%) from VWR International (Paris, France) and isopropanol (HPLC grade) from Fisher Scientific (Loughborough, UK). The mobile phase was composed of HPLC-grade methanol from Fisher Scientific (Loughborough, UK) and deionized water, with a conductivity of 18.2 M$\Omega$cm, supplied by a Milli-Q Plus 185 water-purification system from Millipore (Merck Millipore, MA, USA).

During the experiments, calibration curves for cyclohexanone and cycloheptanone were recorded at 280 nm for several mobile-phase compositions. The column hold-up volume measured with a pycnometer was 1.38 mL. To match the injected amount of solute, the total area under the peaks in the elution profiles was adjusted. Indistinguishable inlet and outlet effects were included in the injection profile, and the injection profile was recorded and used in the calculations.
}

\subsection{Model setting and data generation}
\begin{figure}
  \subfigure[]{
    \includegraphics[width=.47\textwidth]{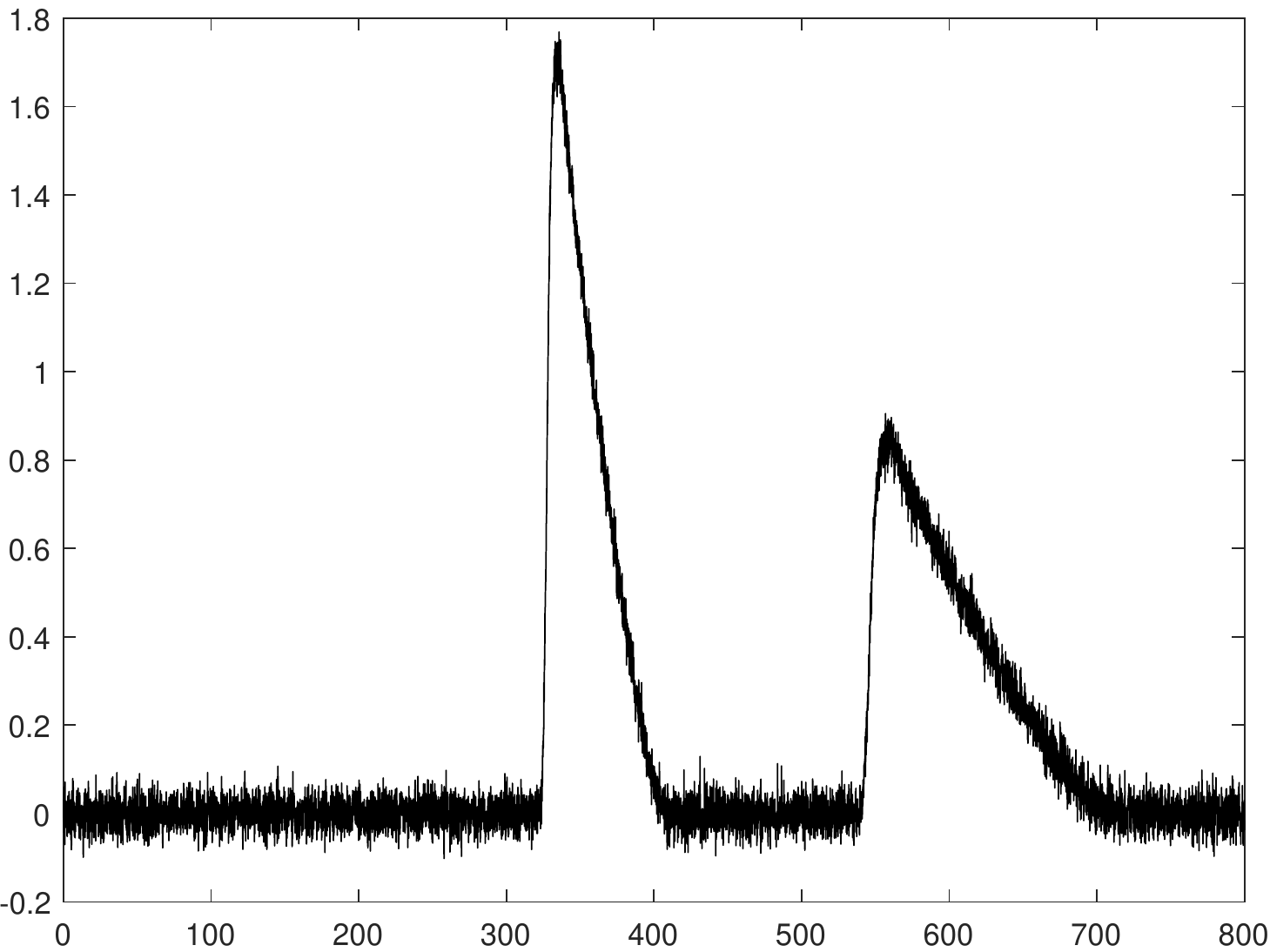}
    \label{fig:RealData:example:noise}
  }
  \subfigure[]{
    \includegraphics[width=.47\textwidth]{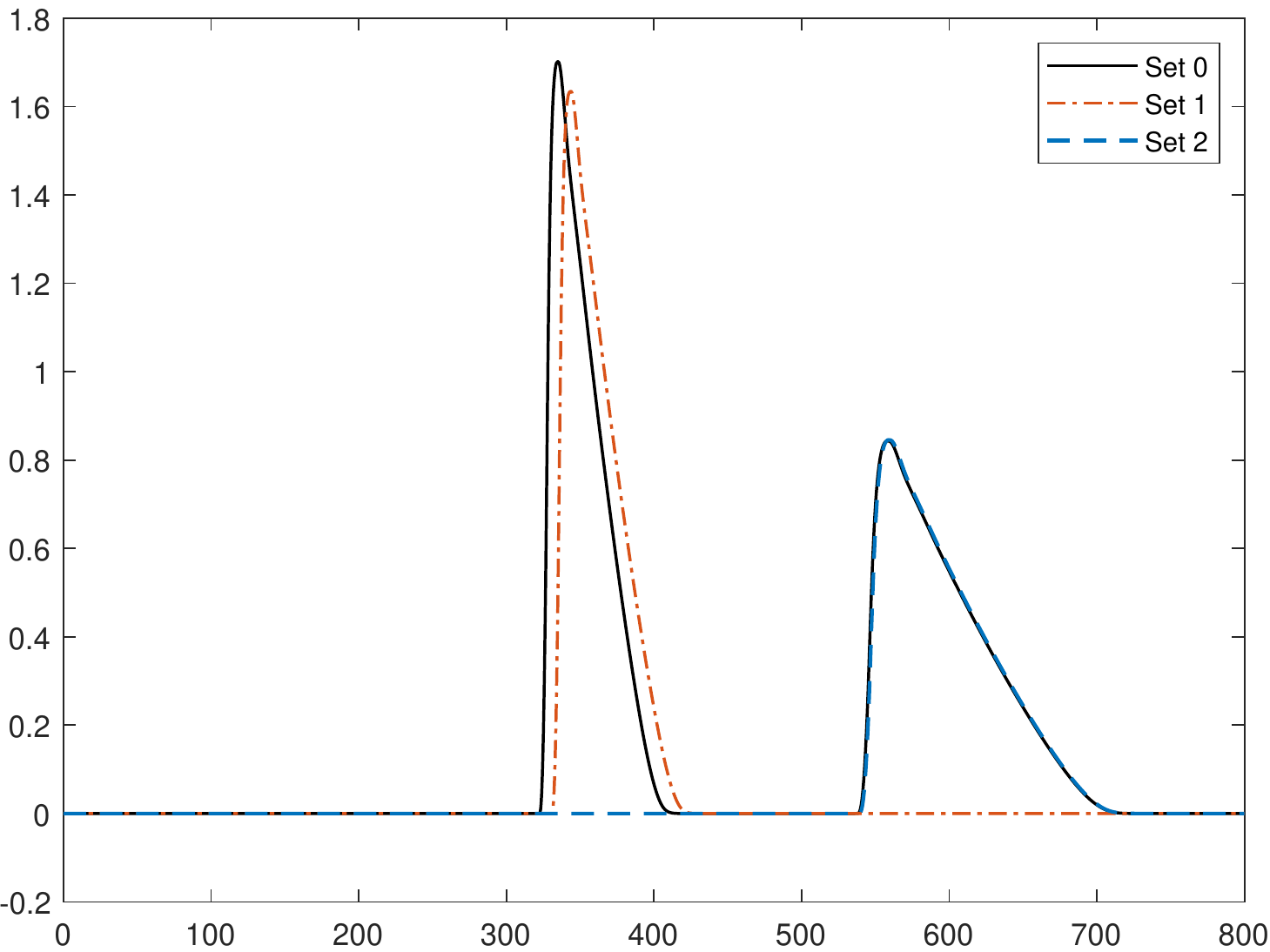}
    \label{fig:RealData:example:dim}
  }
  \label{fig:RealData:example}
  \caption{(a) An example observation with noise from the time-dependent convection-diffusion system. (b) An illustration of the two-peaked data. Set 0 is the clean data with the full set of parameters. Set 1 (or Set 2) stands for the observation with the second (or first) group of parameters set to be $0$.}
\end{figure}

Based on the experiment background, we can consider a chromatography system with a 150 mm $\times$ 4.6 mm column, a flow rate of 0.7 mL/min, and 9000 theoretical plates. The hold time is 1.5 min and 400 $\mu$L samples are introduced using rectangular injection profiles. In this section, we focus on a single observation from
$$R(\vec\xi, t)=\sum_{i=1}^2 C_i(L,t;\,\vec\xi),\quad t\in\cT,$$
where $C_i(L,t;\,\vec\xi)$ is the solution to the time-dependent convection-diffusion system defined in (\ref{eq:solver}); all the other parameters adopted in that PDE system are summarized in Table \ref{tab:RealSolver}. 
In obtaining real data, researchers usually make multiple observations of chromatographic curves with the same parameter settings and average the records to reduce errors. This allows the noise term to have a smaller variance and the averaged noise will be distributed closer to a Gaussian distribution.
Overall, the noise level of observation $\vec\robs$ used in this section is similar to the one shown in Fig. \ref{fig:RealData:example:noise}.

\begin{table*}[htbp]
  \caption{Main parameters adopted in real data solver.}
  \label{tab:RealSolver}
  \begin{tabular}{ll}
  \hline
  Parameter          & Description\\
  \hline
  $u = 0.125$        & Linear velocity [cm/s]\\
  $L = 15$           & Column length [cm]\\
  $T = L/u$          & Dead time [s]\\
  $F = 0.7806$       & Phase ratio\\
  $D_a = 0.00010417$ & Diffusion constant\\
  $g_i(x ) \equiv 0$ & Initial condition\\
  $h = [5,0]$        & Injection profile (mM)\\
  \hline
  \end{tabular}
\end{table*}

To simplify the problem, we silence part of the parameters. The solid black line in Fig. \ref{fig:RealData:example:dim} is an example of a clean bimodal observation from $R(\vec\xi_0, t)$ with full parameter set $\vec\xi_0 = (a_{I,1},a_{II,1},b_{I,1},b_{II,1}, a_{I,2},a_{II,2}, b_{I,2},b_{II,2}) = (2,1,0.1,0.05,4,2,0.2,0.1)$ and $h = [15,15]$. Each peak in that signal roughly corresponds to a set of parameters $(a_{I,i},a_{II,i},b_{I,i},b_{II,i})$, which also represents a component in the sample.
We observe that these peaks can be separated after adjusting injection parameter $h$, and the second one almost has the same shape as that from the observation from $R(\vec\xi_2, t)$, where $\vec\xi_2 = (0,0,0,0,4,2,0.2,0.1)$ and $h = [0,15]$, which is shown as Set 2 in Fig. \ref{fig:RealData:example:dim}.
Therefore, $(a_{I,2},a_{II,2},b_{I,2},b_{II,2})$ can be directly estimated from the signal segment containing only the second peak, and the estimator will help us figure out the remaining parameters.
Given the fact that multiple sets of the parameters will only bring additional calculation difficulties, we focus only on the case of one peak by setting the injection profile of the second group as $0$ and letting the parameter of interest be $\vec \xi = (a_{I},a_{II},b_{I},b_{II})$ in this section.

\subsection{Parameter estimation}
\label{Section:RealData}
After identifying the model and parameter settings, we let the observation of interest be $\vec\robs$ from $\vec R(\vec \xi^*)$, where the noise level estimated from the flat part (i.e. $t\in [0,300]$) is $\widehat{\sigma}^{2}_\epsilon = 0.001$. To estimate $\vec \xi^*$ with the MGDG algorithm, we set $\vec \eta \in \bR^2$ and $\vec\nu \in \bR^2$ as
\begin{align*}
  \eta_1 = \frac{a_{I}}{a_{I}+a_{II}},\quad
  \eta_2 = \frac{b_{I}}{b_{I}+b_{II}},\quad
  \nu_1 = {a_{I}+a_{II}},\quad
  \nu_2 = {b_{I}+b_{II}}.
\end{align*}
To ensure the algorithm MALG produces a valid Markov chain, we adopt the element-wise transformation
$$
\vec \eta = \frac{\tanh(\widetilde{\vec\eta})+1}{2},\quad
\vec\nu = \exp(\widetilde{\vec\nu}),
$$
and sample the elements $(\widetilde{\vec\eta},\widetilde{\vec\nu})$ from the entire real line. 

Since $(a_{I},a_{II},b_{I},b_{II})$ are also non-negative and $\vec \eta \in[0,1]^2$, we still sample $\{\vec \eta_i \}_{i=1}^{600} \sim \text{Uniform}(0,1)^2$ and initialize $\vec \eta^{(0)}$ as the one minimizing $\|\vec R(g(\vec \eta_i,\vec{\hat\nu}(\vec \eta_i)))-\vec \robs\|_2$ with respect to $\vec \eta_i$. The noise variance $\sigma^2_\epsilon$ is initialized with a sampled value from its prior. Together with the other distributions and hyperparameters summarized in Table \ref{tab:RealData}, $K=3000$ posterior samples can be obtained from our algorithms.
\begin{table*}[htbp]
  \caption{Settings in real data application: the prior distributions, proposal distributions, and other parameters.}
  \label{tab:RealData}
  \begin{tabular}{ll}
  \hline
  $\eta_i^\prime \mid \eta_i \sim TN(\eta_i,0.02^2,0,1)$
   & Proposal distribution of $\eta_i$ for $i=1,2$ in MGDG\\
  $\widetilde{\eta}_i^\prime \mid \widetilde{\eta}_i \sim N(\widetilde{\eta}_i,0.05^2)$
   & Proposal distribution of $\widetilde{\eta}_i$ for $i=1,2$ in MALG\\
  $\pi(\vec \eta)\propto \exp\{-\gamma\|\vec R(g(\vec \eta,\vec{\hat\nu}(\vec \eta)))-\vec \robs\|^2_2\}$
   & Prior of $\vec \eta$\\
  $\cT=[0,750]$
   & Recording time\\
  $\cT_n\subseteq[300,500],\,n =100$
   & Equally spaced time points used in calculation\\
  $B = 500$
   & Number of burn-in samples\\
  $\vec\psi = (\alpha,\beta,\gamma)$
    & $(2,\|\vec R(g(\vec \eta^{(0)},\vec{\hat\nu}(\vec \eta^{(0)})))-\vec \robs\|_2^2/n,8)$\\
  $(\tau, m ) = (1\times 10^{-8},20)$
   & Step size and chain length of sampling $\vec\nu$ in MALG\\
  \hline
  \end{tabular}
\end{table*}

\begin{figure}
  \subfigure[$\vec \eta$ from MGDG]{
    \includegraphics[width=.3\textwidth]{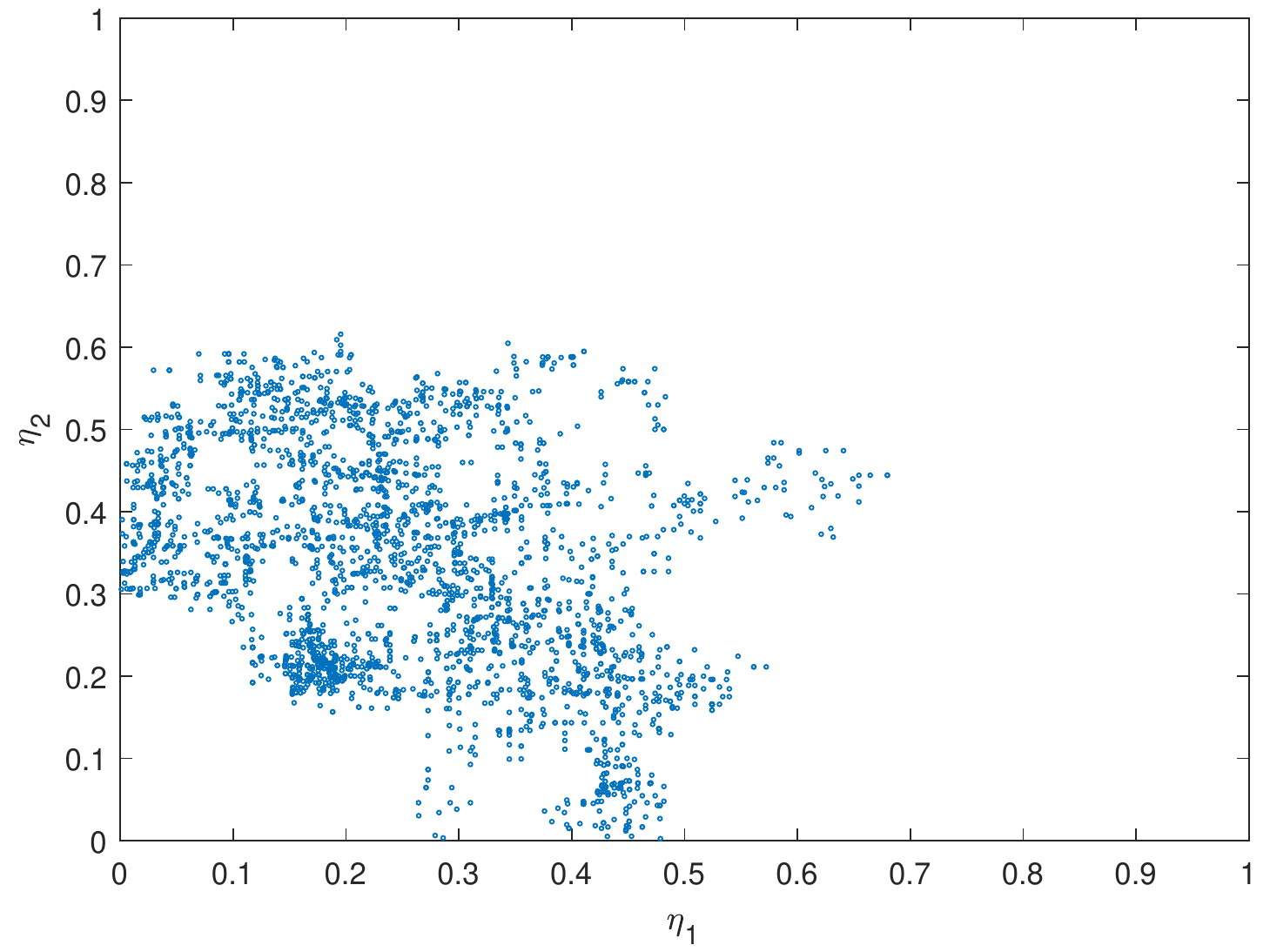}
  }
  \subfigure[$\sigma^2_\epsilon$ from MGDG]{
    \includegraphics[width=.3\textwidth]{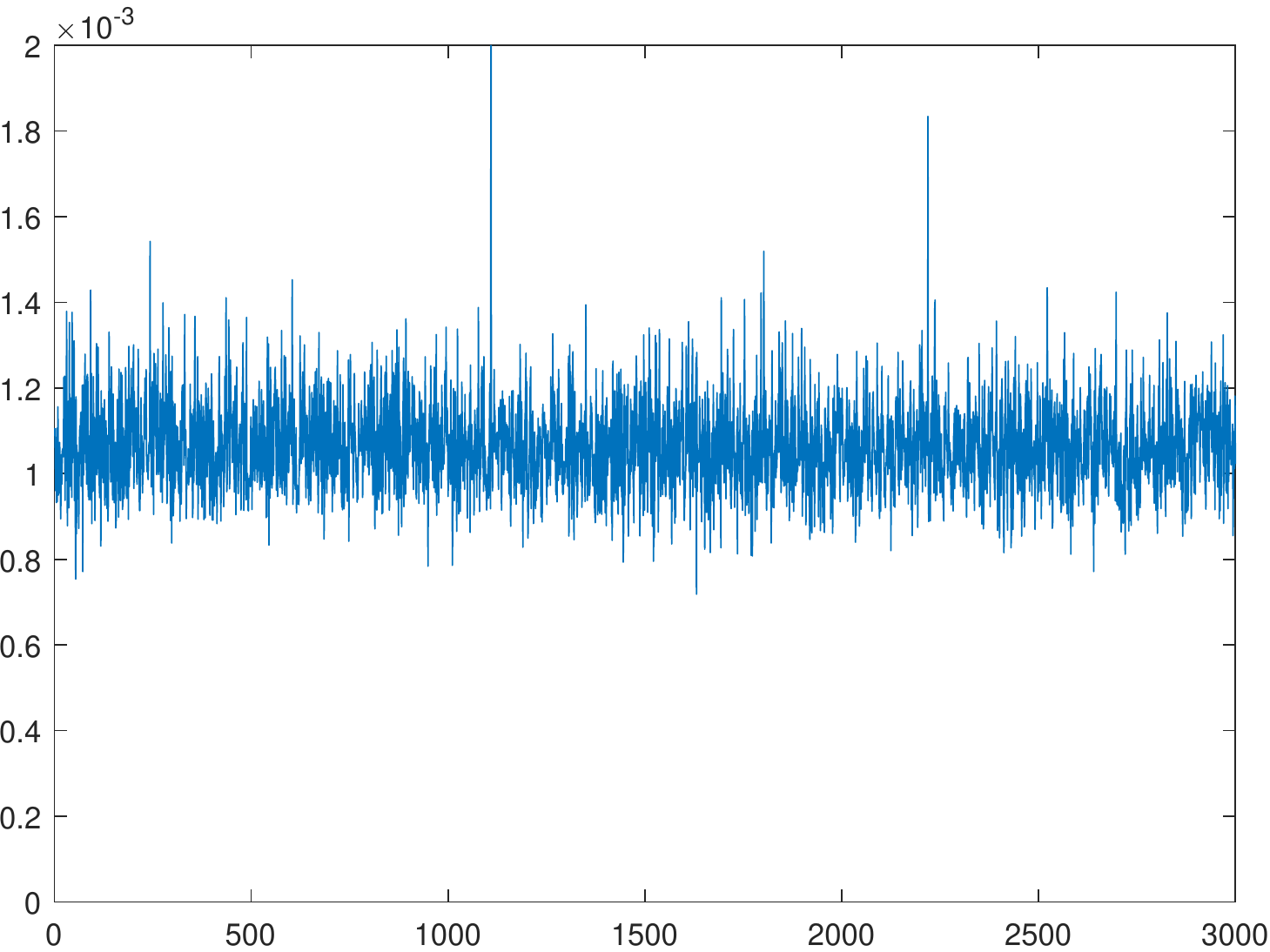}
  }
  \subfigure[$95\%$ CI from MGDG]{
    \includegraphics[width=.3\textwidth]{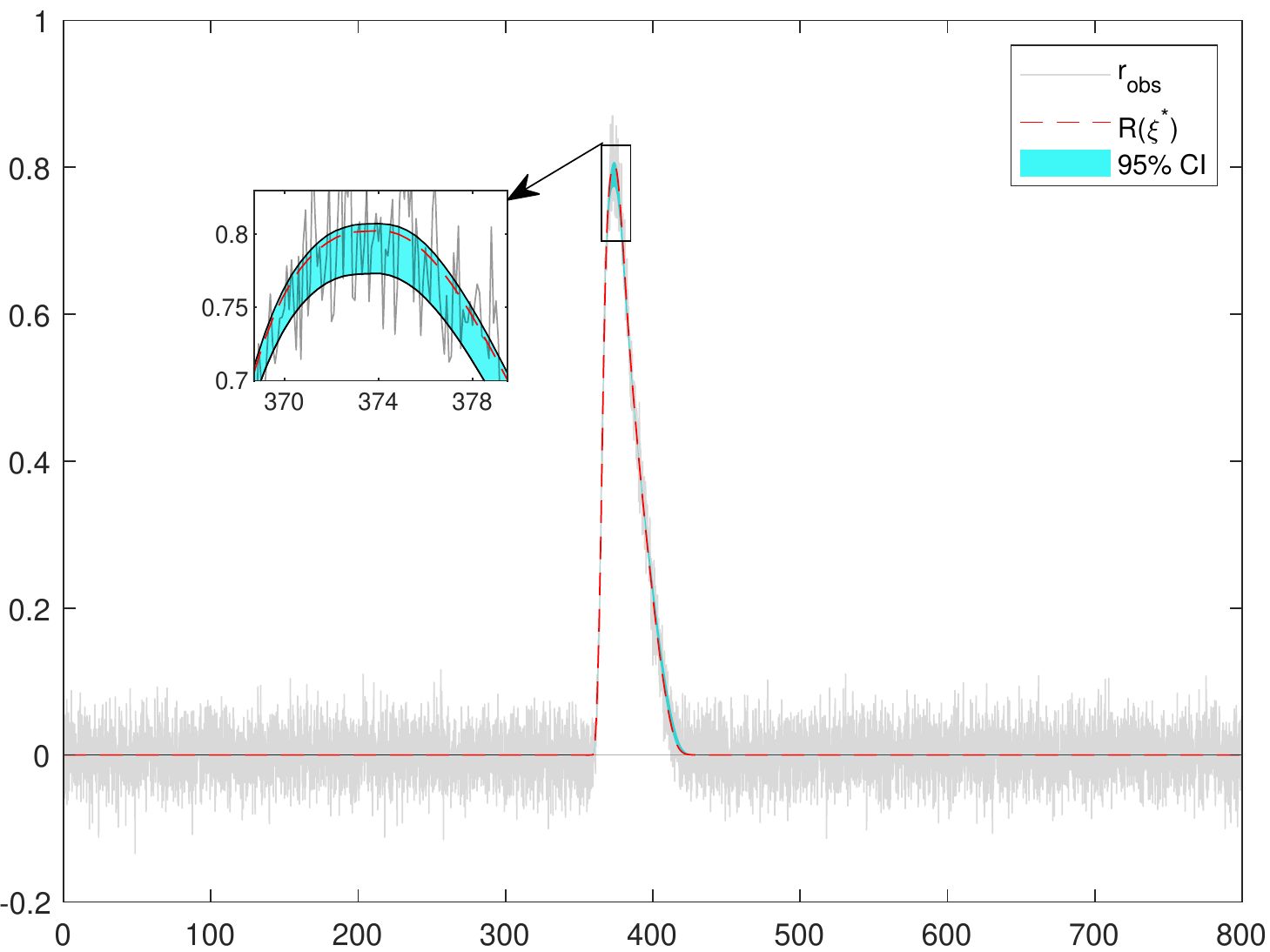}
    \label{fig:RealData:MMHGD_CI}
  }\\
  \subfigure[$\vec \eta$ from MALG]{
    \includegraphics[width=.3\textwidth]{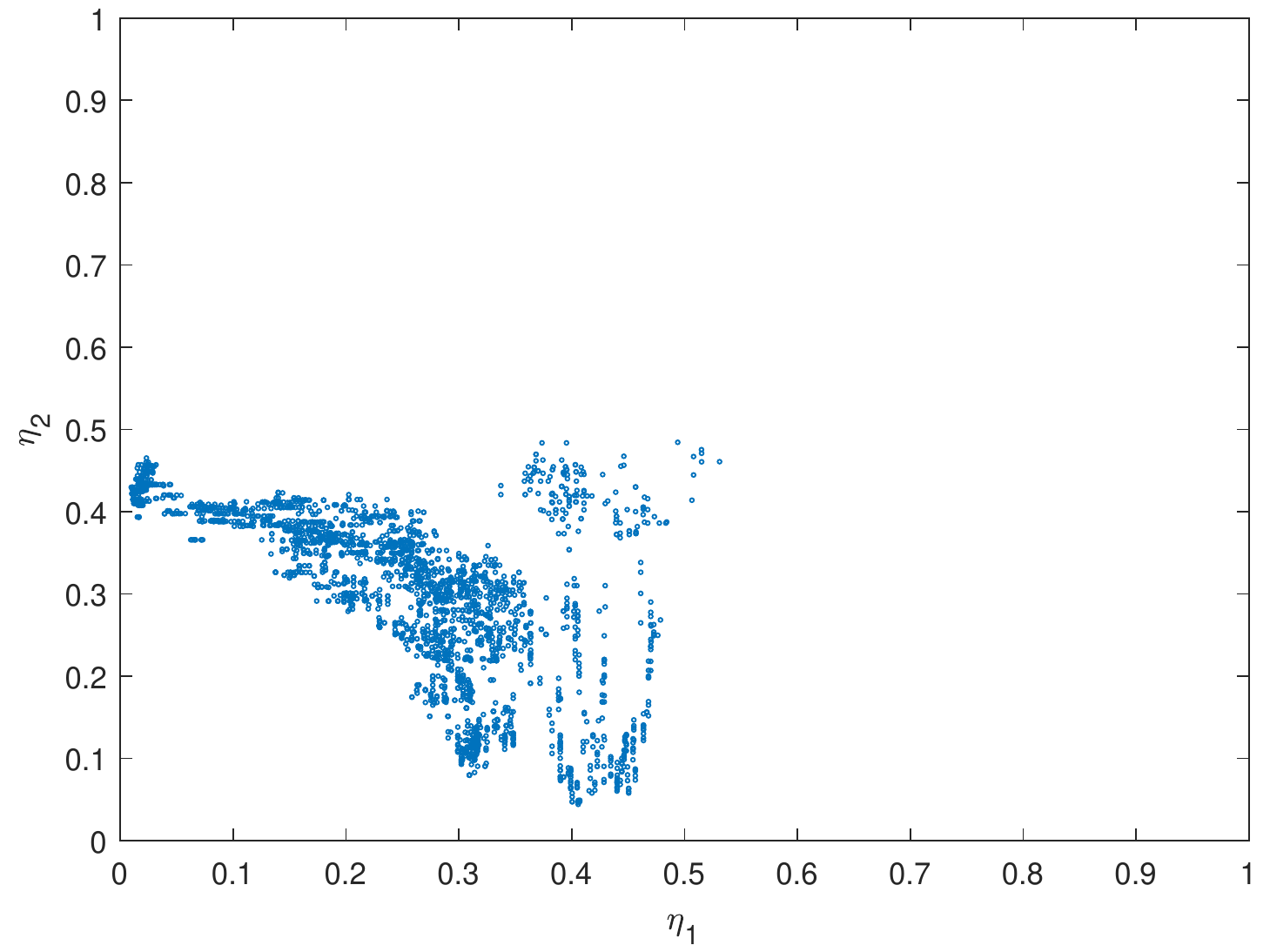}
  }
  \subfigure[$\vec \nu$ from MALG]{
    \includegraphics[width=.3\textwidth]{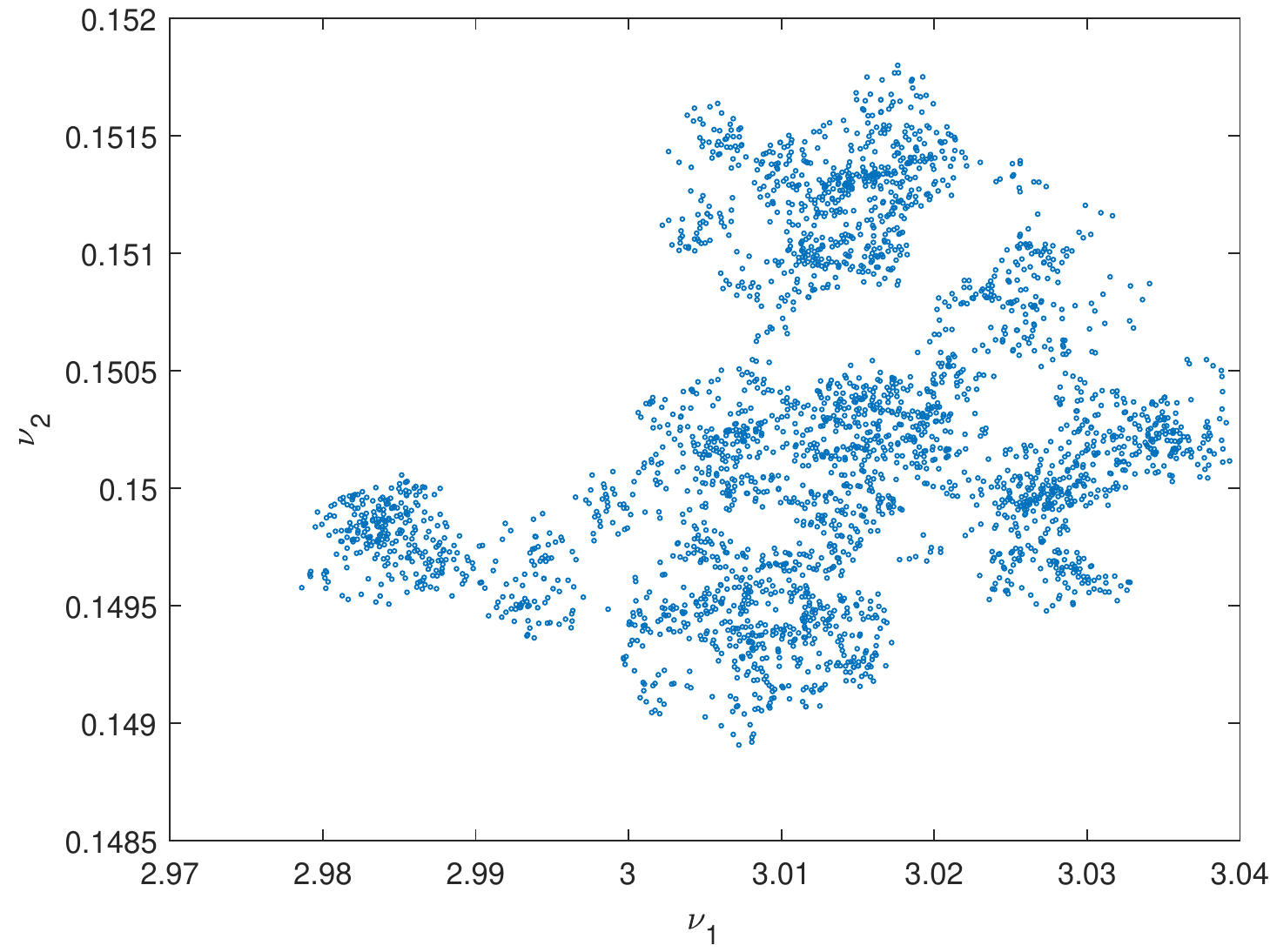}
  }
  \subfigure[$95\%$ CI from MALG]{
    \includegraphics[width=.3\textwidth]{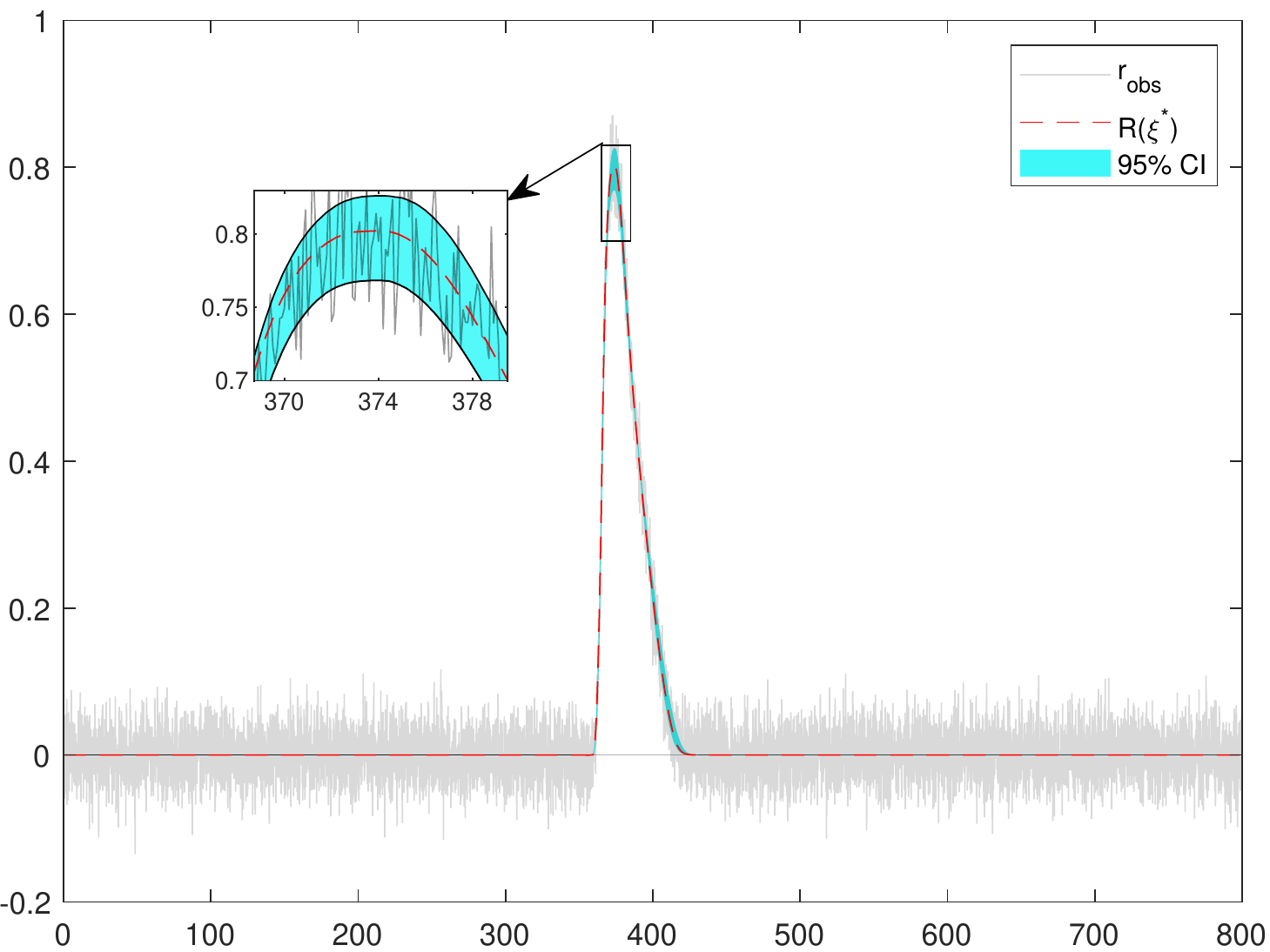}
    \label{fig:RealData:MALG_CI}
  }\\
  \subfigure[$\widetilde{\vec \eta}$ from MALG]{
    \includegraphics[width=.3\textwidth]{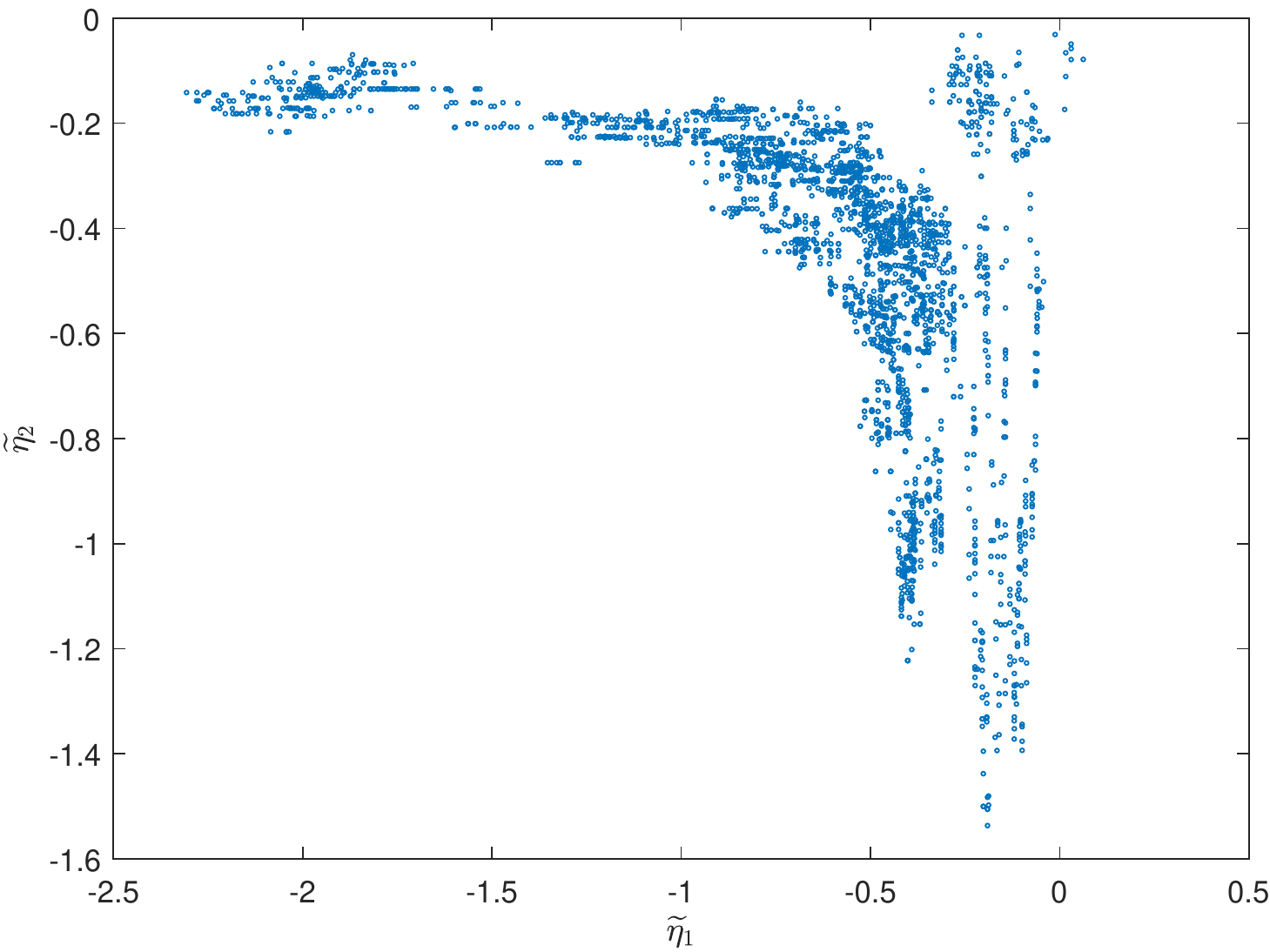}
  }
  \subfigure[$\widetilde{\vec \nu}$ from MALG]{
    \includegraphics[width=.3\textwidth]{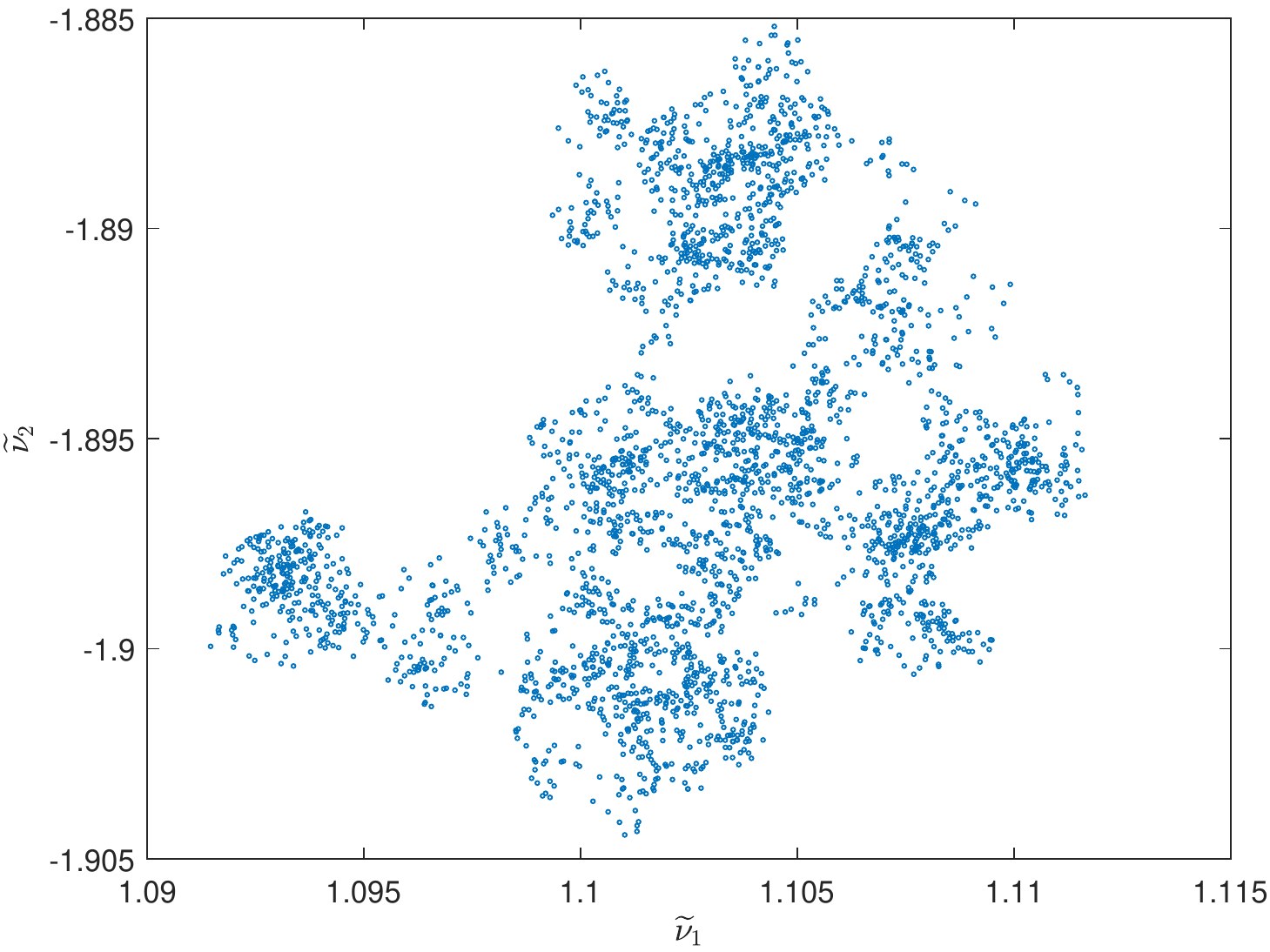}
  }
  \subfigure[$\sigma^2_\epsilon$ from MALG]{
    \includegraphics[width=.3\textwidth]{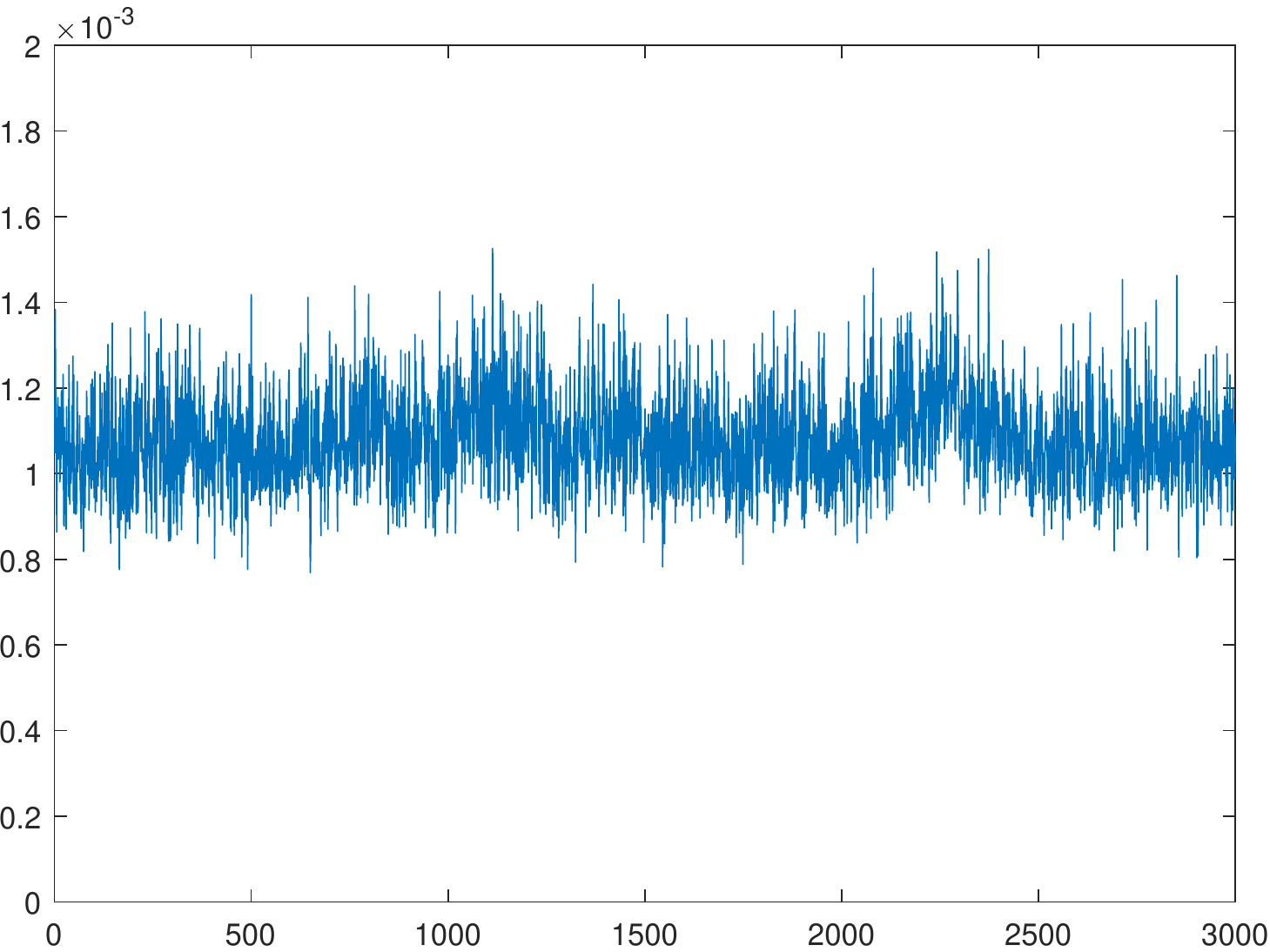}
  }
  \caption{Summary plots of MGDG and MALG with sample size $K = 3000$ in the real data case. (a,d,e,g,h) Scatter plots of variables of interest from the corresponding algorithm. (b,i) Sample trace of $\sigma_\epsilon^2$ from MGDG and MALG. (c,f) The 95\% credible interval (light blue shaded part) from MGDG and MALG, with the input data $\vec r_{obs}$ is drawn in grey solid line and the clean truth $\vec R(\vec \xi^*)$ is drawn as a dotted red line.}
  \label{fig:RealData:Result}
\end{figure}

The sampling result is summarized in Fig. \ref{fig:RealData:Result}, in which the $\vec\eta$ sampled with MGDG is distributed in a wide range, while the samples from MALG are distributed with some pattern and has a higher rejection rate. The acceptance rate is due to the design of the algorithm - a new proposal of $\vec\eta$ is more easily accepted in one MGDG iteration, as we chose an optimal $\vec \nu$ for it. From the samples, an empirical 95\% credible interval (CI) can be constructed from the quantile of $R(\vec\xi^{(i)},t)$ for each $t\in\cT$. Fig. \ref{fig:RealData:MMHGD_CI}\subref{fig:RealData:MALG_CI} present this 95\% CI from one trial of MGDG and MALG, in light blue. The CIs are so narrow that they almost coincide with the clean truth $\vec R(\vec \xi^*)$. Since there are multiple solutions, these samples are evaluated by the relative error between the $\vec R(\vec{\hat \xi})$ and $\vec R(\vec{\xi^*})$, that is,
\begin{equation*}
  RE(\vec{\hat \xi})=\frac {\left\|\vec R(\vec{\hat \xi})-\vec R(\vec{\xi^*}) \right\|_2}{\left\|\vec R(\vec{\xi^*}) \right\|_2},
\end{equation*}
where the lower and upper bound of the 95\% CI in Fig. \ref{fig:RealData:MMHGD_CI} have a relative error of approximately $3.21\%$ and $2.65\%$, while $\vec\robs$ itself has a relative error of approximately $26\%$. Moreover, the two traces of $\sigma_\epsilon^2$ are stationary distributed around $\widehat{\sigma}^{2}_\epsilon = 0.001$. This, on the other hand, confirms the validity of our algorithm with real data. 

\begin{table}[htbp]
  \caption{Comparison of sample means and maximum relative error of 95\% credible interval from MGDG and MALG. Standard deviations computed based on 20 repetitions are shown in parentheses.}
  \label{tab:RealData:RepeatedResult}
  \begin{tabular}{lccccc}
  \hline
& $\bar\eta_1$   & $\bar\eta_2$   & $\bar\nu_1$    & $\bar\nu_2$    & \begin{tabular}[c]{@{}c@{}}max RE within\\ 95\% CI\end{tabular} \\ \hline
  MGDG & 0.4341(0.0944) & 0.4509(0.1313) & 2.9936(0.0000) & 0.1515(0.0000) & 0.0310(0.0020) \\
  MALG & 0.4144(0.1041) & 0.3348(0.1970) & 3.0012(0.0225) & 0.1510(0.0020) & 0.0676(0.0290) \\\hline
  \end{tabular}
  \end{table}

To validate the robustness of our algorithms, the sampling was repeated 20 times, and the result is summarized in Table \ref{tab:RealData:RepeatedResult}. The large standard deviations of $\vec \eta$ can be attributed to multiple solutions from the time-dependent convection-diffusion system in (\ref{eq:solver}). In contrast, the distribution of $\vec \nu$ is very concentrated, and in most cases, the 95\% CIs are very close to $\vec R(\vec \xi^*)$. Overall, these experiments indicate that our algorithms can robustly infer $\vec \xi^*$ based on the observations $\vec\robs$.

\section{Conclusion}
The primary objective of this study was to develop a probabilistic model and estimate the adsorption-isotherm parameters in gradient-elution preparative liquid chromatography from a statistical perspective.

With the aim of estimating the adsorption-isotherm parameters reliably, a statistical observing model with spatial-correlation noise was proposed. Because the estimation was affected by the correlation between the parameters in the preliminary experiment, we designed two modified MCMC algorithms to reduce this effect. These algorithms were verified on several numerical solvers with highly correlated parameters, and they were able to produce approximately normally distributed estimators with acceptable bias. The verification indicates that our method is reliable with correlated parameters. The real data application revealed that the estimation of the parameters is not unique and that our method leads to reasonable results.

Our method has major implications for estimating parameters that are correlated among their elements. On one hand, these algorithms avoid the limitation of the local optimal solution, but on the other hand, they do not require the objective function to be derivable for all variables. The combination of random sampling and optimization methods expands their scope of application. Under certain conditions, the performance of the proposed approach is guaranteed theoretically.

It should be noted that the estimation procedure here has two limitations. First, the gradient of the objective function is calculated numerically. Although it does not consume too much time when implemented in parallel, thousands of repetitions of gradient descent embedded in sampling iterations still require enormous computing resources. Second, the proposed approach can only result in a group of possible estimators when the real data are processed, and we cannot make further evaluations of these results. Because the signals deduced from these estimates are very close to the initial observations, they have almost the same likelihood from a statistical point of view. It is possible that further research on the forward model could reduce the demand for computing resources and make these parameters identifiable. With such a forward model, the efficiency and accuracy of the proposed approach could be further improved.

Overall, a reliable estimation of the adsorption isotherm requires dealing with the correlation efficiently and describing the signal accurately. Our method enables us to explore the combination of sampling and optimization in a more advanced form, to further estimate the adsorption-isotherm parameters.  Besides the considered chromatography application, our statistical approach has potential to be applied in other inverse problems associated with some time-dependent convection-diffusion PDEs such as the hydrogen plasma model in astrophysics (\cite{berryman1978nonlinear}), the autowave model in environmental pollution (\cite{chaikovskii2022convergence}), and the atherosclerosis inflammatory disease model in biology (\cite{hidalgo2014numerical}). We are currently looking into these extensions.

\begin{supplement}
\stitle{Supplementary material for ``A Statistical Approach to Estimating Adsorption-Isotherm Parameters in Gradient-Elution Preparative Liquid Chromatography"}
\slink[doi]{COMPLETED BY THE TYPESETTER}
\sdatatype{.pdf}
\sdescription{We include all materials omitted from the main text.}
\end{supplement}

\section*{Acknowledgements} The authors would also like to thank the Fundamental Separation Science Group (FSSG) under the supervision of Professor Torgny Fornstedt at Karlstad University, Sweden for providing the real data (cyclohexanone and cycloheptanone). Z.Y. would like to thank the Singapore MOE Tier 1 grants R-155-000- 196-114, A-0004826-00-00 and Tier 2 grant A-0008520-00-00 at the National University of Singapore. C.L. would like to thank the Singapore MOE Tier 1 grant A-0004822-00-00 at the National University of Singapore

\bibliographystyle{imsart-nameyear} 
\bibliography{bibliography}       


\end{document}